*Review*

# Observational Constraints on Dynamical Dark Energy Models

Olga Avsajanishvili [1,2,]*, Gennady Y. Chitov [3,4], Tina Kahniashvili [1,2,5], Sayan Mandal [2,5] and Lado Samushia [1,2,6]

[1] E.Kharadze Georgian National Astrophysical Observatory, 47/57 Kostava St., Tbilisi 0179, Georgia;
[2] School of Natural Sciences and Medicine, Ilia State University, 3/5 Cholokashvili Ave., Tbilisi 0162, Georgia;
[3] Bogoliubov Laboratory of Theoretical Physics, Joint Institute for Nuclear Research, Dubna 141980, Russia;
[4] Département de physique, Université de Sherbrooke, Sherbrooke, Québec, J1K 2R1 Canada;
[5] McWilliams Center for Cosmology and Department of Physics, Carnegie Mellon University, Pittsburgh, PA 15213, USA;
[6] Department of Physics, Kansas State University, 116 Cardwell Hall, Manhattan, KS 66506, USA;
olga.avsajanishvili@iliauni.edu.ge
gchitov@theor.jinr.ru
tinatin@andrew.cmu.edu
sayanm@andrew.cmu.edu
ladosamushiaoffice@gmail.com
[*] Correspondence: olga.avsajanishvili@iliauni.edu.ge

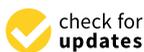

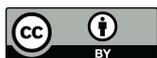

**Abstract:** Scalar field $\phi$CDM models provide an alternative to the standard $\Lambda$CDM paradigm, while being physically better motivated. Dynamical scalar field $\phi$CDM models are divided into two classes: the quintessence (minimally and non-minimally interacting with gravity) and phantom models. These models explain the phenomenology of late-time dark energy. In these models, energy density and pressure are time-dependent functions under the assumption that the scalar field is described by the ideal barotropic fluid model. As a consequence of this, the equation of state parameter of the $\phi$CDM models is also a time-dependent function. The interaction between dark energy and dark matter, namely their transformation into each other, is considered in the interacting dark energy models. The evolution of the universe from the inflationary epoch to the present dark energy epoch is investigated in quintessential inflation models, in which a single scalar field plays a role of both the inflaton field at the inflationary epoch and of the quintessence scalar field at the present epoch. We start with an overview of the motivation behind these classes of models, the basic mathematical formalism, and the different classes of models. We then present a compilation of recent results of applying different observational probes to constraining $\phi$CDM model parameters. Over the last two decades, the precision of observational data has increased immensely, leading to ever tighter constraints. A combination of the recent measurements favors the spatially flat $\Lambda$CDM model but a large class of $\phi$CDM models is still not ruled out.

## 1. Introduction

The accelerated expansion of our universe was first discovered in 1998 on the basis of measurements of the type Ia supernovae (SNe Ia) apparent magnitudes [1–3]. This fact was later confirmed by other cosmological observations, in particular, by measurements of the temperature anisotropy and the polarization in the cosmic microwave background (CMB) radiation [4–13], by studies of the large-scale structure (LSS) of the universe [14–19], by measurements of baryon acoustic oscillations (BAO) peak length scale [20–26], and by measurements of Hubble parameter [27–35].

One possible explanation for this empirical fact is that the energy density of the universe is dominated by dark energy or dark fluid, an energy component with an effective negative pressure (see Refs. [36–43] for reviews). The presence of dark matter in the universe, first discovered through the anomalously high rotation velocity of the outer regions of galaxies [44], is another major mystery of modern cosmology. Different models for dark matter have been proposed including cold dark matter (CDM), consisting of heavy particles with mass $m_{\text{CDM}} \geq 100$ KeV, warm dark matter (WDM), composed of particles with a mass of $m_{\text{WDM}} \approx 3$–30 KeV, and hot dark matter (HDM) consisting of ultrarelativistic particles [45]. Assuming general relativity is the correct description of gravity on cosmological scales, about 95% of the energy in the universe has to be in the "dark" form, i.e., in the form of dark energy and dark matter, to explain available observations. According to the last Planck data release (PR4), our universe consists of 4.86% of ordinary matter, 25.95% of dark matter, and 70.39% of dark energy [46].

The true nature and origin of dark energy and dark matter are still unresolved issues of modern cosmology. The simplest description of dark energy is vacuum energy or the cosmological constant $\Lambda$ (see Refs. [36,39,47] for reviews). The cosmological model based on such a description of dark energy in the spatially flat universe is called the Lambda Cold Dark Matter ($\Lambda$CDM) model, which established itself as the standard or concordance model of the universe in the last two decades (see Ref. [36] for a pioneering work and Refs. [48] for recent review and discussion). In this model, dark matter is presented in the form of non-relativistic cold weakly interacting particles which either have never been in equilibrium with the primordial plasma or have been decoupled from it after becoming non-relativistic at an early stage. A good pedagogical overview of the $\Lambda$CDM model is available in many recent books [49–53] and reviews [37,54–58].

Despite explaining various observations of our universe to a remarkable degree of accuracy, the $\Lambda$CDM model has several unsolved problems and tensions [57,59–65], including the fine-tuning or cosmological constant problem, the coincidence problem, the Hubble and $S_8$ tensions, and the problem of the shape of the universe. A large number of cosmological models that go beyond the standard $\Lambda$CDM scenario with modified dynamics of the expansion of the universe both in early and late times have been considered in order to resolve these tensions. For reviews, see [66–76]. To solve the problems of the $\Lambda$CDM model, models with gravity different from general relativity on cosmological scales in the universe, so-called modified gravity (MG) models, have also been proposed [37,77–88]. For reviews, see [43,89,90] and especially a comprehensive analysis by Ishak [91] on a large class of the MG theories leading to the accelerated universe and the observational constraints on those theories.

The value of the energy density of the cosmological constant $\rho_\Lambda$ following from the quantum field theory estimates is [60] $\rho_\Lambda \sim \hbar M_{\text{pl}}^4 \sim 10^{72}$ Gev$^4 \sim 2 \cdot 10^{110}$ erg/cm$^3$, where $M_{\text{pl}} \sim 10^{19}$ GeV is the Planck mass and $\hbar$ is the reduced Planck constant, while cosmological observations of the cosmological constant (like dark energy) show a very different result [60]: $\rho_\Lambda^{\text{obs}} \sim 10^{-48}$ Gev$^4 \sim 2 \cdot 10^{-10}$ erg/cm$^3$. This discrepancy in 120 orders of magnitude between the predicted and observed values of the energy scale of the cosmological constant is called the cosmological constant problem or the fine-tuning problem [54–56,92,93]. An alternative point of view compatible with Einstein's equations of general relativity is to abandon the attempts to explain the minuscule value of the cosmological constant due to some "magic" cancellations of the quantum field theory vacuum terms and to assume its pure geometric origin. The drawback is that in such case the trivial space–time without sources would be the de Sitter universe with an intrinsic curvature [93].

The coincidence is that, based on precise cosmological observations [13,94], the energy density in dark energy (68.7%) is comparable (within an order of magnitude) to that of non-relativistic matter (31.3%) at present. This problem can also be presented as the why now problem, namely: "Why did the acceleration occur in the present epoch of cosmic evolution?" (surely any earlier event would have prevented the formation of structures in the universe) [36,57,59,61]. This fact is an enigma [52,56,93,95–101], because, in the $\Lambda$CDM model, the energy density of the cosmological constant does not depend on time, $\rho_\Lambda = $ const, while the energy density of matter varies over time as $\rho_{\text{DM}} \sim a^{-3}(t)$ ($a(t)$ and $t$ are the scale factor and cosmic time, respectively), so the ratio of these quantities is time-dependent, $\rho_{\text{DM}}/\rho_\Lambda \propto a^{-3}(t)$. Since the vacuum energy does not change over time, it was insignificant during both the radiation domination epoch and the matter domination epoch, but it has become the dominant component recently, at a scale factor $a \approx 0.77$ (or a redshift $z \approx 0.3$), according to Planck 2018 data [13], and it will be the only component in the universe in the future. The energy density of matter and the energy density of the cosmological constant are comparable for a very short period of time, so the following question arises: "Why did it happen that we live in this short (by the cosmological scale) period of time?" After all, this fact is in contradiction to the Copernican principle, since this coincidence implies that the present epoch is a special time, between the matter- and dark-energy-dominated epochs, and may hint at some physical mechanism at play which ensures these energy densities are similar.

The anthropic principle [102,103] can explain the cosmological constant problems and answer the questions: "Why is the energy density of the cosmological constant so small?" and "Why has the accelerated expansion of the universe started recently?" According to the anthropic principle, the energy density of the cosmological constant observed today should be suitable for the evolution of intelligent beings in the universe [92,104–106]. For a more detailed discussion and approaches to solve this problem, see Ref. [107].

The Hubble tension problem is that there is a discrepancy at the level of $\sim 5\sigma$ between the value of the Hubble parameter at the present epoch $H_0 = 100h$ km c$^{-1}$ Mpc$^{-1}$, where $h$ is a dimensionless normalized Hubble constant, obtained by the local measurements, and CMB temperature, polarization, and lensing anisotropy data [13,108–113]. In particular, supernova measurements give $H_0 = 73.04 \pm 1.04$ km/s/Mpc [114], while CMB measurements (TT,TE,EE + lowE + lensing) lead to $H_0 = 67.36 \pm 0.54$ km/s/Mpc [13].

The $S_8$ tension problem is that there is a discrepancy at the level of $\sim 2\sigma$ - $3\sigma$ confidence level between the primary CMB temperature anisotropy measurements by the Planck satellite [13] in the strength of matter clustering compared to lower redshift measurements such as the weak gravitational lensing and galaxy clustering [66,68,115–118]. This tension is quantified using the weighted amplitude of the matter fluctuation parameter $S_8 = \sigma_8 (\Omega_\mathrm{m}/0.3)^{0.5}$, which modulates the amplitude of weak lensing measurements; here, $\sigma_8$ is an amplitude of mass fluctuations on scales of $8h^{-1}$ Mpc; $\Omega_\mathrm{m} = \Omega_\mathrm{m0} a^{-3}/E^2(a)$ is a matter density parameter; $\Omega_\mathrm{m0}$ is a matter density parameter at the present epoch; $E(a) = H(a)/H_0$ is a normalized Hubble parameter; $H(a) = \dot{a}/a$ is a Hubble parameter; and $\dot{a}$ is a derivative of the scale factor $a$ with respect to cosmic time.

The problem with the shape of the universe is that the CMB anisotropy power spectra measured by the Planck space telescope show a preference for a spatially closed universe at a more than $3\sigma$ confidence level [13,119]. This fact contradicts expectations from the simplest inflationary models [63,66,120] and is interpreted by the cosmological community as a possible crisis of modern cosmology [120–125].

One of the alternatives to the $\Lambda$CDM model, during the period of time when the accelerated expansion of the universe is governed by the cosmological constant $\Lambda$, is dynamical dark energy scalar field $\phi$CDM models [126–134]. In these models, dark energy is described through the equation of state (EoS) parameter, $w_\phi(t)$, which depends on time: $w_\phi(t) \equiv p_\phi/\rho_\phi$, $p_\phi$ is the scalar field pressure, $\rho_\phi$ is the scalar field energy density; whereas, in the $\Lambda$CDM model, the EoS parameter is a constant, $w_\Lambda = -1$. At the same time, at the present epoch, the value of the time-dependent EoS parameter in scalar field models becomes approximately equal to minus one, $w_\phi \approx -1$; thus, dynamical dark energy mimics the cosmological constant and becomes almost indistinguishable from it. However, dark energy is a dynamic parameter related to the current value of the scalar field potential, while the universe evolves towards its true vacuum with zero energy, i.e., the zero cosmological term.

Depending on the value of the EoS parameter at the present epoch, $\phi$CDM scalar field models are divided into quintessence models, with $-1 < w_\phi(t) < -1/3$ [95,98, 135–138], see, e.g., Refs. [36,43] (for a review), and phantom models, with $w_\phi(t) < -1$ [139–145]. Quintessence models are divided into two classes: tracker (freezing) models, in which the scalar field evolves slower than the Hubble expansion rate, and thawing models, in which the scalar field evolves faster than the Hubble expansion rate [95,136, 142,146,147]. In quintessence tracker models, the energy density of the scalar field first tracks the radiation energy density and then the matter energy density, while it remains subdominant [148]. Only recently does the scalar field become dominant and begins to behave as a component with negative pressure, which leads to the accelerated expansion of the universe [137,149,150]. For certain forms of potentials, the quintessence tracker models have an attractor solution that is insensitive to initial conditions [148].

The interaction between dark energy and dark matter, namely their transformation into each other, is considered in the interacting dark energy (IDE) models [151–155]. In these models, the coincidence problem of the standard $\Lambda$CDM model as well as the Hubble constant $H_0$ tension can be alleviated [67,70,73,156–161].

In the standard $\Lambda$CDM cosmological scenario, one assumes the existence of two epochs of accelerated expansion in the universe. The first is inflation [162–170], which happens in the very early universe, and the second is the dark-energy-dominated epoch observed today [50,51,171,172]. Inflation is a theory of the exponential expansion of space in the early universe, which is believed to have lasted approximately from $10^{-36}$ to $10^{-33}$–$10^{-32}$ seconds after the Big Bang, the exact times being dependent on the microphysics of the model describing inflation. The inflationary models explain the quantum origin of tiny primordial density fluctuations in the universe, which must have been present at very early epochs, as the seeds both for the CMB anisotropies and for the structure formation in the later evolution of the universe. The exponential expansion during inflation comes to an end when a phase transition transforms the vacuum energy into radiation and matter, after which the radiation-dominated epoch begins. This phase transition is called the reheating and its governing dynamics is still debated. A successful inflationary model requires a smooth transition to the decelerated epoch (in which inflation rules the universe as if it were dominated by non-relativistic matter) because, otherwise, the homogeneity of the universe would be violated [173,174]. Inflation resolves several problems in cosmology, namely, the horizon problem, associated with the lack of causal relationship between different regions in the early universe before the recombination epoch (this is an epoch of forming the electrically neutral hydrogen atoms, which began at $t_{\rm rec} \approx 350{,}000$ years after the Big Bang), and the flatness problem, related to the fine tuning of the spatial flatness of the universe in the early epoch so that the spatial flatness of the universe is preserved at the present epoch). The evolution of the universe from the inflationary epoch to the present dark energy epoch is investigated in quintessential inflation models too [175–180]. In these models, a single scalar field plays a role of both the inflaton field at the inflationary epoch and of the quintessence scalar field at the present epoch; thereby, the origin of dark energy at the present epoch is also explained within the same model.

The running vacuum models (RVMs) describe dark energy as a quantum vacuum, the energy density of which slowly evolves with the expansion of the universe [181]. RVM models, like the scalar field $\phi$CDM models, are associated with scalar fields, but describe dark energy as a quantum vacuum not just the vacuum of a classical scalar field [182]. The EoS parameter of the running vacuum is moderately dynamic in the late universe, $w_{\rm vac} \geqslant -1$, mimicking the quintessence scalar fields [183]. In contrast to classical scalar fields which depend on an arbitrary potential, running cosmic vacuum arises from quantum effects and can be derived from explicit calculations of quantum field theory (QFT) both in the spatially flat and non-flat hypersurfaces (see Ref. [184] for reviews). The latest cosmological data are in good agreement with the RVMs [185,186], while confirming the results of earlier constraints of these models [187–189]. The cosmological constant problem of the $\Lambda$CDM model can be resolved in the RVMs [190,191], and the $H_0$ and $\sigma_8$ tensions weaken in these models, as can be seen from the data constraints presented in [185,186].

It has also been suggested that the current accelerated expansion of the universe can be explained by modifications of the general theory of relativity. Several such modifications of general relativity have been proposed, see Refs. [88–90,192], to explain a host of cosmological observations. Although the current observational constraints are still too large to exclude some of the MG theories [91], it seems to be premature at this point to consider such theories as a comprehensive and viable alternative to the minimal model of dark matter and/or dark energy based on Einstein's general relativity, in order to explain the observations. We will therefore not discuss theories of modified gravity in our review.

In this paper we reviewed and analyzed, to the best of our knowledge of current literature, the most relevant studies of the observational constraints on dynamical dark energy models over the past twenty years, in particular, scalar field $\phi$CDM models, quintessential inflation scalar field $\phi$CDM models, and IDE models both in the spatially flat and non-flat hypersurfaces. The research effort on the complication, refinement of cosmological data, and the increase in the variety of methods for studying dynamical dark energy models lead to more accurate constraints on the values of the cosmological parameters in these models. Despite the refinement of various observational data and the complication of methods for studying dark energy in the universe, current observational data still favor the standard spatially flat $\Lambda$CDM model, while not excluding dynamical dark energy models and spatially closed hyperspaces [193–207]. At the same time, recent studies showed that the currently available observational datasets favor the IDE model at a more than $2\sigma$ confidence level [67,159,160,208,209].

This paper is organized as follows: the different cosmological dynamical dark energy models are described in Section 2, observational constraints on dynamical dark energy models by various observational data are presented in Section 3, the main results are summarized in Section 4, the ongoing and upcoming missions are listed in Section 5, and conclusions are presented in Section 6. In this paper, we used the natural system of units: $c = \hbar = k_B = 1$.

## 2. Cosmological Dark Energy Models
### 2.1. $\Lambda$CDM Model

As highlighted in the Introduction, the Lambda Cold Dark Matter ($\Lambda$CDM) model is the standard or concordance model of a spatially flat universe. In this model, dark energy is represented by the cosmological constant $\Lambda$, and its energy density is constant

$$\rho_\Lambda = \frac{\Lambda M_{\text{pl}}^2}{8\pi} = \rho_{\text{vac}} = \text{const}, \quad (1)$$

where $\Lambda = 4.33 \cdot 10^{-66}$ eV$^2$. The pressure and the energy density in the $\Lambda$CDM model are related as

$$p_\Lambda = -\rho_\Lambda = \text{const}, \quad (2)$$

leading to the constant EoS parameter

$$w_\Lambda = -1. \quad (3)$$

The action with the cosmological constant $\Lambda$ is [43]

$$S = -\frac{M_{\text{pl}}^2}{16\pi} \int d^4x \sqrt{-g}(R + 2\Lambda) + S_M, \quad (4)$$

where $g \equiv \det(g_{\mu\nu})$ is the determinant of the metric tensor $g_{\mu\nu}$, $R$ is the Ricci scalar, and $S_M$ is the action of the matter. The spatially flat $\Lambda$CDM model is typically characterized by six independent parameter [210]: the physical baryon density parameter, $\Omega_b h^2$; the dark matter physical density parameter, $\Omega_c h^2$; the age of the universe, $t_0$; the scalar spectral index, $n_s$; the amplitude of the curvature fluctuations, $\Delta_R^2$; and the optical depth during the reionization period for $z \in (6, 20)$, $\tau$. In addition to these parameters, the $\Lambda$CDM model is described by six extended fixed parameters: the total density parameter, $\Omega_{\text{tot}}$; the EoS parameter, $w_\Lambda$; the total mass of three types of neutrinos, $\sum m_\nu$; the effective number of the relativistic degrees of freedom, $N_{\text{eff}}$; the tensor/scalar ratio, $r$; and the scalar spectral index running, $a_s$.

The extension of the spatially flat ΛCDM model to spatially non-flat hypersurfaces is the *o*CDM model. The first Friedmann equation describing the evolution of the universe for the spatially non-flat *o*CDM model (for $\Omega_{k0} = 0$ this equation is applicable for the spatially flat ΛCDM model) has a form

$$E(a) = (\Omega_{r0}a^{-4} + \Omega_{m0}a^{-3} + \Omega_{k0}a^{-2} + \Omega_{\Lambda})^{1/2}, \tag{5}$$

where $\Omega_{r0} = \rho_{r0}/\rho_{cr}$, $\Omega_{m0} = \rho_{m0}/\rho_{cr}$, and $\Omega_{\Lambda} = \rho_{\Lambda}/\rho_{cr}$ are density parameters at the present epoch for radiation, matter, and vacuum, respectively, where $\rho_{r0}$ and $\rho_{m0}$ are energy densities for radiation and matter at the present epoch, respectively. The value of the critical energy density at the present epoch is equal to $\rho_{cr} = 3M_{pl}^2 H_0^2/8\pi = 1.8791 h^2 \cdot 10^{-29}$ g cm$^{-3}$; $\Omega_{k0} = -k/H_0^2$ is a spatial curvature density parameter at the present epoch, $\rho_{k0} = -3k M_{pl}^2/8\pi$ is a spatial curvature density, k is a curvature parameter, and $E(a) \equiv H(a)/H_0$ is a normalized Hubble parameter.

Observational constraints on the cosmological parameters $\Omega_{m0}$ and $\Omega_{\Lambda}$, obtained from different cosmological datasets for the ΛCDM model and for the *o*CDM model, are represented in Figure 1.

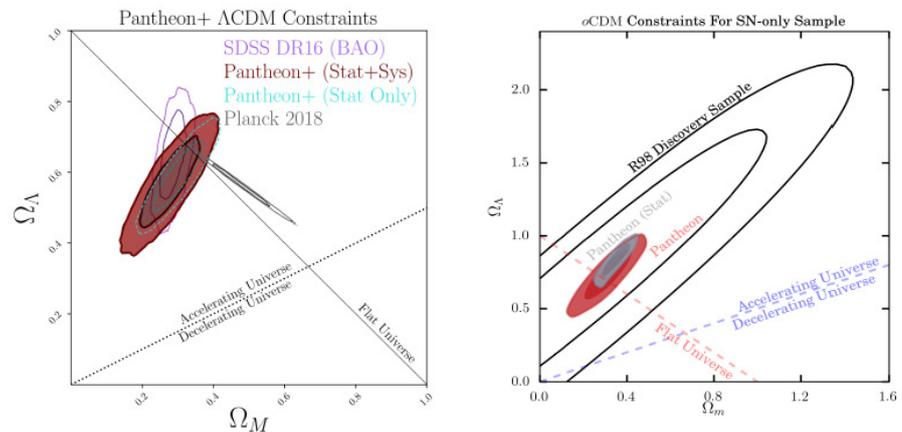

**Figure 1.** The 1 $\sigma$ and 2$\sigma$ confidence level contour constraints on $\Omega_m$ and $\Omega_\Lambda$ parameters. (Left panel) In the standard spatially flat ΛCDM model from the SNe Ia Pantheon dataset, as well as from the combined BAO peak length scale and Planck datasets. (The figure is adapted from [211].) (Right panel) In the spatially non-flat *o*CDM model using discovery sample of Riess et al. [212] and the full Pantheon sample of Scolnic et al. [213]. Pantheon constraints with systematic uncertainties are presented in red, while only statistical uncertainties are denoted in gray. (The figure is adapted from [213].)

As we mentioned above, the ΛCDM model is the fiducial model against which all alternative models are compared regarding their fit to observational data. Its predictions agree with the observational data pertaining to the accelerated expansion of the universe, the statistical distribution of LSS, the CMB temperature and polarization anisotropies, and the abundance of light elements in the universe [94,214].

*2.2. Dynamical Dark Energy Scalar Field ϕCDM Models*

There are numerous physically motivated alternative models for the ΛCDM model [37,43,77–87]. One of the prominent alternatives to the ΛCDM model are the dynamical scalar field *ϕ*CDM models, in which the scalar field can interact with gravity both minimally [36,127,128] and non-minimally via different coupling terms (the so-called extended scalar–tensor models) [215–223]. We will concentrate on the minimally coupled models as the simplest and more natural choice.

In models with minimal interaction with gravity, the role of dark energy is played by a slowly varying uniform self-interacting scalar field $\phi$. These $\phi$CDM models involving a dynamical scalar field do not suffer from the fine-tuning problem of the $\Lambda$CDM model, and have a more natural explanation for the observed low-energy scale of dark energy. When the energy density of the scalar field begins to dominate over the energy density of both radiation and matter, the universe begins the stage of the accelerated expansion. At early times during the evolution of the universe, the behavior of the dynamical scalar field is different from that of the cosmological constant $\Lambda$, but is almost indistinguishable from that of the cosmological constant during later times.

The dynamical scalar field $\phi$CDM models are divided into two classes: the quintessence models [148] and phantom models [139,224]. These two classes of models differ from each other by the following attributes:

(i) The EoS parameter—For quintessence fields, $-1 < w_\phi < -1/3$, while for phantom fields, $w_\phi < -1$.
(ii) The sign of the kinetic term—For quintessence fields, the kinetic term in the Lagrangian has a positive sign, while it is negative for phantom fields.
(iii) The dynamics of the scalar fiel—The quintessence field rolls gradually to the minimum of its potential, while the phantom field rolls to the maximum of its potential.
(iv) Temporal evolution of dark energy—For quintessence fields, the dark energy density remains almost unchanging with time, while it increases for phantom fields.
(v) Forecasting the future of the universe—The quintessence models predict either an eternal expansion of the universe or a repeated collapse, depending on the spatial curvature of the universe. On the other hand, the phantom models predict the destruction of any gravitationally related structures in the universe. Depending on the asymptotic behavior of the Hubble parameter $H(t)$, the future scenarios of the universe are divided into a *big rip* for which $H(t) \to \infty$ for a finite future time $t = \mathrm{const}$; a little rip for which $H(t) \to \infty$ at an infinite future time $t \to \infty$, and a pseudo rip for which $H(t) \to \mathrm{const}$ for an infinite future time $t \to \infty$.

The action describing a scalar field in the presence of matter is given by

$$S = \int d^4x \sqrt{-g} \left( -\frac{M_{\mathrm{pl}}^2}{16\pi} R + \mathcal{L}_\phi \right) + S_{\mathrm{M}}, \tag{6}$$

where $\mathcal{L}_\phi$ is the Lagrangian density of the scalar field, the form of which depends on the type of the chosen model. We describe the form of $\mathcal{L}_\phi$ for the quintessence and the phantom fields below.

2.2.1. Quintessence Scalar Field

The dynamics of the quintessence scalar field is described by the Lagrangian density

$$\mathcal{L}_\phi = \frac{1}{2} g^{\mu\nu} \partial_\mu \phi \, \partial_\nu \phi - V(\phi), \tag{7}$$

where $V(\phi)$ is a scalar field potential. There are various quintessence potentials discussed in the literature but there is currently no observational constraint to prefer one of these over the others. A list of some of the quintessence potentials is presented in Table 1.

The EoS parameter for the quintessence scalar field is given by

$$w_\phi \equiv \frac{p_\phi}{\rho_\phi} = \frac{\dot\phi^2/2 - V(\phi)}{\dot\phi^2/2 + V(\phi)}, \tag{8}$$

where $p_\phi = \frac{1}{2}\dot\phi^2 - V(\phi)$ and $\rho_\phi = \frac{1}{2}\dot\phi^2 + V(\phi)$ are, respectively, the pressure and energy density of the quintessence field. Here, the overdots denote derivatives with respect to the cosmic time $t$. The equation of motion for the quintessence scalar field can be obtained by varying the action in Equation (6), along with the Lagrangian in Equation (7),

$$\ddot{\phi} + 3H\dot{\phi} + V'(\phi) = 0, \tag{9}$$

with the prime denoting a derivative with respect to the scalar field $\phi$. The first Friedmann's equation for a $\phi$CDM model in a spatially non-flat spacetime has the form

$$E(a) = (\Omega_{\rm r0}a^{-4} + \Omega_{\rm m0}a^{-3} + \Omega_{\rm k0}a^{-2} + \Omega_\phi(a))^{1/2}, \tag{10}$$

where $\Omega_\phi(a)$ is the dark energy (scalar field) density parameter.

Depending on the shape of potentials, quintessence models are further subdivided into thawing models and freezing (tracking) models [43,136]. In the $w_\phi - dw_\phi/d\ln a$ plane, thawing and freezing scalar field models are located in strictly designated zones for each of them [136], see Figure (2) (Left panel).

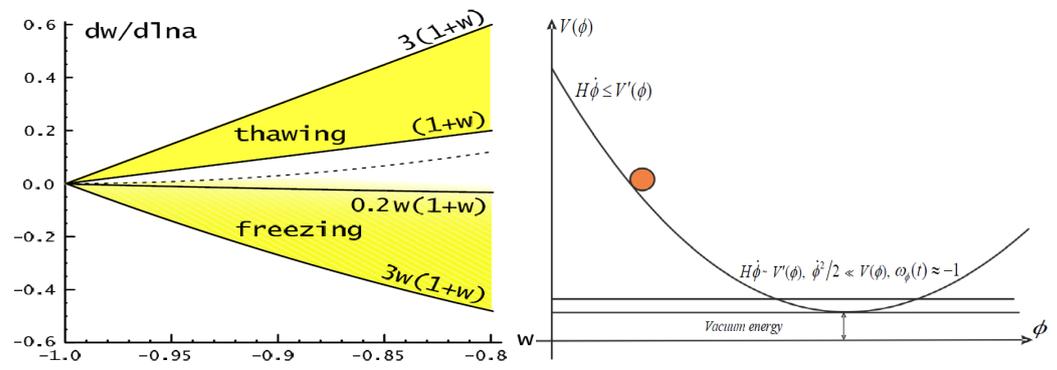

**Figure 2.** (Left panel) The location of thawing and freezing scalar fields in the $w_\phi - dw_\phi/d\ln a$ plane. (The figure is adapted from [136].) (Right panel) Regimes of the quick rolling down and slow rolling down of the scalar field to the minimum of its potential.

(a) In the thawing models, the scalar field was too suppressed by the retarding effect of the Hubble expansion, represented by the term $3H\dot{\phi}$ in Equation (9), until recently. This results in a much slower evolution of the scalar field compared to the Hubble expansion and the thawing scalar field manifests itself as the vacuum energy, with the EoS parameter $w_\phi \sim -1$. The Hubble expansion rate $H(t)$ decreases with time and, after it falls below $\sqrt{V''(\phi)}$, the scalar field begins to roll to the minimum of its potential, see Figure 2 (Right panel). The value of the EoS parameter for the scalar field thus increases over the time and becomes $w_\phi > -1$.

(b) In the freezing models, the scalar field is always suppressed (it is damped), i.e., $H(t) > \sqrt{V''(\phi)}$. Freezing scalar field models have so-called tracking solutions. According to tracking solutions, the quintessence component tracks the background EoS parameter (radiation in the radiation-dominated epoch and matter in the matter-dominated epoch) and eventually only recently grows to dominate the energy density in the universe. This leads to the accelerated expansion of the universe at late times, since the scalar field has a negative effective pressure. The tracker behavior allows the quintessence model to be insensitive to initial conditions. But this requires fine tuning of the potential energy, since $\sqrt{V''(\phi)} \sim H_0 \sim 10^{-33}$ eV.

Table 1. Scalar field quintessence potentials.

| Name | Form | Reference |
|---|---|---|
| Ratra-Peebles | $V(\phi) \propto \phi^{-\alpha}$; ($\alpha = $ const $> 0$) | Ratra and Peebles [127] |
| Exponential | $V(\phi) \propto \exp(-\lambda\phi/M_{\rm pl})$; ($\lambda = $ const $> 0$) | Wetterich [128], Ratra and Peebles [126], Lucchin and Matarrese [169], Ferreira and Joyce [225] |
| Zlatev-Wang-Steinhardt | $V(\phi) \propto (\exp(M_{\rm pl}/\phi) - 1)$ | Zlatev et al. [148] |
| Sugra | $V(\phi) \propto \phi^{-\chi}\exp(\gamma\phi^2/M_{\rm pl}^2)$; ($\chi, \gamma = $ const $> 0$) | Brax and Martin [226] |
| Sahni-Wang | $V(\phi) \propto (\cosh(\varsigma\phi) - 1)^g$; ($\varsigma = $ const $> 0$, $g = $ const $< 1/2$) | Sahni and Wang [227] |
| Barreiro-Copeland-Nunes | $V(\phi) \propto (\exp(\nu\phi) + \exp(\upsilon\phi))$; ($\nu, \upsilon = $ const $\geq 0$) | Barreiro et al. [228] |
| Albrecht-Skordis | $V(\phi) \propto ((\phi - B)^2 + A)\exp(-\mu\phi)$; ($A, B = $ const $\geq 0$, $\mu = $ const $> 0$) | Albrecht and Skordis [229] |
| Urēna-López-Matos | $V(\phi) \propto \sinh^m(nM_{\rm pl}\phi)$; ($n = $ const $> 0$, $m = $ const $< 0$) | Urena-Lopez and Matos [230] |
| Inverse exponent potential | $V(\phi) \propto \exp(M_{\rm pl}/\phi)$ | Caldwell and Linder [136] |
| Chang-Scherrer | $V(\phi) \propto (\exp(-\tau\phi) + 1)$; ($\tau = $ const $> 0$) | Chang and Scherrer (2016) [231] |

In 1988, Ratra and Peebles introduced a tracker $\phi$CDM model comprising a scalar field with an inverse power-law potential of the form $V(\phi) = \kappa/2M_{\rm pl}^2\phi^{-\alpha}$, for a model parameter $\alpha > 0$ [126,127]. For $\alpha = 0$, this $\phi$CDM Ratra–Peebles (RP) model reduces to the $\Lambda$CDM model. The positive parameter $\kappa$ relates to the mass scale of the particles, $M_\phi$, as $M_\phi \sim (\kappa M_{\rm pl}^2/2)^{\frac{1}{\alpha+4}}$. The RP $\phi$CDM model is a typical representative of the behavior of tracker quintessence scalar field $\phi$CDM models.

2.2.2. Phantom Scalar Field

The Lagrangian density describing a phantom scalar field has the form

$$\mathcal{L}_\phi = -\frac{1}{2}g^{\mu\nu}\partial_\mu\phi\partial_\nu\phi - V(\phi), \tag{11}$$

where the negative sign of the kinetic energy term is required to ensure the dark energy EoS parameter is less than $-1$, i.e., $w_\phi < -1$, and the energy density increases over time [144]. A phantom or ghost scalar field suffers from quantum instability because its energy density is not limited from below [132]. An incomplete list of phantom potentials is given in Table 2.

Analogous to Equation (8), the EoS parameter for the phantom scalar field is given by

$$w_\phi \equiv \frac{p_\phi}{\rho_\phi} = \frac{-\dot\phi^2/2 - V(\phi)}{-\dot\phi^2/2 + V(\phi)}, \tag{12}$$

where $p_\phi = -\frac{1}{2}\dot\phi^2 - V(\phi)$ and $\rho_\phi = -\frac{1}{2}\dot\phi^2 + V(\phi)$ are, respectively, the pressure and the energy density of the phantom field. The equation of motion for the phantom scalar field has the form

$$\ddot\phi + 3H\dot\phi - V'(\phi) = 0. \tag{13}$$

**Table 2.** Scalar field phantom potentials.

| Name | Form | Reference |
| --- | --- | --- |
| Fifth power | $V(\phi) \propto \phi^5$ | Scherrer and Sen [141] |
| Inverse square power | $V(\phi) \propto \phi^{-2}$ | Scherrer and Sen [141] |
| Exponent | $V(\phi) \propto \exp(\beta\phi); (\beta = \mathrm{const} > 0)$ | Scherrer and Sen [141] |
| Quadratic | $V(\phi) \propto \phi^2$ | Dutta and Scherrer [142] |
| Gaussian | $V(\phi) \propto (1 - \exp(\phi^2/\sigma^2));$ $(\sigma = \mathrm{const})$ | Dutta and Scherrer [142] |
| Pseudo-Nambu–Goldstone boson (pNGb) | $V(\phi) \propto (1 - \cos(\phi/\kappa));$ $(\kappa = \mathrm{const} > 0)$ | Frieman et al. [232] |
| Inverse hyperbolic cosine | $V(\phi) \propto \cosh^{-1}(\psi\phi);$ $(\psi = \mathrm{const} > 0)$ | Dutta and Scherrer [142] |

*2.3. Parameterized Dark Energy Models*

2.3.1. wCdm Parameterization

In dynamical dark energy models, one can use the $w$CDM parameterization where the EoS parameter can be expressed as $p = w(a)\rho$. Dark energy models are sometimes characterized only by the EoS parameter and corresponding cosmological models are called $w$CDM models [233]. This parameterization has no physical motivation, but is commonly used as an ansatz in data analysis to quantify differences and distinguish between dark energy models. The $w$CDM parameterization in particular makes it possible to differentiate, at the present epoch, the $\Lambda$CDM model from other dark energy models.

The time-dependent EoS parameter in the $w$CDM models is often characterized by the Chevallier–Polarsky–Linder (CPL) $w_0 - w_a$ parameterization [234,235]

$$w(a) = w_0 + w_a(1 - a), \tag{14}$$

where $w_0 = w(a_0)$ and $w_a = 2dw(a)/d\ln(1+z)|_{z=1} = -2dw(a)/d\ln a|_{a=1/2}$, with $z$ being the cosmological redshift defined as $z = 1/a - 1$ and $a_0$ being the scale factor at the present time, conventionally normalized as $a_0 = 1$. Although the CPL parameterization is simple and flexible enough to accurately describe EoS parameters in most dark energy models, it cannot describe arbitrary dark energy models with good accuracy (up to a few percent) over a wide redshift range [235]. The dynamical dark energy models where the EoS parameter is expressed through the CPL parameterization are called the $w_0w_a$CDM models.

The normalized Hubble parameter for the $w_0w_a$CDM model for the spatially flat universe has the form

$$E(a) = \left[\Omega_{\mathrm{r}0}a^{-4} + \Omega_{\mathrm{m}0}a^{-3} + (1 - \Omega_{\mathrm{m}0})a^{-3(1+w_0+w_a)}e^{-3w_a(1-a)}\right]^{1/2}. \tag{15}$$

The $1\sigma$ and $2\sigma$ confidence level contour constraints on the cosmological parameters $w_0$ and $w_a$ in the $w_0w_a$CDM model from different combinations of datasets—the SNe Ia apparent magnitude (including measurements of the Hubble Space Telescope (HST)), the CMB temperature anisotropy, and the BAO peak length scale—are presented in Figure 3 (left panel).

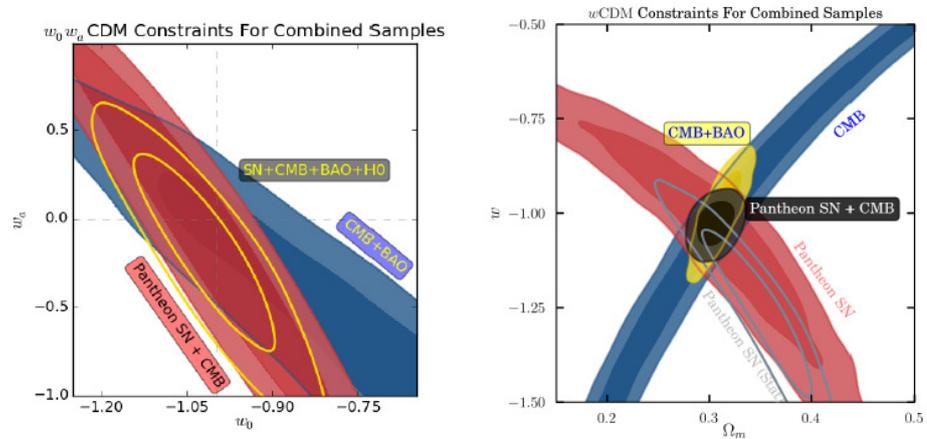

**Figure 3.** (Left panel) The $1\sigma$ and $2\sigma$ confidence level contour constraints on cosmological parameters $w_0$ and $w_a$ in the $w$CDM model from various datasets: SNe Ia apparent magnitude+CMB temperature anisotropy+BAO peak length scale+HST (yellow), BAO peak length scale+CMB temperature anisotropy (blue), SNe Ia apparent magnitude+CMB temperature anisotropy (red). (Right panel) The $1\sigma$ and $2\sigma$ confidence level contours constraints on cosmological parameters $\Omega_m$ and $w$ in the $w$CDM model from various datasets: SNe Ia apparent magnitude+CMB temperature anisotropy (black), CMB temperature anisotropy (blue), SNe Ia apparent magnitude (red) (with systematic uncertainties), SNe Ia apparent magnitude (gray line) (with only statistical uncertainties). The figure is adapted from [213].

2.3.2. XCDM Models

Cosmological dark energy models with a constant value of the EoS parameter are called XCDM models. These models are defined both in the spatially flat and spatially non-flat hyperspaces. The normalized Hubble parameter expressed through the dark energy EoS parameter $w_X$ has the form

$$E(a) = \left[\Omega_{r0}a^{-4} + \Omega_{m0}a^{-3} + \Omega_{k0}a^{-2} + (1 - \Omega_{m0})a^{-3(1+w_X)}\right]^{1/2}. \quad (16)$$

The case $w_X = -1$ is equivalent to the standard spatially flat $\Lambda$CDM model with the same matter–energy density parameter $\Omega_{m0}$ and zero spatial curvature, $\Omega_{k0} = 0$, at the present epoch.

The $1\sigma$ and $2\sigma$ confidence level contour constraints on the cosmological parameter $w_X$ in the XCDM model from different combinations of datasets—the SNe Ia apparent magnitude, the CMB temperature anisotropy, and the BAO peak length scale—are presented in Figure 3 (right panel).

*2.4. Quintessential Inflation Models*

Quintessential inflation models describe the evolution of the universe from the inflation epoch till the present dark energy epoch. In these models, a single field $\phi$ plays the double role of the inflaton field at the inflation epoch and the quintessence scalar field at the present epoch.

The general form of the action for quintessential inflation models reads:

$$S = \int d^4x \sqrt{-g}\left(-\frac{M_{pl}^2}{16\pi}R + \frac{1}{2}g^{\mu\nu}\partial_\mu\phi\partial_\nu\phi - V(\phi)\right) + S_M + S_I(g_{\mu\nu}, \phi, \psi, \chi, B_\mu), \quad (17)$$

where $S_{\rm I}$ is the action describing the interactions of the inflaton field with the fermion ($\psi$), scalar ($\chi$), and vector ($B_\mu$) degrees of freedom in the Standard Model and beyond.

To maintain inflation over a long period of time, it is necessary that the acceleration caused by the inflaton field be sufficiently small compared to its velocity over the Hubble time. Under these conditions, the first Friedmann's and Klein–Gordon's equations for the inflaton in the spatially flat universe take the form [169,236,237]

$$\rho = \dot\phi^2/2 + V(\phi) \xrightarrow{\dot\phi^2/2 \ll V(\phi)} \rho = \frac{3M_{\rm Pl}^2}{8\pi}H^2 \simeq V(\phi),$$
$$\ddot\phi + 3H\dot\phi + V'(\phi) = 0 \xrightarrow{|\ddot\phi| \ll 3H|\dot\phi|} -V'(\phi) \simeq 3H\dot\phi. \qquad (18)$$

The slow-roll regime of the inflaton field is provided by the potential $V(\phi)$ with certain shapes: exponential [178], power-law [175,176], and plateau-like [179,180]. The slow-roll parameters, which determine the curvature and slope of the potential, should remain small for some period of time to sustain the inflationary behavior:

$$\epsilon = \frac{M_{\rm pl}^2}{2}\left(\frac{V'(\phi)}{V(\phi)}\right)^2 \ll 1, \quad \eta = M_{\rm pl}^2 \frac{V''(\phi)}{V(\phi)} \ll 1, \quad \theta = M_{\rm pl}^2 \frac{V'(\phi)V'''(\phi)}{V(\phi)^2} \ll 1. \qquad (19)$$

The scalar spectral index ($n_s$), tensor spectral index ($n_t$), scalar spectral index running ($a_s$), and tensor-to-scalar ratio ($r$) are defined, respectively, as [178]

$$n_s - 1 = -6\epsilon + 2\eta, \quad n_t = -2\epsilon, \quad a_s \equiv {\rm d}n_s/{\rm d}\ln k = 16\epsilon\eta - 24\epsilon^2 - 2\theta, \quad r = 16\epsilon. \qquad (20)$$

During the inflationary epoch of the universe, scalar and tensor perturbations are created from quantum vacuum fluctuations and are spatially stretched to superhorizon scales, where they become classical, and the almost scale-invariant tilted primordial power spectrum is formed [238]. The tilted primordial scalar $\mathcal{P}_s(k)$ and tensor $\mathcal{P}_t(k)$ power spectra for spatially flat tilted quintessential inflation models are defined in terms of the wave number $k$ as [169,236,237]

$$\mathcal{P}_s(k) = A_s\left(\frac{k}{k_0}\right)^{n_s-1}, \quad \mathcal{P}_t(k) = A_t\left(\frac{k}{k_0}\right)^{n_t}, \qquad (21)$$

where $A_s$ and $A_t$ are the curvature perturbations amplitude and tensor amplitude at the pivot scale $k_0 = 0.05 \, {\rm Mpc}^{-1}$ [51].

The untilted primordial power spectrum for untilted spatially non-flat quintessential inflation models is defined as [177,239]

$$P(q) \propto \frac{(q^2 - 4K)^2}{q(q^2 - K)}, \qquad (22)$$

where $q = \sqrt{k^2 + K}$ is the wavenumber for scalar perturbations. In the spatially flat limit $K = 0$, $P(q)$ reduces to the $n_s = 1$ spectrum.

## 2.5. Interacting Dark Energy Models

As mentioned above, one of the major unresolved problems of modern cosmology is the so-called coincidence problem, i.e., the energy densities of dark energy and dark matter are of the same order of magnitude at the present epoch. One way to resolve this problem is to assume that these components somehow interact with each other. IDE models consider the transformation of dark energy and dark matter into each other, with their interaction described by the following modified continuity equations for dark energy and matter, respectively

$$\dot{\rho}_m + 3H\rho_m = \delta_{\text{couple}}, \tag{23}$$
$$\dot{\rho}_\phi + 3H(\rho_\phi + p_\phi) = -\delta_{\text{couple}}, \tag{24}$$

where $\rho_m$ is the matter energy density and $\delta_{\text{couple}}$ is the interaction function. In IDE models, the following forms of the coupling coefficient $\delta_{\text{couple}}$ are typically used [153,154]

$$\delta_{\text{couple}} = nQ\rho_m\dot{\phi}, \tag{25}$$
$$\delta_{\text{couple}} = \beta H(\rho_m + \rho_\phi), \tag{26}$$

where $n = \sqrt{8\pi/M_{\text{pl}}^2}$, and $\beta$ and $Q$ are dimensionless constants. The IDE models are subdivided into two types, as described below [121,153,154].

2.5.1. Coupling of The First Type

The IDE models of the first type are characterized by the exponential potential for the scalar $\phi$ and the linear interaction determined by the coupling coefficient given by Equation (25), as discussed in [153]. The coupled quintessence scalar field equation is given as

$$\ddot{\phi} + 3H\dot{\phi} + V'(\phi) = -nQ\rho_m\dot{\phi}, \tag{27}$$

where $V(\phi) = V_0 e^{-n\lambda\phi}$ is the scalar field potential and $\lambda$ is a model parameter. The coupled continuity equation for dark energy is

$$\dot{\rho}_\phi + 3H(\rho_\phi + p_\phi) = -nQ\rho_m\dot{\phi}. \tag{28}$$

The matter energy density evolves according to

$$\dot{\rho}_m + 3H\rho_m = nQ\rho_m, \tag{29}$$

leading to

$$\rho_m = \rho_{m0}a^{-3}e^{nQ\phi}. \tag{30}$$

2.5.2. Coupling of The Second Type

For the second type of IDE models, the scalar potential, and hence the dynamics of the interaction between dark energy and matter is constructed with the requirement that the coincidence parameter $r = \rho_m/\rho_\phi$ takes an analytic expression and for $z \to 0$ becomes a constant, thereby alleviating the coincidence problem of the $\Lambda$CDM model [121,154].

The equation of motion for $\phi$, Equation (24), can be written as

$$\dot{\phi}\left[\ddot{\phi} + 3H\dot{\phi} + V'(\phi)\right] = -\delta_{\text{couple}}. \tag{31}$$

The coupling coefficient $\delta_{\text{couple}}$ is constrained by the requirement that the solution to Equation (24) be compatible with a constant relationship between $\rho_m$ and $\rho_\phi$ energy densities. It is convenient to introduce the quantities $\Pi_m$ and $\Pi_\phi$ by

$$\delta_{\text{couple}} = -3H\Pi_m = 3H\Pi_\phi, \tag{32}$$

by introducing these quantities, the continuity equations for dark energy and matter (Equation (24)) will have the form

$$\dot{\rho}_m + 3H(\rho_m + \Pi_m) = 0, \qquad \dot{\rho}_{\text{Œ}} + 3H(\rho_{\text{Œ}} + p_{\text{Œ}} + \Pi_\phi) = 0. \tag{33}$$

The quantities $\Pi_m$ and $\Pi_\phi$ are related as

$$\Pi_m = -\Pi_\phi = \frac{\rho_m \rho_\phi}{\rho}(\gamma_\phi - 1), \tag{34}$$

$$\gamma_\phi = \frac{p_{\text{Œ}} + \rho_\phi}{\rho_\phi} = \frac{\dot{\phi}^2}{\rho_\phi} \tag{35}$$

where $\rho = \rho_m + \rho_\phi$.

Assuming $\gamma_\phi$ is a constant, the value of which $\gamma_\phi \in (0,2)$, it can be found

$$\rho_m \propto \rho_\phi \propto \rho \propto a^{-\nu}, \text{ for } \nu = 3\frac{\gamma_\phi + r}{r+1}, \tag{36}$$

where $r$ is a coincidence parameter, which takes an analytic expression for $r$ and becomes a constant, thereby alleviating the coincidence problem. The solution of the second Friedmann's equation $3M_{\text{pl}}H^2 = 8\pi\rho$ for the result obtained in Equation (36), has the form $a \propto t^{2/\nu}$. Thus, the Hubble parameter is defined as

$$H = \frac{2}{\nu t} = \frac{2(r+1)}{3(\gamma_\phi + r)}\frac{1}{t}. \tag{37}$$

The energy density parameter is defined as $\Omega_\phi = \frac{8\pi M_{\text{pl}}^2}{3H^2}\rho_\phi$, as well as $\Omega_\phi = \frac{1}{r+1}$. Equating these equations and inserting Equation (37), we have

$$\rho_\phi = \frac{M_{\text{pl}}^2}{6\pi}\frac{1+r}{(\gamma_\phi + r)^2}\frac{1}{t^2}. \tag{38}$$

The combination with Equation (35) gives

$$\dot{\phi} = \sqrt{\frac{M_{\text{pl}}^2 \gamma_\phi (1+r)}{6\pi}}\frac{1}{(\gamma_\phi + r)^2}\frac{1}{t}, \tag{39}$$

thus, the consequence of the condition $\rho_\phi \sim \rho_m$ is the logarithmic evolution of the scalar field $\phi$ with time.

Applying the equation for the energy density for the scalar field and Equation (35) yields

$$\rho_\phi = \frac{2V(\phi)}{2 - \gamma_\phi} = \frac{\dot{\phi}^2}{\gamma_\phi}, \tag{40}$$

which together with Equations (38) and (39) leads to

$$V(\phi) = \frac{M_{\text{pl}}^2}{6\pi}\left(1 - \frac{\gamma_\phi}{2}\right)\frac{1+r}{(\gamma_\phi + r)^2}\frac{1}{t^2} \Rightarrow \frac{\partial V(\phi)}{\partial \phi} = -\lambda V(\phi), \tag{41}$$

where $\lambda = \sqrt{\frac{24\pi}{M_{\text{pl}}^2 \gamma_\phi (1+r)}}$. Equation (41) implies that the potential has the exponential form

$$V(\phi) = V_0 e^{-\lambda(\phi - \phi_0)}. \tag{42}$$

A significant drawback of this model is the absence of a convincing explanation for the onset of the interaction of dark energy and matter at the epoch of transition from the decelerated to the accelerated expansion of the universe. The thermal quantum-field theory treatment of the quintessence dark energy coupled to the matter field shows that one can consistently recover different expansion regimes of the universe, including the late-time acceleration; however, more work is needed to relate the matter field to a viable dark matter candidate [240,241].

## 3. Constraints from Observational Data

### 3.1. Type Ia Supernovae

The observed magnitudes of type Ia supernovas are among the best data for constraining the distance–redshift relationship through the determination of the luminosity distance. In the $\phi$CDM models, the distances tend to be smaller compared to the $\Lambda$CDM predictions at the same redshift. This provides an opportunity for differentiating these models from each other.

One of the first studies in this direction was performed in Podariu and Ratra [242]. They used three datasets of SNe Ia apparent magnitude versus redshift—(i) R98 data [212], both including and excluding the unclassified SNe Ia 1997ck at $z = 0.97$ (with 50 and 49 SNe Ia apparent magnitude data, respectively), (ii) P99 data [2], and (iii) a third set with the corrected/effective stretch factor magnitudes for the 54 Fit C SNe Ia of P99 apparent magnitude data—and obtained constraints on the $\phi$CDM model with RP potential ($\phi$CDM-RP model) (see Figure 4).

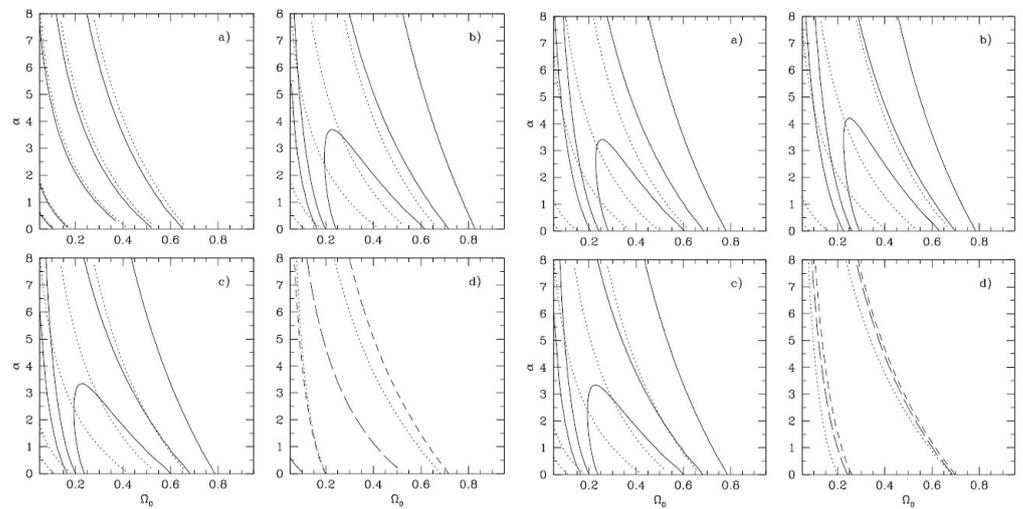

**Figure 4.** The $1\sigma$, $2\sigma$, and $3\sigma$ confidence level contour constraints on parameters of the scalar field $\phi$CDM model with the RP potential. (Left panel) (**a**) For all R98 SNe Ia apparent magnitude data, (**b**) for R98 SNe Ia apparent magnitude data excluding the $z = 0.97$ measurement, (**c**) for P99 Fit C SNe Ia apparent magnitude dataset, (**d**) for three datasets: all R98 SNe Ia apparent magnitude (long-dashed lines) data, R98 SNe Ia apparent magnitude data excluding the $z = 0.97$ measurement (short-dashed lines), and P99 Fit C SNe Ia apparent magnitude data (dotted lines). (Right panel) (**a**) For all the R98 SNe Ia apparent magnitude data, (**b**) for R98 SNe Ia apparent magnitude data excluding the $z = 0.97$ measurement, (**c**) for P99 Fit C SNe Ia apparent magnitude data, (**d**) for the $H_0$ and $t_0$ constraints used in conjunction with all R98 SNe Ia (long-dashed lines) apparent magnitude data, R98 SNe Ia apparent magnitude data excluding the $z = 0.97$ measurement (short-dashed lines), and the P99 Fit C SNe Ia apparent magnitude data (dotted lines). The figure is adapted from [242].

Caresia et al. [215] obtained constraints on the parameters of the $\phi$CDM with the RP and Sugra potentials [226,243] and also of the extended quintessence models with the inverse power-law RP potential [244,245] from the datasets of apparent magnitude versus redshift measurements of 176 SNe Ia [2,212,246], and the data from the SuperNovae Acceleration Probe (SNAP) satellite [247]. [1] The obtained constraints on the model parameters are shown in Figures 5 and 6. No useful constraints on the model parameters were found for the Sugra potential, while $1\sigma$ constraints of $\alpha < 0.8$ and $\alpha < 0.6$, for both the extended and ordinary quintessence models using the RP potential, were obtained using the SNe Ia apparent magnitude and SNAP satellite data, respectively.

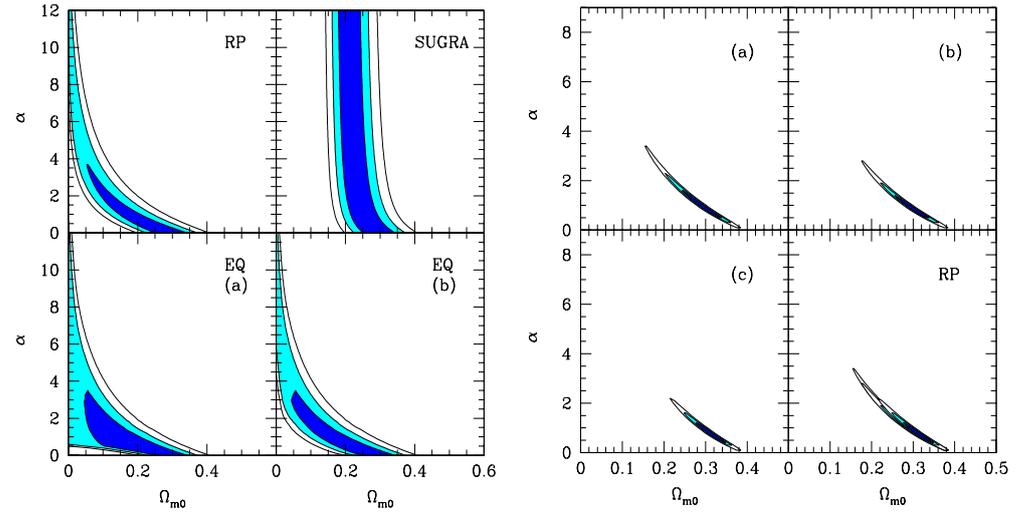

**Figure 5.** (Left panel) The 1 $\sigma$, 2$\sigma$, and 3$\sigma$ confidence level contour constraints on $\Omega_{m0}$ and $\alpha$ parameters by using a sample of 176 SNe Ia apparent magnitude data. (Upper left sub-panel) For the ordinary quintessence with the inverse power-law RP potential, (upper right sub-panel) for ordinary quintessence with the Sugra potential, (lower left sub-panel) for extended quintessence with the inverse power-law RP potential, (lower right sub-panel) for extended quintessence with the inverse power-law RP potential when upper limits on the time variation of the gravitational constant are satisfied. (Right panel) The 1$\sigma$, 2$\sigma$, and 3$\sigma$ confidence level contour constraints on $\Omega_{m0}$ and $\alpha$ parameters for the ordinary $\phi$CDM quintessence model with the inverse power-law RP potential by using SNAP sample data. Upper left sub-panel corresponds to constraints obtained by assuming the exact EoS parameter, upper right sub-panel corresponds to the linear approximation of the EoS parameter, lower left sub-panel corresponds to the constant approximation of the EoS parameter, lower right sub-panel corresponds to the superposition of above-mentioned three cases. The figure is adapted from [215].

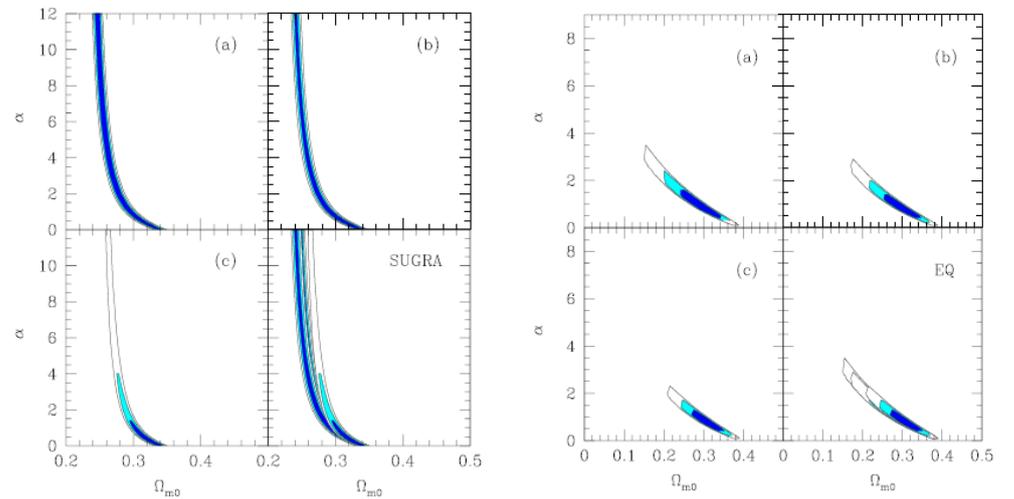

**Figure 6.** (Left panel) As Figure 11, but for the ordinary $\phi$CDM quintessence model with the Sugra potential. (Right panel) As Figure 11, but for the extended quintessence model with the inverse power-law RP potential. The results are obtained by imposing the upper bound on the time variation of the gravitational constant. The figure is adapted from [215].

Doran et al. [248] considered a DE model parameterized as [248]

$$\Omega_\phi(a) = \frac{e^{R(a)}}{1+e^{R(a)}}, \quad w(a) = w_0 \ln a/(1-b\ln a)^2, \tag{43}$$

where $R(a) = \ln(\Omega_\phi(a)/(1-\Omega_\phi(a)))$ corresponds to

$$R(a) = R_0 - \frac{2w_0 \ln a}{1 - b \ln a}, \tag{44}$$

the constant $b$ is defined by the EoS parameter at the present epoch $w_0$, the dark energy density parameter at the present epoch $\Omega_{\phi 0}$, and the parameter $\Omega_e$ characterizing the amount of dark energy at early times to which it asymptotes for very large redshifts, as

$$b = -3w_0 \left( \ln \frac{1-\Omega_e}{\Omega_e} + \ln \frac{1-\Omega_{\phi 0}}{\Omega_{\phi 0}} \right). \tag{45}$$

Using a combination of datasets from SNe Ia [212], Wilkinson Microwave Anisotropy Probe (WMAP) [6], Cosmic Background Imager (CBI) [249], Very Small Array (VSA) [250], SDSS [251], and HST [252], the authors find $w_0 < -0.8$ and $\Omega_e < 0.03$ at the $1\sigma$ confidence level; the contours are shown in Figure 7. It should be noted that the SNe Ia apparent magnitude data are most sensitive to $w_0$, while the CMB temperature anisotropies and the LSS growth rate are the best constraints of $\Omega_e$ (see Figure 8).

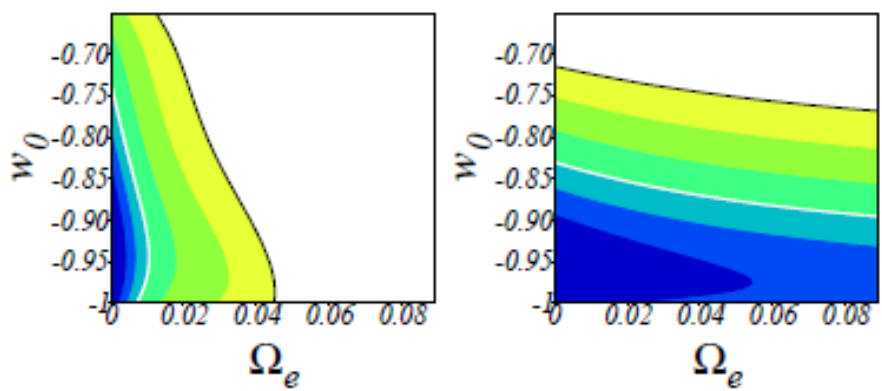

**Figure 7.** Constraints on parameters $\Omega_e$ and $w_0$. The left picture depicts the distribution from WMAP+CBI+VSA+SDSS+HST data and the right picture is that of SNe Ia apparent magnitude versus redshift data alone. The regions of $1\sigma$ ($2\sigma$) confidence level are enclosed by a white (black) line. The figure is adapted from [248].

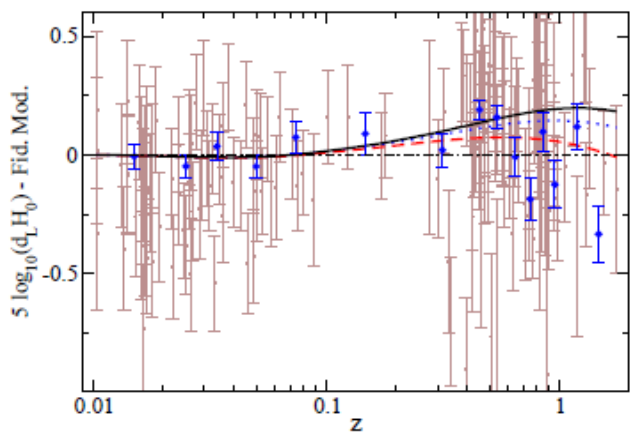

**Figure 8.** SNe Ia apparent magnitude versus redshift data [212] as data points with thin (brown) error bars. The authors plotted the logarithm of the luminosity distance minus a fiducial model for which $d_L H_0 = (1+z)\ln(1+z)$. The solid (black) line is for the spatially flat $\Lambda$CDM model, the dotted (blue) line is for $\Omega_e = 10^{-4}$, and the dashed (red) line is for $\Omega_e = 10^{-1}$. For all models, $w_0 = -1$. The figure is adapted from [248].

Pavlov et al. [253] found that, for the *ϕ*CDM-RP model in a spacetime with nonzero spatial curvature, the dynamical scalar field has an attractor solution in the curvature dominated epoch, while the energy density of the scalar field increases relative to that of the spatial curvature. In the left panel of Figure 39, we see that the values $\Omega_{m0} = 0.27$ and $\alpha = 3$ are consistent with these constraints for a range of values $\Omega_{k0}$ and for a set of dimensionless time parameter values of $H_0 t_0 = H_0 \int_0^{a_0} da/\dot{a}(t)$, where $t_0$ is the age of the universe. The right panel of Figure 9 shows a similar analysis for several values of the cosmological test parameter $\triangle(\Omega_{m0}, \Omega_{k0}, \alpha) = \delta(t_0)/(1 + z_i)\delta(t_i)$, where $\delta(t_0)$ and $\delta(t_i)$ are the values of the matter density contrast at, respectively, the present time $t_0$ and an arbitrary time $t_i$ such that $a(t_i) \ll a(t_0)$, i.e., a time when the universe is well approximated by the Einstein–de Sitter model in the matter-dominated epoch.

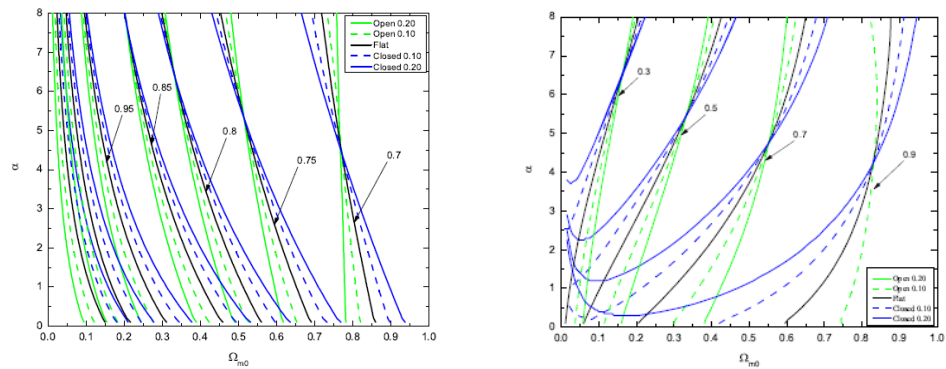

**Figure 9.** (Left panel) The 2*σ* contours of the fixed time parameter $H_0 t_0$ as a function of values of the matter density parameter at the present epoch $\Omega_{m0}$ and space curvature density parameter at the present epoch $\Omega_{k0}$, as well as the model parameter *α* in the scalar field *ϕ*CDM model with the RP potential. The results obtained for larger values of free parameters $(\Omega_{m0}, \Omega_{k0}, \alpha)$ and for $H_0 t_0 = [0.7, 0.75, 0.8, 0.85, 0.95, 1.05, 1.15]$. (Right panel) The 2*σ* contours of the factor by which the growth of matter perturbations falls lower than in the Einstein–de Sitter model. The cosmological test parameter values $\triangle(\Omega_{m0}, \Omega_{k0}, \alpha)$ obtained for the larger values of free parameters $(\Omega_{m0}, \Omega_{k0}, \alpha)$. The figure is adapted from [253].

Fuzfa and Alim [254] studied the *ϕ*CDM model with the RP and Sugra potentials in a spatially closed universe. The estimated values of $\Omega_{m0}$ and $\Omega_{\phi 0}$, using SNe Ia apparent magnitude data from the SNLS collaboration [255], are quite different from those for the standard spatially flat ΛCDM model (Figure 10). Such a result is expected due to the different cosmic acceleration and dark matter clustering predicted between quintessence models and the standard ΛCDM model, arising from the differences in cosmological parameters, even at $z = 0$. The quintessence scalar field creates more structures outside the filaments, lighter halos with higher internal velocity dispersion, as seen from N-body simulations performed to study the influence of quintessence on the distribution of matter at large scales.

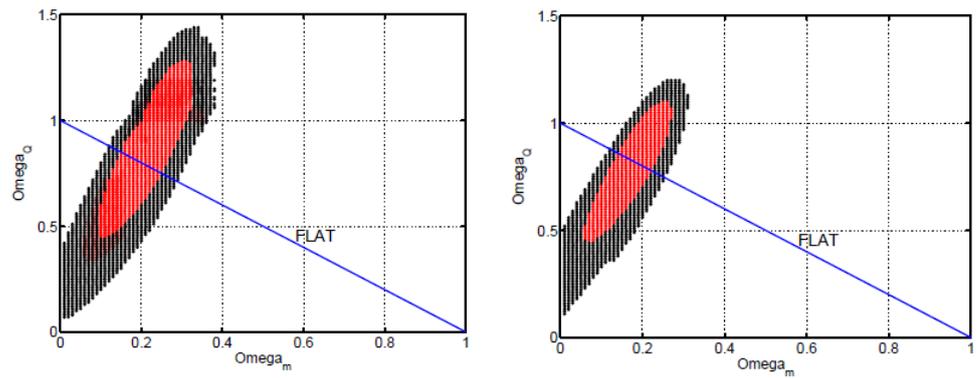

**Figure 10.** The 1 $\sigma$ and 2$\sigma$ confidence level contour constraints on the matter density parameter at the present epoch $\Omega_{m0}$ and the dark energy density (quintessence) parameter at the present epoch $\Omega_{Q0}$ for scalar field $\phi$CDM models. (Left panel) With the inverse power-law PR potential. (Right panel) With Sugra potential $V(\phi) \propto \phi^{-\alpha} \exp(4\pi\phi^2)$. The blue lines accord to the flat universe. The figure is adapted from [254].

Farooq et al. [256] constrained the $\phi$CDM-RP model in a spacetime with non-zero spatial curvature, as well as the XCDM model, using the Union2.1 compilation of the 580 SNe Ia apparent magnitude measurements of Suzuki et al. [257], Hubble parameter observations [28,30,258,259], and the $0.1 \leq z \leq 0.75$ BAO peak length scale measurements [21,22,260] (see Figure 11). They constrain the spatial curvature density parameter today to be $|\Omega_{k0}| \leq 0.15$ at a 1$\sigma$ confidence level and more precise data are required to tighten the bounds on the parameters.

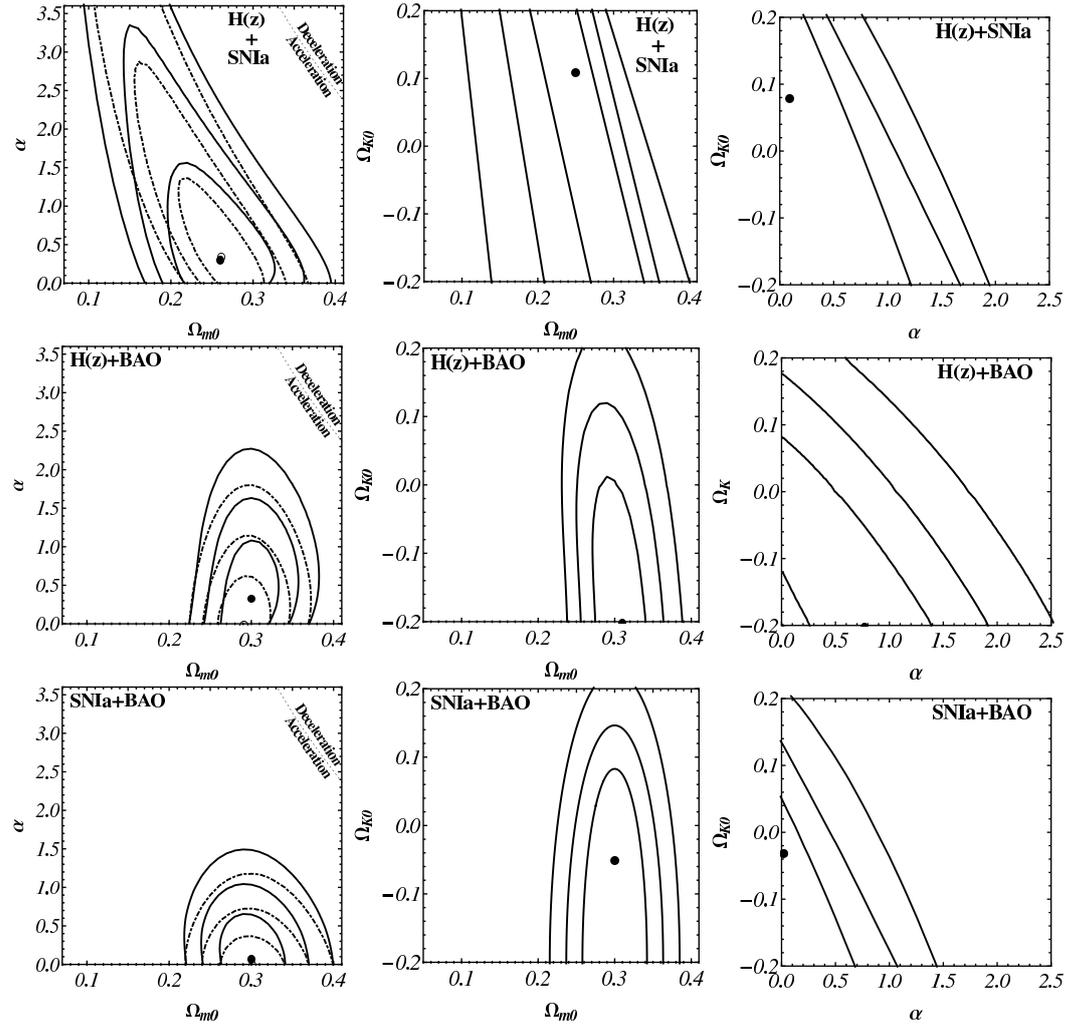

**Figure 11.** The 1 $\sigma$, 2$\sigma$, and 3$\sigma$ confidence level contour constraints on parameters of the spatially non-flat scalar field $\phi$CDM model with the RP potential from compilations of data: $H(z)$+SNe Ia apparent magnitude (first row), $H(z)$+BAO peak length scale (second row), and BAO peak length scale+SNe Ia apparent magnitude (third row). Filled circles denote best-fit points. The dot–dashed lines in the first column panels are 1$\sigma$, 2$\sigma$, and 3$\sigma$ confidence level contours obtained by Farooq et al. [261] for the spatially flat $\phi$CDM model (open circles denote best-fit points). Dotted lines separate the accelerating and decelerating models (at zero space curvature). The horizontal axis with $\alpha = 0$ corresponds to the standard spatially flat $\Lambda$CDM model. First, second, and third columns correspond to marginalizing over $\Omega_{k0}$, $\alpha$, and $\Omega_{m0}$, respectively. The figure is adapted from [256].

Assuming that the Hubble constant $H_0$ tension of the $\Lambda$CDM model is actually a tension on the SNe Ia absolute magnitude $M_B$, Nunes and Di Valentino [159] assessed the $M_B$ tension by comparing the spatially flat $\Lambda$CDM model, $w$CDM, and IDE models using a compilation of two datasets: the SNe Ia Pantheon sample [213] + BAO [22,24,25,262,263] + big bang nucleosynthesis (BBN) [264] and the SNe Ia Pantheon sample + BAO [22,24,25,262,263] + BBN [264] + $M_B$ [265] (see Figure 12). They found that the IDE model can alleviate both the $M_B$ and $H_0$ tensions with a coupling different from zero at a 2$\sigma$ confidence level with a preference for a compilation of the Pantheon + BAO + BBN + $M_B$ datasets.

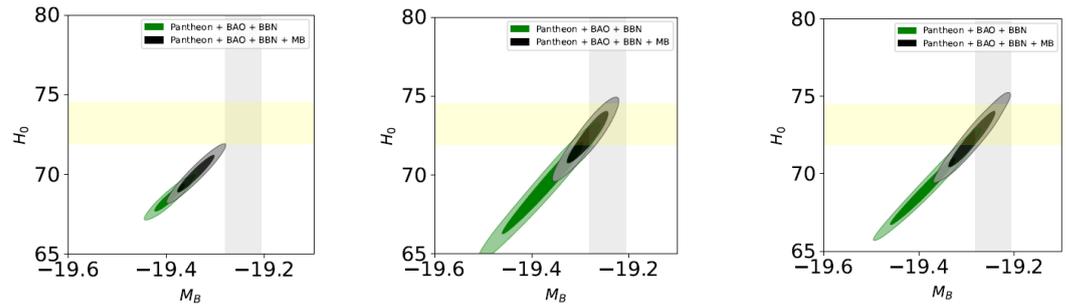

**Figure 12.** The $1\sigma$ and $2\sigma$ confidence level contour constraints on $M_B$ and $H_0$ values for $\Lambda$CDM (left panel), $w$CDM (middle panel), and IDE (right panel) models, obtained from compilations of Pantheon + BAO + BBN and Pantheon + BAO + BBN + $M_B$ datasets. The figure is adapted from [159].

*3.2. Cosmic Microwave Background Radiation Data*

The CMB provides a very accurate determination of the angular diameter distance $d_A$ at a redshift of $z\sim 1000$. This measurement is sensitive to the entire expansion of the universe over this wide range of redshifts. As pointed out before, the $\phi$CDM models tend to predict smaller distances and can therefore be constrained with the CMB geometric measurements.

In one of the first such studies, Doran et al. [266] used the CMB temperature anisotropy data from the BOOMERANG and MAXIMA experiments [267,268] to distinguish quintessential inflation models with different EoS parameters, described by a kinetic term $k(\phi)$ of the cosmon field this model is described by a Lagrangian of the form $\mathcal{L}_\phi = \frac{1}{2}(\partial_\mu \phi)^2 k^2(\phi) + V(\phi)$: (i) the RP potential with $k(\phi) = 1$, (ii) the leaping kinetic term model where $V(\phi) = M_{\bar{\text{pl}}}^4 \exp(-\phi/M_{\bar{\text{pl}}})$, $M_{\bar{\text{pl}}} = \sqrt{8\pi} M_{\text{pl}}$ is the reduced Plank mass, $k(\phi) = k_{\min} + \tanh[(\phi-\phi_1)/M_{\bar{\text{pl}}}] + 1$, $\phi_1 \approx 277$ eV, and $k_{\min} = [0.05, 0.1, 0.2, 0.26]$ [269], and (iii) the exponential potential with $V(\phi) = M_{pl}^4 \exp(-\sqrt{2}\alpha\phi/M_{\bar{\text{pl}}})$, $\alpha = \sqrt{3/2\Omega_{\phi 0}}$, and $k(\phi) = 1$ [151]. The dark energy density parameters today and at the last scattering epoch, $\Omega_{\phi 0}$ and $\Omega_{\phi \text{ls}}$, respectively, and the averaged EoS parameter of the field $\phi$ are used to parameterize the separation of peaks in CMB temperature anisotropies, which can be used to measure the value of $\Omega_\phi$ before the last scattering (see Figure 13).

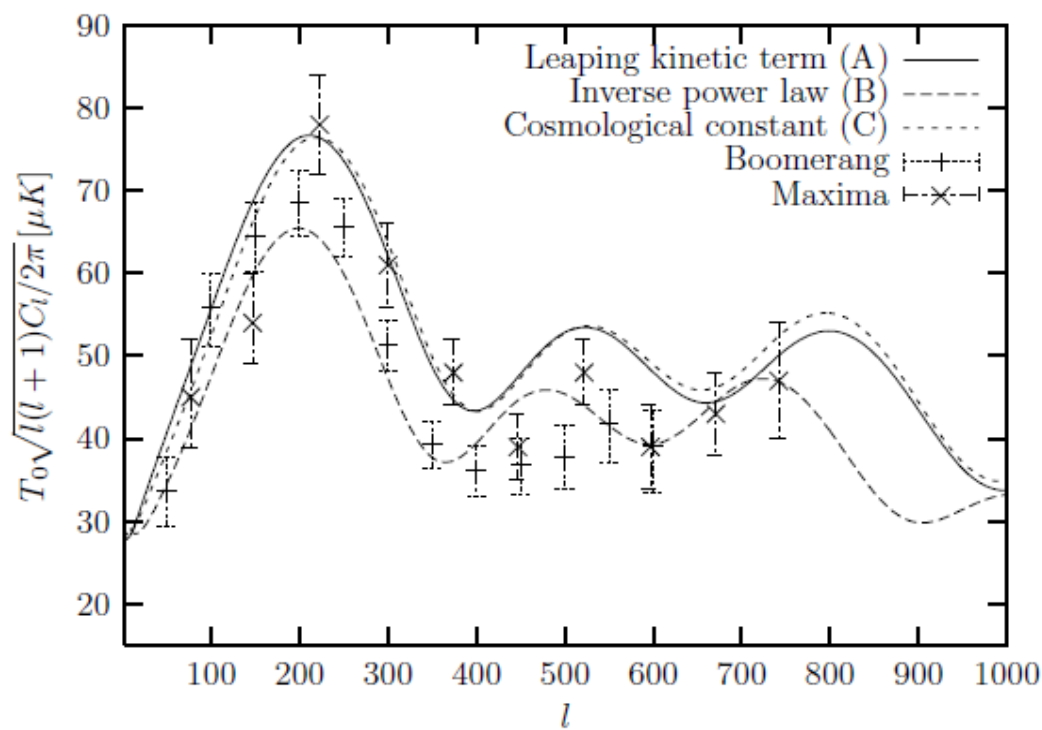

**Figure 13.** The CMB temperature anisotropy spectrum for different quintessence scalar field $\phi$CDM models: with the leaping kinetic term (model A), with the inverse power-law RP potential (model B) (here the dark energy density parameter at the present epoch $\Omega_{\phi 0} = 0.6$), and for the $\Lambda$CDM model (model C). Data points from the BOOMERANG [267] and MAXIMA [268] experiments are shown for reference. The figure is adapted from Ref. [266].

Caldwell et al. [270] investigated how early quintessence dark energy, i.e., a non-negligible quintessence energy density during the recombination and structure formation epochs, affects the baryon–photon fluid and the clustering of dark matter, and thus the CMB temperature anisotropy and the matter power spectra. They showed that early quintessence is characterized by a suppressed ability to cluster at small length scales, as suggested by the compilation of data from WMAP [271,272], CBI [273,274], Arcminute Cosmology Bolometer Array Receiver (ACBAR) [275], the LSS growth rate dataset of Two degree Field Galaxy Redshift Survey (2dFGRS) [276–278], and $L_{y-\alpha}$ forest [279,280]; these are shown in Figure 14.

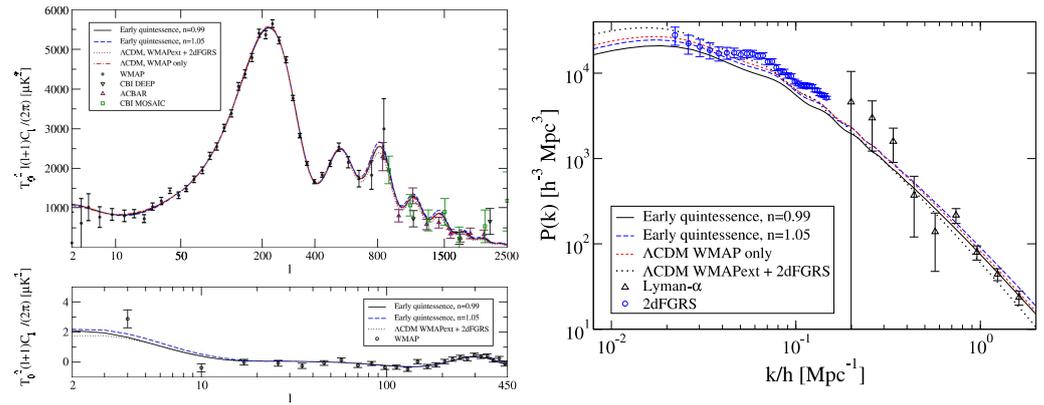

**Figure 14.** (Left panel) Polarization (TE) and temperature (TT) as functions of the multipole *l*. Two quintessential inflation models with $n_s = 0.99$ and $n_s = 1.05$ are presented with WMAP data from [271,272]. WMAP-normalized spectra for the best fit for the $\Lambda$CDM model (no $L_{y-\alpha}$ data) with the constant spectral index $n = 0.97$ [6] and the $\Lambda$CDM model with the running spectral index $n_s = 0.93$, $dn_s/d \ln k = -0.031$ are shown for comparison. For large *l*, CBI data and ACBAR data are used. (Right panel) The CDM power spectrum at the present epoch as a function of $k/h$. The linear spectrum for two quintessential inflation models with spectral indices $n_s = 0.99$ and $n_s = 1.05$ are plotted. The best fit for the $\Lambda$CDM model with running spectral index $n_s = 0.93$, $dn_s/d \ln k = -0.031$ [6]), normalized to WMAP data (no $L_{y-\alpha}$ data), is shown. 2dFGRS measurements and $L_{y-\alpha}$ data are converted to $z = 0$. The figure is adapted from [270].

Pettorino et al. [216] studied a class of the extended $\phi$CDM models, where the scalar field is exponentially coupled to the Ricci scalar and is described by the RP potential. The projection of the ISW effect on the CMB temperature anisotropy [2] is found to be considerably larger in the exponential case with respect to a quadratic non-minimal coupling as seen in Figure 15. This reflects the fact that the effective gravitational constant depends exponentially on the dynamics of the scalar field.

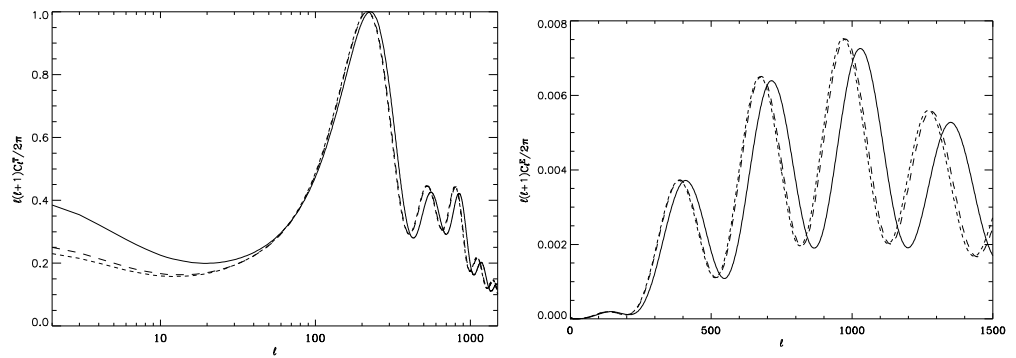

**Figure 15.** The spectra are in arbitrary units, normalized to unity at the first acoustic peak. (Left panel) CMB angular total intensity power spectra for the scalar field $\phi$CDM model with the inverse power-law RP potential (dotted), quadratic (dashed) and exponential coupling extended quintessence (solid) with $\omega_{\text{JBD0}} = 50$. (Right panel) CMB angular polarization power spectra for the $\phi$CDM model (dotted), quadratic (dashed), and exponential coupling extended quintessence (solid) with $\omega_{\text{JBD0}}$. The figure is adapted from [216].

Mukherjee et al. [281] conducted a likelihood analysis of the Cosmic Background Explorer—Differential Microwave Radiometers (COBE-DMR) sky maps to normalize the $\phi$CDM-RP model in flat space [3]. As seen from Figure 16, this model remains an observationally viable alternative to the standard spatially flat $\Lambda$CDM model.

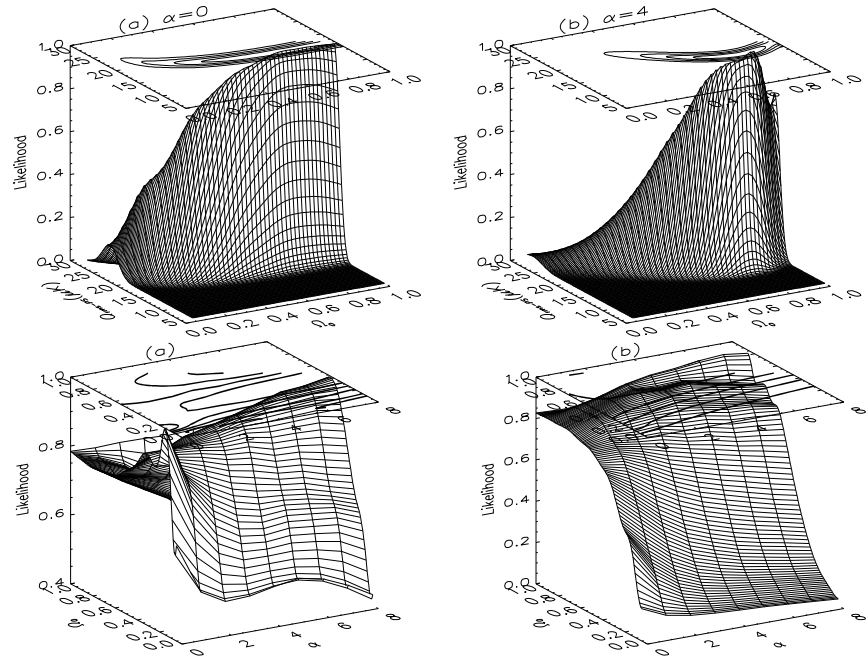

**Figure 16.** The model with $t_0 = 13$ Gyr and $\Omega_b h^2 = 0.014$. Likelihood functions $L(Q_{rms-PS}, \alpha, \Omega_0)$ (arbitrarily normalized to unity at the highest peak). (Left panel) Derived from a simultaneous analysis of DMR 53 and 90 GHz galactic-frame data. The faint high-latitude foreground galactic emission is corrected and the quadrupole moment in the analysis is included: (**a**) for $\alpha = 0$ and (**b**) for $\alpha = 4$. (Right panel) Derived by marginalizing $L(Q_{rms-PS}, \alpha, \Omega_0)$ over $Q_{rms-PS}$ with a uniform prior: (**a**) the correction for the faint high-latitude foreground galactic emission is ignored and the quadrupole moment from the analysis is excluded, (**b**) for the faint high latitude foreground galactic emission is corrected and the quadrupole moment in the analysis is included. The figure is adapted from [281].

Samushia and Ratra [282] constrained model parameters of the $\Lambda$CDM model, the XCDM model, and the $\phi$CDM-RP model using galaxy cluster gas mass fraction data [283]; for this, they introduced an auxiliary random variable as opposed to integrating over nuisance parameters of the Markov Chain Monte Carlo (MCMC) method. Two different sets of priors were chosen to study the influence of the type of priors on the obtained results—one set has [7] $h = 0.73 \pm 0.03$, $\Omega_b h^2 = 0.0223 \pm 0.0008$ ($1\sigma$ errors) and the other set has $h = 0.68 \pm 0.04$ [284,285], and $\Omega_b h^2 = 0.0205 \pm 0.0018$ [286]. The obtained constraints on the $\phi$CDM model with the RP potential are shown in Figure 17. We see that $\Omega_m$ is better constrained than $\alpha$, whose best-fit value is $\alpha = 0$, corresponding to the standard spatially flat $\Lambda$CDM model; however, the scalar field $\phi$CDM model is not excluded.

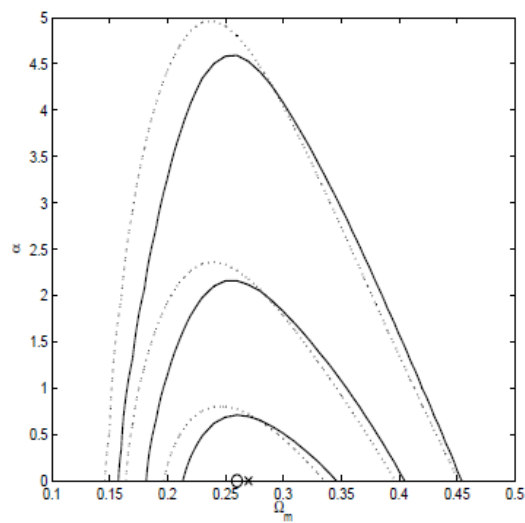

**Figure 17.** The 1 $\sigma$, 2$\sigma$, and 3$\sigma$ confidence level contour constraints on parameters of the scalar field $\phi$CDM model with the inverse power-law RP potential using cluster gas mass fraction data. Solid lines correspond to WMAP prior while dashed lines correspond to the alternate prior. The cross matches the best fit at $\Omega_{m0} = 0.27$ and $\alpha = 0$. The circle denotes the best fit at $\Omega_{m0} = 0.26$ and $\alpha = 0$. The horizontal axis for which $\alpha = 0$ corresponds to the spatially flat $\Lambda$CDM model. The figure is adapted from [282].

Chen et al. [287] constrained the scalar field $\phi$CDM-RP model and the $\Lambda$CDM model with massive neutrinos assuming two different neutrino mass hierarchies in both the spatially flat and non-flat universes, using a joint dataset comprising the CMB temperature anisotropy data [12,288], the BAO peak length scale data from the 6dF Galaxy Survey (6dFGS), SDSS—Main Galaxy Sample (MGS), Baryon Oscillation Spectroscopic Survey (BOSS)-LOWZ (galaxies within the redshift range $0.2 < z < 0.43$), BOSS CMASS-DR11 (galaxies within the redshift range $0.43 < z < 0.7$) [23], the joint light–curve analysis (JLA) compilation from SNe Ia apparent magnitude measurements [289], and the Hubble Space Telescope $H_0$ prior observations [29]. Assuming three species of degenerate massive neutrinos, they found the 2$\sigma$ upper bounds of $\sum m_\nu < 0.165\ (0.299)$ eV and $\sum m_\nu < 0.164\ (0.301)$ eV, respectively, for the spatially flat (spatially non-flat) $\Lambda$CDM model and the spatially flat (spatially non-flat) $\phi$CDM model (Figure 18). The inclusion of spatial curvature as a free parameter leads to a significant expansion of the confidence regions for $\sum m_\nu$ and other parameters in spatially flat $\phi$CDM models, but the corresponding differences are larger for both the spatially non-flat $\Lambda$CDM and spatially non-flat $\phi$CDM models.

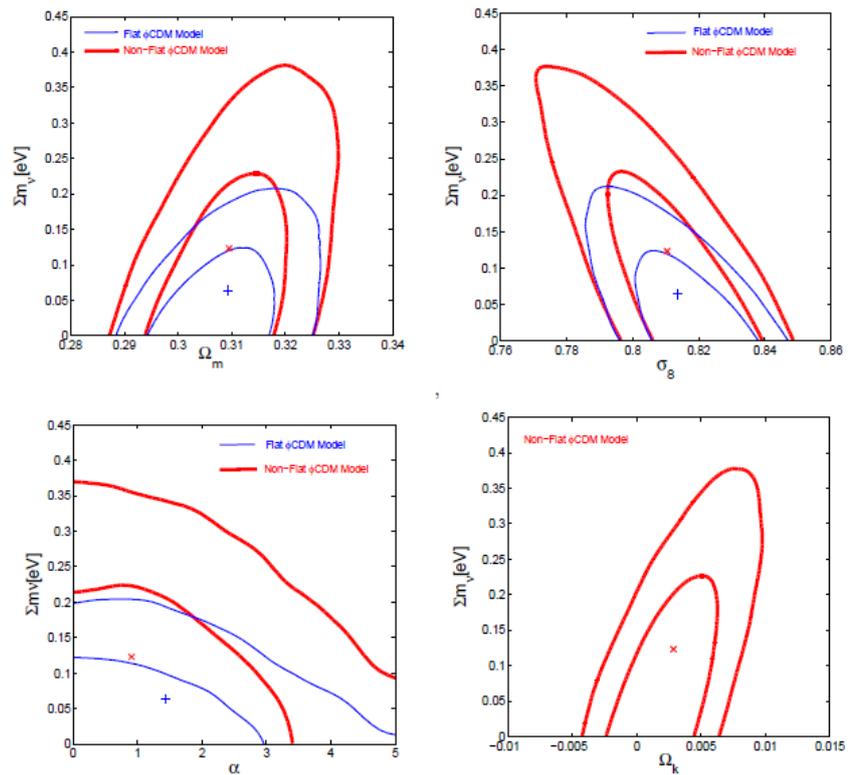

**Figure 18.** The $1\sigma$ and $2\sigma$ confidence level contour constraints on parameters of the spatially flat and spatially non-flat scalar field $\phi$CDM model with the inverse power-law potential from a joint analysis using the HST $H_0$ prior in the scenario with three species of degenerate massive neutrinos. (Upper left, upper right, and lower left panels) Contours are presented in the $\Omega_{\rm m} - \sum m_\nu$, $\sigma_8 - \sum m_\nu$, and $\alpha - \sum m_\nu$ planes. The thin blue (thick red) lines correspond to constraints in the spatially flat (spatially non-flat) universe. The "+" ("x") denotes the mean values of the pair in the spatially flat (spatially non-flat) universe. (Lower right panel) Contours are in the $\Omega_{\rm k} - \sum m_\nu$ plane for the spatially non-flat universe. The "x" denotes the mean values of the $(\Omega_{\rm k}, \sum m_\nu)$ pair. The figure is adapted from [287].

Park and Ratra [290] constrained the spatially flat tilted and spatially non-flat untilted $\phi$CDM-RP inflation model by analyzing the CMB temperature anisotropy angular power spectrum data from the Planck 2015 mission [291], the BAO peak length scale measurements [26], a Pantheon collection of 1048 SNe Ia apparent magnitude measurements over the broader redshift range of $0.01 < z < 2.3$ [213], Hubble parameter observation [21,25,28,30–34,258,292], and LSS growth rate measurements [25] (Figures 19 and 20). Constraints on parameters of the spatially non-flat model were improved from a $1.8\sigma$ to a more than $3.1\sigma$ confidence level by combining CMB temperature anisotropy data with other datasets. Present observations favor a spatially closed universe with the spatial curvature contributing about two-thirds of a percent of the current total cosmological energy budget. The spatially flat tilted $\phi$CDM model is a $0.4\sigma$ better fit to the observational data than the standard spatially flat tilted $\Lambda$CDM model, i.e., current observational data allow for the possibility of dynamical dark energy in the universe. The spatially non-flat tilted $\phi$CDM model better fits the DES bounds on the root mean square (rms) amplitude of mass fluctuations $\sigma_8$ as a function of the matter density parameter at the present epoch $\Omega_{\rm m0}$, but it does not provide such good agreement with the larger multipoles of Planck 2015 CMB temperature anisotropy data as the spatially flat tilted $\Lambda$CDM model.

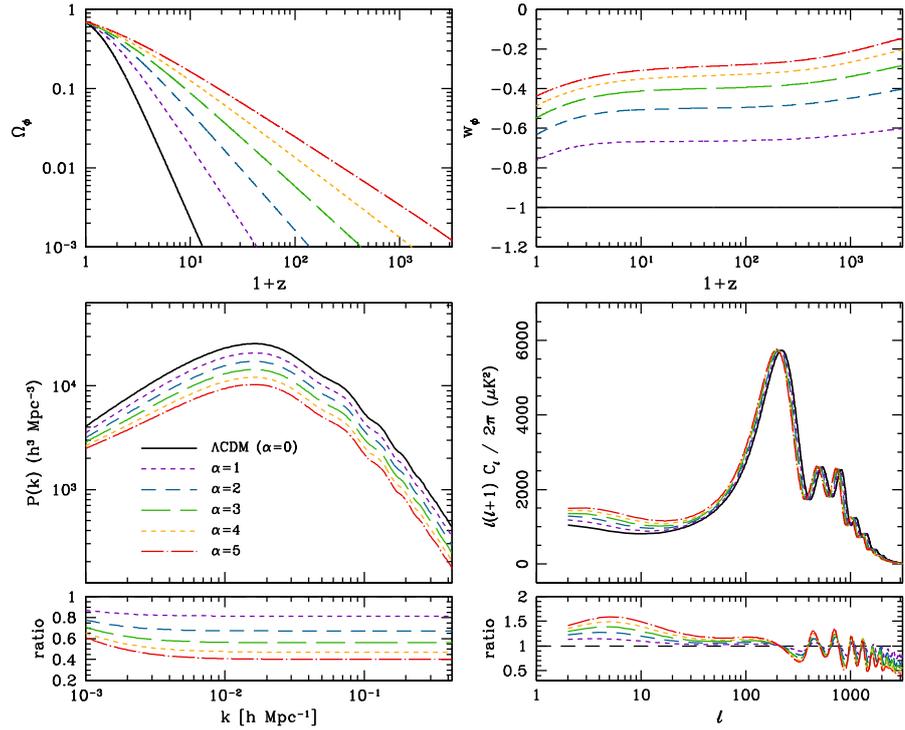

**Figure 19.** (Upper panels) Evolution of the EoS parameter $w_\phi$ and dark energy density parameter $\Omega_\phi$ in the tilted spatially flat $\phi$CDM model for the range of values of $\alpha$ parameter $\alpha \in (1,5)$. The black solid curve accords to the $\Lambda$CDM model, which corresponds to reduced $\phi$CDM model with $\alpha = 0$. (Middle panels) Theoretical predictions for matter density and CMB temperature anisotropy angular power spectra for the $\phi$CDM model depending on parameter $\alpha$. (Lower panels) Ratios of the $\phi$CDM model power spectra relative to the $\Lambda$CDM model. The figure is adapted from [290].

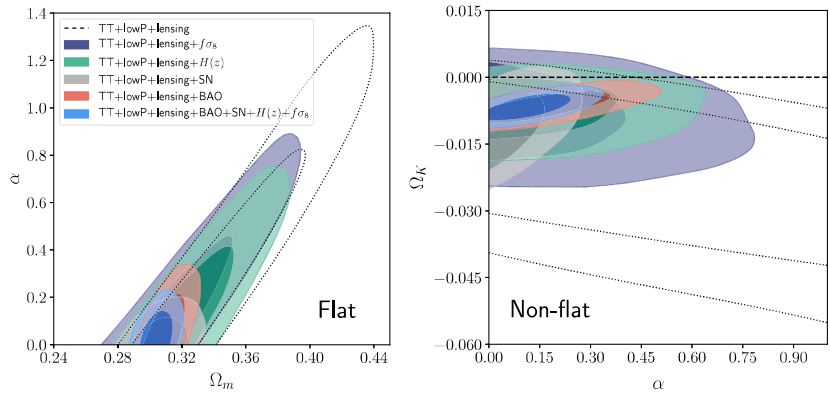

**Figure 20.** The 1 $\sigma$ and 2$\sigma$ confidence level contours. (Left panel) In the $\Omega_m - \alpha$ plane for the tilted spatially flat scalar field $\phi$CDM model. (Right panel) In the $\alpha - \Omega_k$ plane for the untilted spatially non-flat scalar field $\phi$CDM model. Constraints are derived from Planck CMB TT + lowP + lensing and non-CMB datasets. The horizontal dashed line indicates the spatially flat curvature with $\Omega_k = 0$. For the spatially non-flat $\phi$CDM model constrained with TT + lowP + lensing, the $h > 0.45$ prior has been used. The figure is adapted from [290].

Constraints on model parameters of the XCDM and $\phi$CDM-RP (spatially flat tilted) inflation models using the compilation of CMB [291] and BAO data [22–24,293–295] were derived by Ooba et al. [293]. The authors calculated the angular power spectra of the CMB temperature anisotropy using the CLASS code of Blas et al. (2011) [294] and executed the MCMC analysis with Monte Python (Audren et al. [295]). Having one additional parameter compared to the standard spatially flat ΛCDM model, both $\phi$CDM and XCMB models better fit the TT + lowP + lensing + BAO peak length scale data than does the standard spatially flat ΛCDM model (Figure 21). For the $\phi$CDM model, $\triangle\chi^2 = -1.60$ and, for the XCDM model, $\triangle\chi^2 = -1.26$ relative to the ΛCDM model. The improvement over the standard spatially flat ΛCDM model in 1.3$\sigma$ and in 1.1$\sigma$ for the XCDM model are not significant, but these dynamical dark energy models cannot be ruled out. Both the $\phi$CDM and XCMB dynamical dark energy models reduce the tension between the Planck 2015 (Aghanim et al. [291]) CMB temperature anisotropy and the weak lensing constraints of the rms amplitude of mass fluctuations $\sigma_8$.

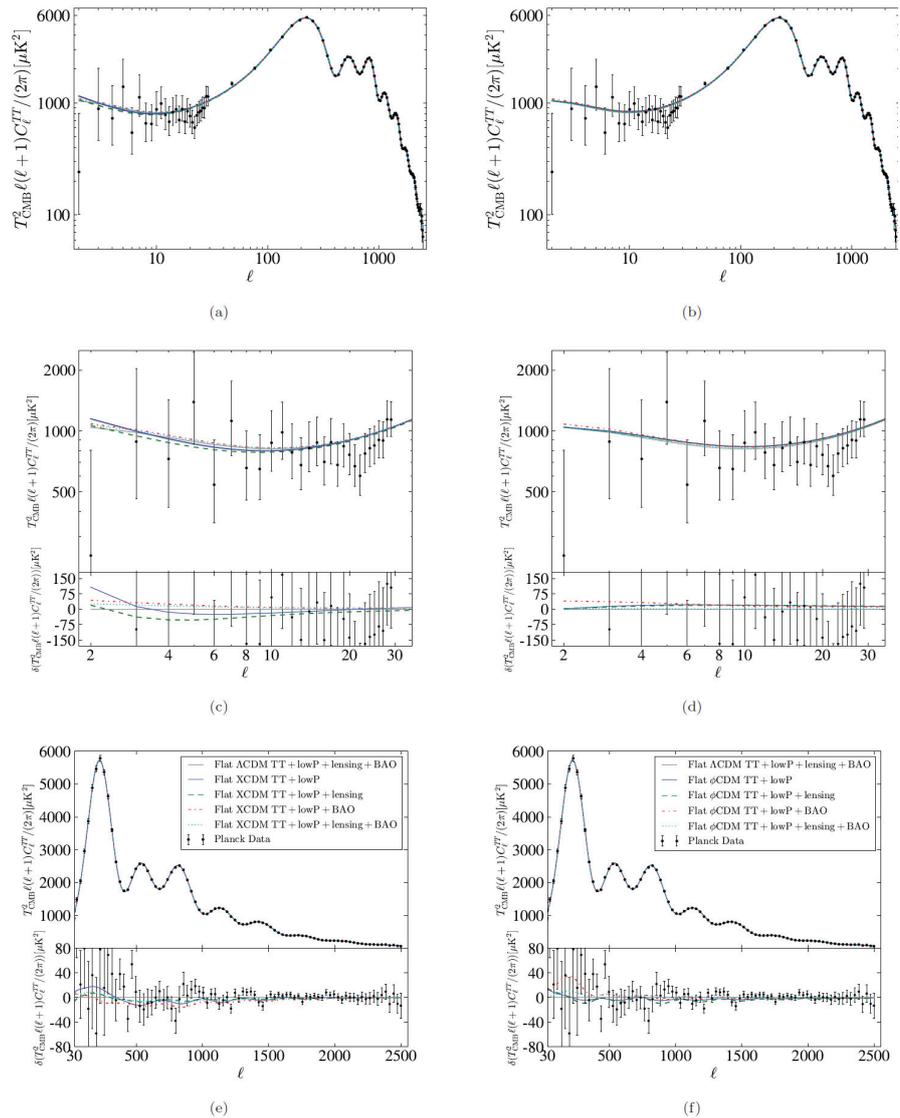

**Figure 21.** (Left panels (**a**,**c**,**e**)) The comparison of the spatially flat tilted ΛCDM model (gray solid line) with the best−fit $C_l$'s for the XCDM model and the $\phi$CDM model. (Right panels (**b**,**d**,**f**)) The comparison of the spatially flat tilted ΛCDM model (gray solid line) with the best-fit $C_l$'s for the $\phi$CDM model. The all−$l$ region is shown on top panels. The low-$l$ region $C_l$ and residuals are represented on middle panels. The high−$l$ region $C_l$ and residuals are demonstrated on bottom panels. The figure is adapted from [293].

Mitra et al. [195] studied the influence of dynamical dark energy and spatial curvature on cosmic reionization. For this aim, the authors examined reionization in the **tilted spa-** **tially flat** and **untilted spatially non-flat XCDM and $\phi$CDM-RP quintessential inflation models**. Statistical analysis was performed based on a principal component analysis and the MCMC analysis using a compilation of the lower-redshift reionization data by Wyithe & Bolton [296], Becker & Bolton [297] to estimate uncertainties in the model reionization histories. The obtained constraints for the tilted spatially flat and untilted spatially non-flat dynamical inflation $\phi$CDM model with the RP potential are shown in Fig. (22). The authors found that regardless of the nature of dark energy, there are significant differences between the reionization histories of the spatially flat and spatially non-flat cosmological models. Although both the flat and non-flat models fit well the low-redshift $z \leq 6$ reionization observations, there is a discrepancy between high-redshifts $z \geq 7$ Lyman-$\alpha$ emitter data by Songaila & Cowie [298], Prochaska *et al.* [299] and the predictions from spatially non-flat models.

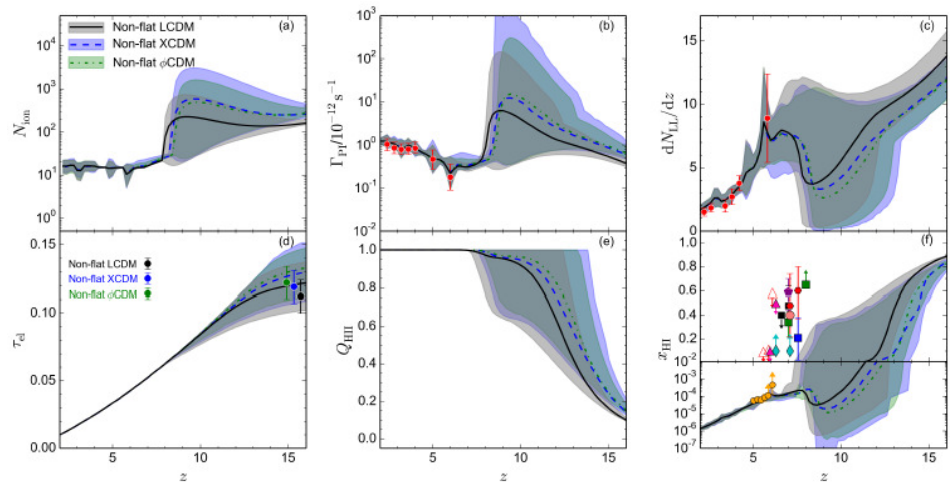

**Figure 22.** Constraints on various quantities related to reionization obtained from the MCMC analysis for the untilted spatially non-flat $\Lambda$CDM, XCDM, and $\phi$CDM quintessential inflation models that fit best the dataset: Planck 2015 TT + lowP + lensing and SNe Ia apparent magnitude, BAO peak length scale, $H(z)$, and LSS growth rate data. The thick central lines along with surrounding shaded regions correspond to best−fit models and their $2\sigma$ uncertainty ranges. (Upper panels) (**a**) A number of ionizing photons in the IGM per baryon in stars, (**b**) photoionization rates for hydrogen along with observational data from Wyithe and Bolton [296] and Becker and Bolton [297], (**c**) a specific number of Lyman−limit systems with data points from Songaila and Cowie [298] and Prochaska et al. [299]. (Lower panel) (**d**) Electron scattering optical depths along with their values from Park and Ratra [290], (**e**) volume filling factor of ionized regions, (**f**) global neutral hydrogen fraction with different present observational limits. The figure is adapted from [195].

The compilation of the South Pole Telescope polarization (SPTpol) CMB temperature anisotropy data [300], alone and in combination with the Planck 2015 CMB temperature anisotropy data [291] and the non-CMB temperature anisotropy data, consisting of the Pantheon Type SNe Ia apparent magnitude measurements [213], the BAO peak length scale measurements [22,24–26,292], the Hubble parameter $H(z)$ data [21,28,30–34,258], and the LSS growth rate data [25], was used by Park and Ratra [301] to obtain constraints on parameters of the spatially flat and untilted spatially non-flat $\Lambda$CDM and XCDM scalar field $\phi$CDM-RP quintessential inflation models. In each dark energy model, constraints on the cosmological parameters from the SPTpol measurements, the Planck CMB temperature anisotropy, and the non-CMB temperature anisotropy measurements are largely consistent with one another. Smaller angular scale SPTpol measurements (used jointly with only the Planck CMB temperature anisotropy data or with the combination of the Planck CMB temperature anisotropy data and the non-CMB temperature anisotropy data) favor the untilted spatially closed models.

Di Valentino et al. [121] explored the IDE models to find out whether these models can resolve both the Hubble constant $H_0$ tension problem of the standard spatially flat ΛCDM model and resolve the contradictions between observations of the Hubble constant in high and low redshifts in the spatially non-flat ΛCDM scenario.

The authors constrained parameters of the spatially flat IDE and ΛCDM models as well as the spatially non-flat IDE and ΛCDM models applying the CMB Planck 2018 data [13], BAO [22,24,25] measurements, 1048 data points in the redshift range $z \in (0.01, 2.3)$ of the Pantheon SNe Ia luminosity distance data [213], and a Gaussian prior of the Hubble constant ($H_0 = 74.03 \pm 1.42$ km s$^{-1}$Mpc$^{-1}$ at $1\sigma$ CL), obtained from a reanalysis of the HST data by the SH0ES collaboration [112].

Based on the results of this observational analysis, it was found that the Planck 2018 CMB data favor spatially closed hypersurfaces at more than 99% CL (with a significance of $5\sigma$), while a larger value of the Hubble constant, i.e., alleviation of the Hubble constant tension (with a significance of $3.6\sigma$) has been obtained for the spatially non-flat IDE models. The authors concluded that searches for other forms of the interaction coupling parameter and the EoS for the dark energy component in IDE models are needed, which may further ease the tension of the Hubble constant. The $1\sigma$ and $2\sigma$ confidence level contours on parameters of the spatially non-flat IDE model are shown in Figure 23.

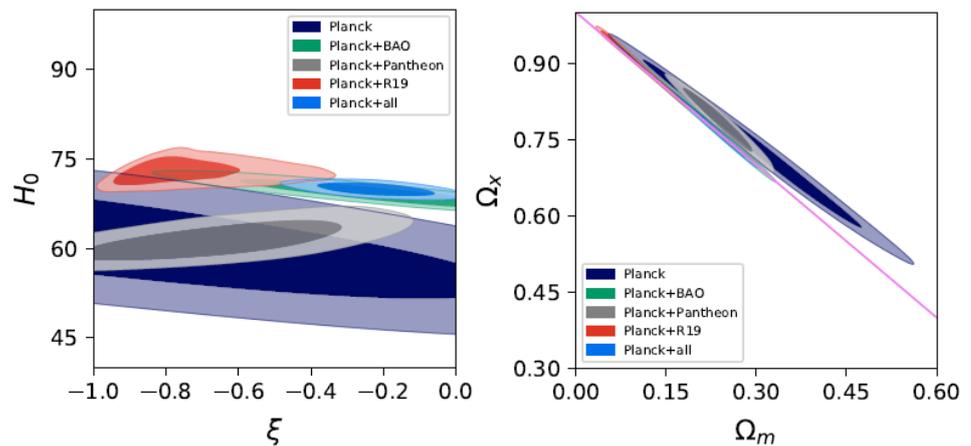

**Figure 23.** The $1\sigma$ and $2\sigma$ confidence level contours on parameters of the spatially non-flat IDE model, presented in ($H_0, \xi$) plane (left panel) and in ($\Omega_\Lambda, \Omega_m$) plane (right panel). Here $\xi$ is the dimensionless coupling parameter which characterizes the strength of the interaction between the dark sectors. The figure is adapted from [121].

Investigating both the minimally and non-minimally coupled to gravity the spatially-flat scalar field $\phi$CDM-RP and the extended quintessence models, Davari et al. [222] applied the following dataset: the Pantheon SNe Ia luminosity distance data [213], BAO (6dFGS, SDSSLRG, BOSS-MGS, BOSS-LOWZ, WiggleZ, BOSS-CMASS, BOSS-DR12), CMB [302], $H(z)$ [112], and redshift space distortion (RSD) [303]. According to their results, the ΛCDM model has a strong advantage when local measurements of the Hubble constant $H_0$ [13] are not taken into account and, conversely, this statement is weakened when local measurements of $H_0$ are included to the data analysis.

### 3.3. Large-Scale Structure Growth Rate Data

Another potentially powerful probe of the $\phi$CDM signatures is the growth rate in the low redshift LSS. The growth rate is expected to be stronger in the $\phi$CDM models compared to their ΛCDM counterparts.

Pavlov et al. [304] constrained the spatially flat $\phi$CDM-RP, the XCDM, the $w$CDM, and the ΛCDM models from future LSS growth rate data, by considering that the full sky space-based survey will observe $H_\alpha$-emitter galaxies over 15,000 deg$^2$ of the sky. For the bias and density of observed galaxies, they applied the predictions of Orsi et al. [305] and Geach et al. [306], respectively. They also assumed that half of the galaxies would be detected within the reliable redshift range, which roughly reflects the expected outcomes of proposed space missions, such as the ESA's Euclidean Space Telescope (Euclid) mission and the NASA's Nancy Grace Roman Space Telescope mission. The obtained results are shown in Figure 24, where we see that measurements of the LSS growth rate in the near future will be able to constrain scalar field $\phi$CDM models with an accuracy of about 10%, considering the fiducial spatially flat $\Lambda$CDM model, an improvement of almost an order of magnitude compared to those from currently available datasets [260,307–312]. Constraints on the growth index parameter $\gamma$ are more restrictive in the $\Lambda$CDM model than in other models. For the $\phi$CDM model, constraints on the growth index parameter $\gamma$ are about a third tighter than for the $w$CDM and XCDM models.

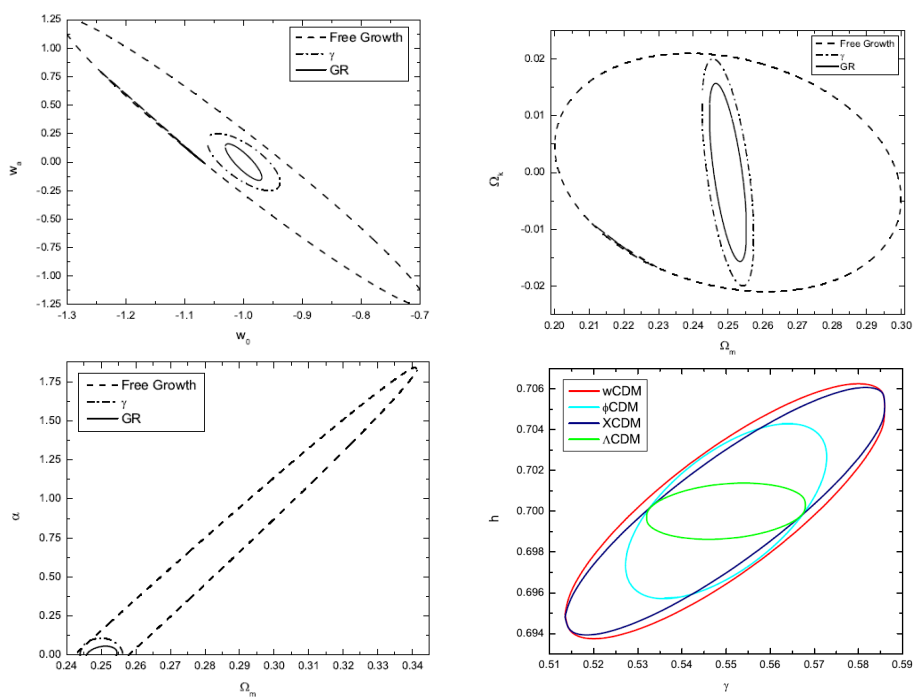

**Figure 24.** (Left upper panel) The $1\sigma$ confidence level contour constraints on parameters $w_a$ and $w_0$ of the $w$CDM model. (Right upper panel) The $1\sigma$ confidence level contour constraints on parameters $\Omega_k$ and $\Omega_m$ of the $w$CDM model. (Left lower panel) The $1\sigma$ confidence level contour constraints on parameters $\alpha$ and $\Omega_m$ of the scalar field spatially flat $\phi$CDM model with the inverse power−law RP potential. (Right lower panel) The $1\sigma$ confidence level contour constraints on the normalized Hubble constant $h$ and the parameter $\gamma$ describing deviations from general relativity for various dark energy models. The figure is adapted from [304].

Pavlov et al. [313] also obtained constraints on the above DE models from the Hubble parameter $H(z)$ observations [28,30,258,259], the Union2.1 compilation of 580 SNe Ia apparent magnitude measurements [257], and a compilation of 14 independent LSS growth rate measurements within the redshift range $0.067 \leq z \leq 0.8$ [21,22,260,314]. The authors performed two joint analyses, first for the combination of the $H(z)$ and SNe Ia apparent magnitude data, and second for measurements of the LSS growth rate, the Hubble parameter $H(z)$, and the SNe Ia apparent magnitude. The results of these analyses are presented in Figure 25. Constraints on cosmological parameters of the spatially flat $\phi$CDM model from LSS growth rate data are quite restrictive. In combination with the SNe Ia apparent magnitude versus the redshift data and the Hubble parameter measurements, the LSS growth rate data are consistent with the standard spatially flat $\Lambda$CDM model, as well as with the spatially flat $\phi$CDM model.

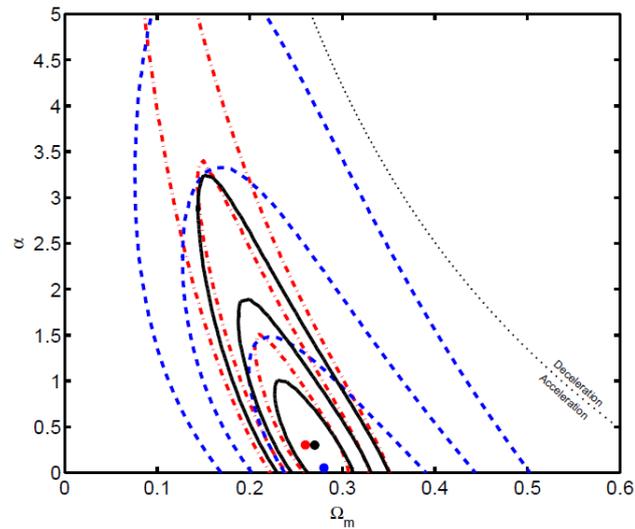

**Figure 25.** The 1 $\sigma$, 2$\sigma$, and 3$\sigma$ confidence level contour constraints on parameters of the spatially flat scalar field $\phi$CDM model with the inverse power-law RP potential from LSS growth rate measurements (blue dashed lines with blue filled circle at best fit $(\Omega_\mathrm{m}, \alpha) = (0.28, 0.052)$, $\chi^2_\mathrm{min}/\mathrm{dof} = 8.62/12$); SNe Ia apparent magnitude+$H(z)$ data (red dot–dashed lines with red filled circle at best fit $(\Omega_\mathrm{m}, \alpha) = (0.26, 0.302)$, $\chi^2_\mathrm{min}/\mathrm{dof} = 562/598$); and a combination of all datasets (black solid lines and black filled circle at the best fit $(\Omega_\mathrm{m}, \alpha) = (0.27, 0.300)$, $\chi^2_\mathrm{min}/\mathrm{dof} = 570/612$). The horizontal axis with $\alpha = 0$ corresponds to the standard spatially flat $\Lambda$CDM model and the curved dotted line denotes zero-acceleration models. The figure is adapted from [313].

Avsajanishvili et al. [315] constrained the parameters $\Omega_\mathrm{m}$ and $\alpha$ of the spatially flat $\phi$CDM-RP model. Applying only measurements of the LSS growth rate [316], the authors obtained a strong degeneracy between the model parameters $\Omega_\mathrm{m}$ and $\alpha$, Figure 26 (left panel). This was followed by obtaining constraints from a compilation of data from the LSS growth rate measurements [316], and the distance–redshift ratio of the BAO peak length scale observations, and prior distance from the CMB temperature anisotropy [317], which eliminated the degeneracy between $\Omega_\mathrm{m}$ and $\alpha$, giving $\Omega_\mathrm{m} = 0.30 \pm 0.04$ and $0 \leq \alpha \leq 1.30$ at a $1\sigma$ confidence level (the best-fit value for the model parameter $\alpha$ is $\alpha = 0$). Constraints on $\Omega_\mathrm{m}$ and $\alpha$ from the data compilation of Gupta et al. (2012) [316] and Giostri et al. (2012) [317] are presented in Figure 41 (right panel).

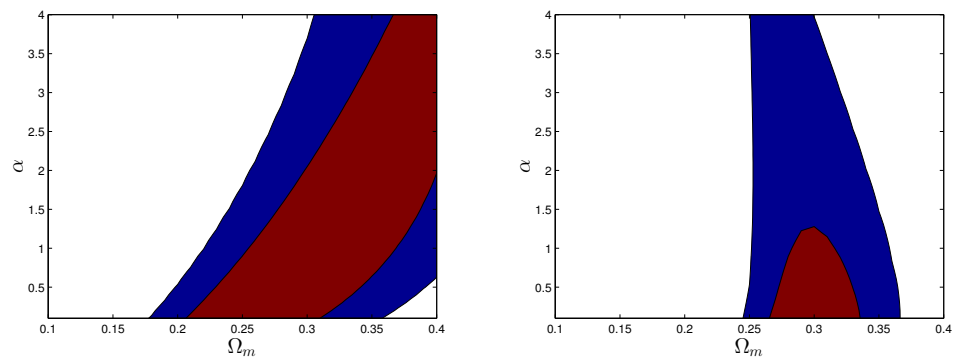

**Figure 26.** The 1 $\sigma$ and 2$\sigma$ confidence level contour constraints on parameters $\Omega_\mathrm{m}$ and $\alpha$ in the scalar field $\phi$CDM model with the inverse power-law RP potential. (Left panel) Constraints are obtained from the LSS growth rate data [316]. (Right panel) Constraints are obtained from the data compilation of Gupta et al. (2012) [316] and Giostri et al. (2012) [317]. The figure is adapted from [315].

Avsajanishvili et al. [318] also constrained various quintessence and phantom scalar field $\phi$CDM models, presented in Tables 1 and 2, using observational data predicted for the Dark Energy Spectroscopic Instrument (DESI) [292]. The parameters of these models were constrained using the MCMC methods by comparing measurements of the expansion rate of the universe $H(z)$, the angular diameter distance $d_A$, and the LSS growth rate, predicted for the standard spatially flat ΛCDM model with corresponding values calculated for the $\phi$CDM models. Results of constraints for the Zlatev–Wang–Steinhardt potential, the phantom pNGb potential, and the inverse power-law RP potential are shown in Figures 27, 28 and 29. To compare quintessence and phantom models, Bayesian statistical tests were conducted, namely, the Bayes factor, and the $AIC$ and $BIC$ information criteria were calculated. The $\phi$CDM scalar field models could not be unambiguously preferred, from the DESI predictive data, over the standard ΛCDM spatially flat model, the latter still being the most preferred dark energy model. The authors also investigated how the $\phi$CDM models can be approximated by the CPL parameterization, by plotting the CPL-ΛCDM $3\sigma$ confidence level contours, using MCMC techniques, and displayed on them the largest ranges of the current EoS parameters for each $\phi$CDM model. These ranges were obtained for different values of model parameters or initial conditions from the prior ranges. The authors classified the scalar field models based on whether they can or cannot be distinguished from the standard spatially flat ΛCDM model at the present epoch, as seen in Figure 30. They found that all studied models can be divided into two classes: models that have attractor solutions and models whose evolution depends on initial conditions.

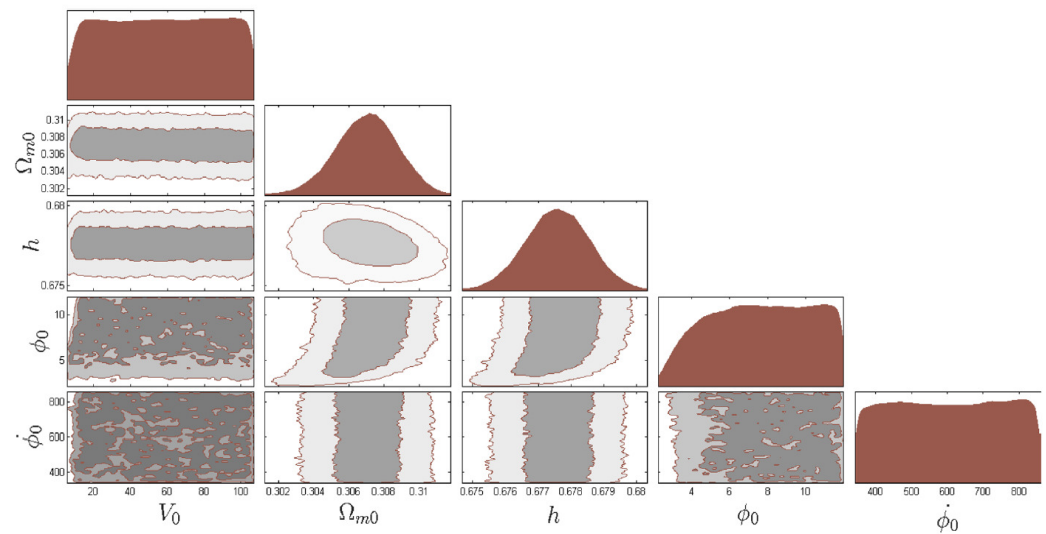

**Figure 27.** The $1\sigma$ and $2\sigma$ confidence level contour plots for various pairs of free parameters ($V_0$, $\Omega_{m0}$, $h$, $\phi_0$, and $\dot{\phi}_0$), for which the spatially flat $\phi$CDM model with the Zlatev–Wang–Steinhardt potential is in the best fit with the standard spatially flat ΛCDM model. The figure is adapted from [318].

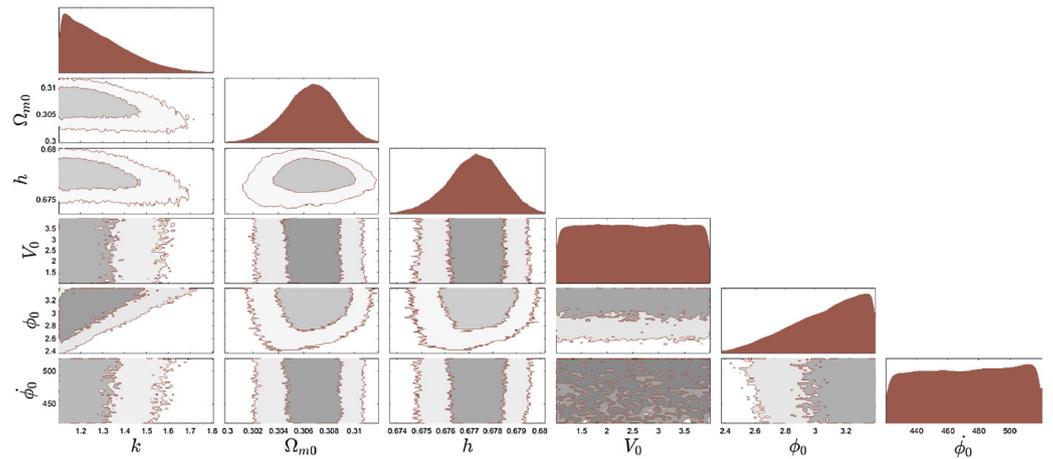

**Figure 28.** The 1$\sigma$ and 2$\sigma$ confidence level contour plots for various pairs of free parameters ($k$, $\Omega_{m0}$, $h$, $V_0$, $\phi_0$, and $\dot{\phi}_0$), for which the spatially flat $\phi$CDM model with the phantom PNGB potential is in the best fit with the standard spatially flat $\Lambda$CDM model. The figure is adapted from [318].

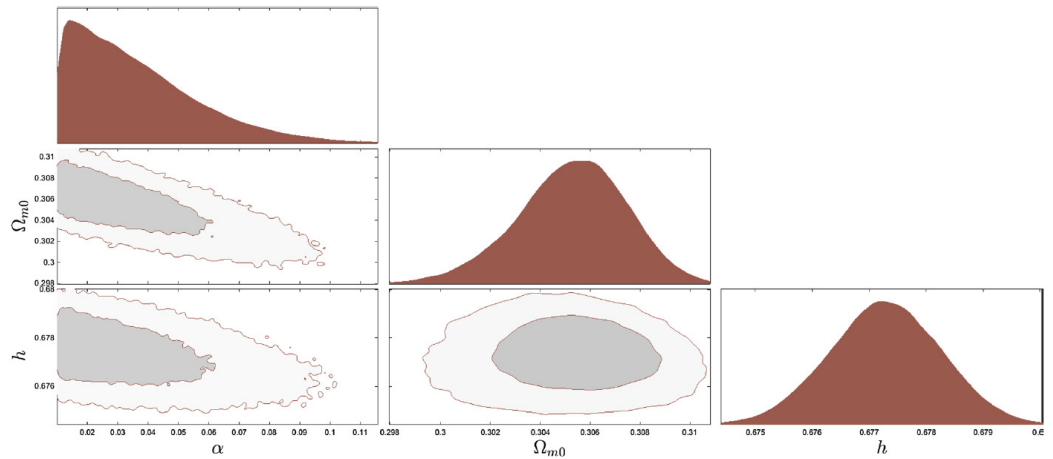

**Figure 29.** The 1 $\sigma$ and 2$\sigma$ confidence level contour plots for various pairs of free parameters ($\alpha$, $\Omega_{m0}$, $h$), for which the spatially flat $\phi$CDM model with the RP potential is the best fit with the standard spatially flat $\Lambda$CDM model. The figure is adapted from [318].

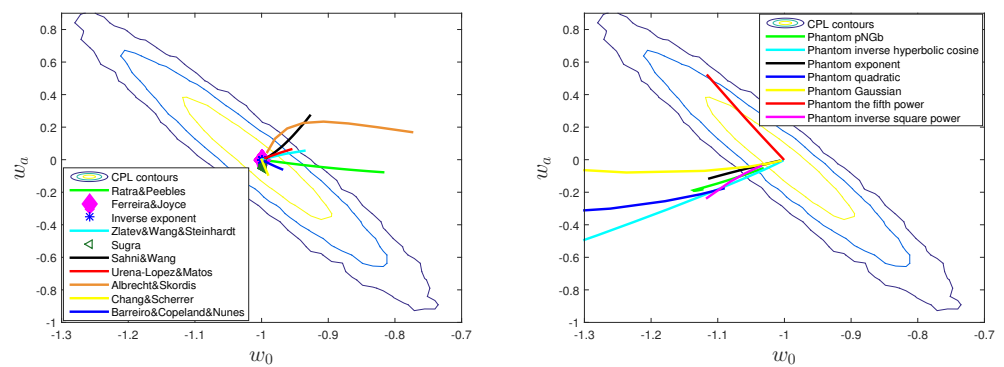

**Figure 30.** (Left panel) The comparison of the possible $w_0$ and $w_a$ values for quintessence dark energy potentials in the spatially flat scalar field $\phi$CDM models with the CPL$-\Lambda$CDM 1$\sigma$, 2$\sigma$, and 3$\sigma$ confidence level contours. (Right panel) The comparison of possible $w_0$ and $w_a$ values for phantom dark energy potentials in the spatially flat scalar field $\phi$CDM models with the CPL$-\Lambda$CDM 1$\sigma$, 2$\sigma$, and 3$\sigma$ confidence level contours. The figure is adapted from [318].

Peracaula et al. [319] constrained the spatially flat ΛCDM, XCDM, and $\phi$CDM-RP models by constructing three datasets: DS1/SP consisting of SNe Ia apparent magnitude + $H(z)$ + BAO peak length scale + LSS growth rate + CMB temperature anisotropy data with matter power spectrum SP; DS1/BSP consisting of SNe Ia apparent magnitude + $H(z)$ + BAO peak length scale + LSS growth rate+CMB temperature anisotropy data with both matter power spectrum and bispectrum; and DS2/BSP, which involves BAO peak length scale + LSS growth rate + CMB temperature anisotropy data with both matter power spectrum and bispectrum. These datasets include 1063 SNe Ia apparent magnitude data [110,213], 31 measurements of $H(z)$ from cosmic chronometers [35,258], 16 BAO peak length scale data [320,321], LSS growth rate data, specifically 18 points from data [21,321,322], one point from the weak lensing observable $S_8$ [323], and full CMB likelihood from Planck 2015 TT + lowP + lensing [12]. The obtained constraints are shown in Figures 31 and 32. The authors tested the effect of separating the expansion history data (SNe Ia apparent magnitude + $H(z)$) from the CMB temperature anisotropy characteristics and the LSS formation data (BAO peak length scale + LSS), where LSS includes RSD and weak lensing measurements, and found that the expansion history data are not particularly sensitive to the dynamic effects of dark energy, while the data compilation of BAO peak length scale + LSS + CMB temperature anisotropy is more sensitive. Also, the influence of the bispectral component of the matter correlation function on the dynamics of dark energy is studied. For this the BAO peak length scale + LSS data were considered, including both the conventional power spectrum and the bispectrum. As a result, when the bispectral component is excluded, the results obtained are consistent with previous studies by other authors, which means that no clear signs of dynamical dark energy have been found in this case. On the contrary, when the bispectrum component was included in the BAO peak length scale + LSS growth rate dataset for the $\phi$CDM model, a significant dynamical dark energy signal was achieved at a $2.5 - 3\sigma$ confidence level. The bispectrum can therefore be a very useful tool for tracking and examining the possible dynamical features of dark energy and their influence on the LSS formation in the linear regime.

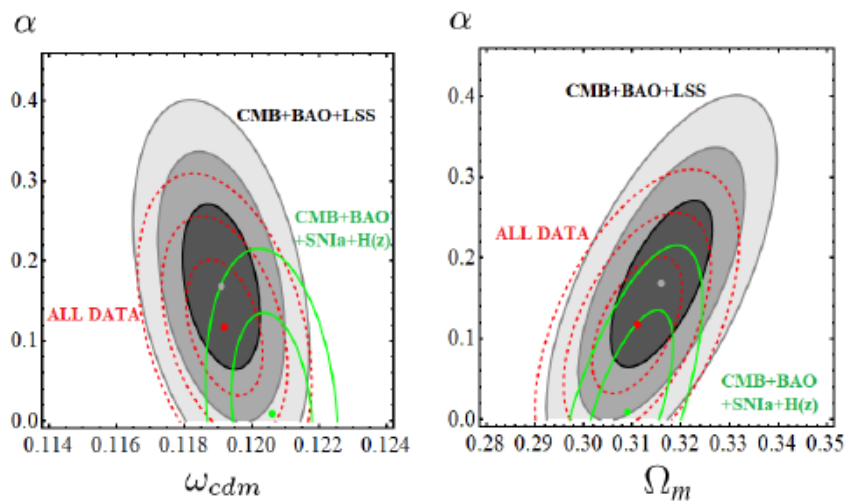

**Figure 31.** The $1\sigma$, $2\sigma$, and $3\sigma$ confidence level contour constraints on parameters of the spatially flat scalar field $\phi$CDM model with the inverse power-law RP potential using different combinations of datasets and the compressed Planck 2018 data [13]. The results obtained from DS2/BSP dataset CMB temperature anisotropy + BAO peak length scale+LSS growth rate (gray contours), DS1/BSP dataset: SNe Ia apparent magnitude + $H(z)$ + BAO peak length scale + LSS growth rate + CMB temperature anisotropy (dashed red contours), and DS1/BSP dataset without LSS growth rate data CMB temperature anisotropy + BAO peak length scale + SNe Ia apparent magnitude + $H(z)$ (solid green contours). The results are presented in the $w_{\text{cdm}} - \alpha$ plane (left panel) and in the $w_{\text{cdm}} - \Omega_{\text{m}}$ plane (right panel), where $w_{\text{cdm}} = \Omega_{\text{m}} h^2$ is a physical matter density parameter. The figure is adapted from [319].

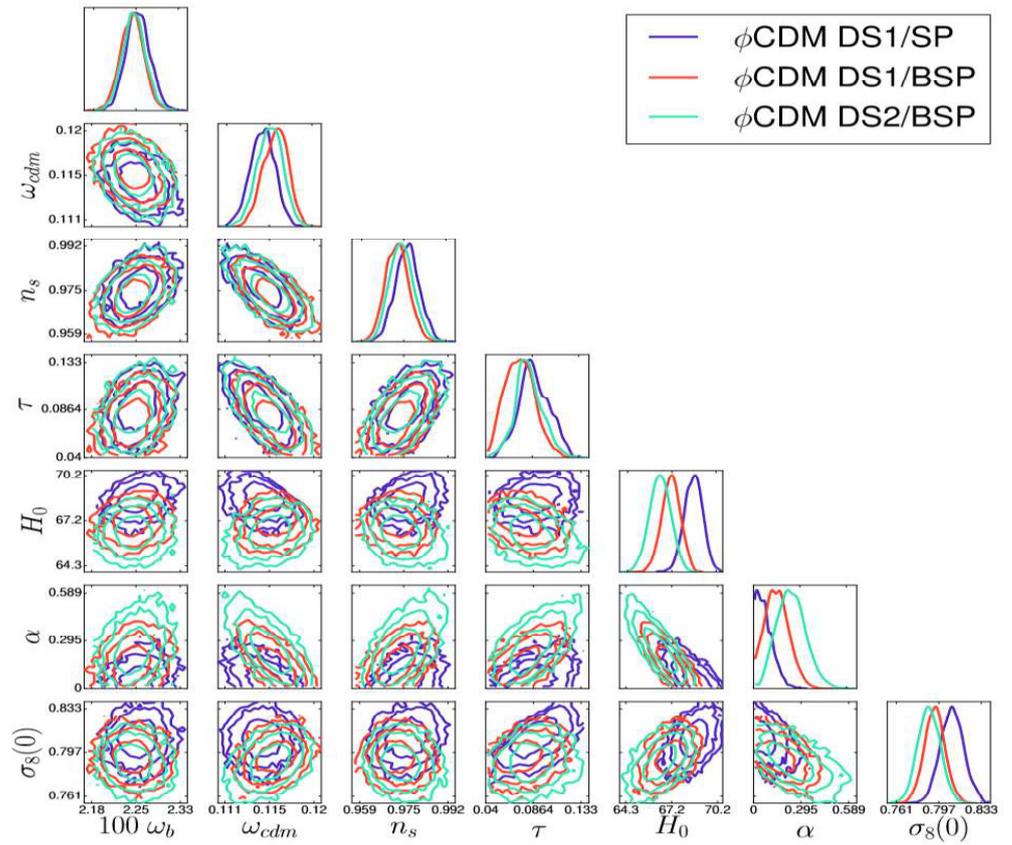

**Figure 32.** The 1 $\sigma$, 2$\sigma$, and 3$\sigma$ confidence level contour constraints on parameters of the spatially flat scalar field $\phi$CDM model with the inverse power-law RP potential from DS1/SP, DS1/BSP, and DS2/BSP datasets. The figure is adapted from [319].

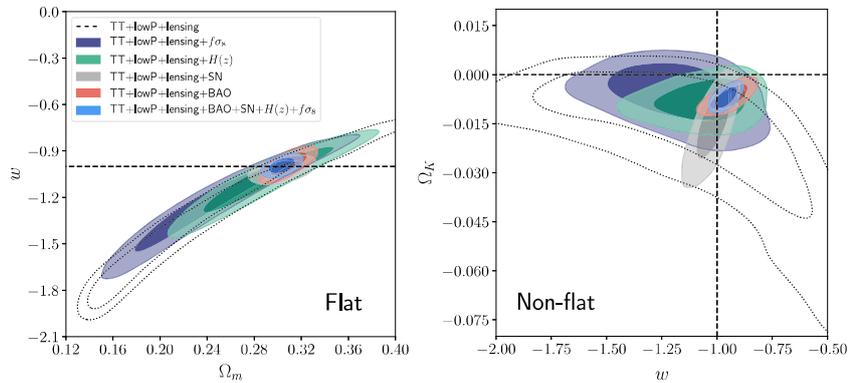

**Figure 33.** The 1 $\sigma$ and 2$\sigma$ confidence level contour for the tilted spatially flat XCDM model (left panel) and for the untilted spatially non-flat XCDM model (right panel), constrained by Planck CMB TT + lowP + lensing and non-CMB datasets. The horizontal and vertical dashed lines indicate the standard spatially flat $\Lambda$CDM model (with $w = -1$ and $\Omega_k = 0$). The figure is adapted from [324].

Park and Ratra [324] constrained the tilted spatially flat and untilted spatially non-flat XCDM models by applying the Planck 2015 CMB temperature anisotropy data [291], BAO peak length scale measurements [26], a Pantheon collection of 1048 SNe Ia apparent magnitude measurements over the broader redshift range $0.01 < z < 2.3$ [213], Hubble parameter observations [21,25,28,30–34,258,292], and LSS growth rate measurements [25], and obtained results as shown in Figures 33 and 34. These data slightly favor the spatially closed XCDM model over the spatially flat $\Lambda$CDM model at a 1.2$\sigma$ confidence level, while also being in better agreement with the untilted spatially flat XCDM model than with the spatially flat $\Lambda$CDM model at the 0.3$\sigma$ confidence level. Current observational data are unable to rule out dynamical dark energy models. The dynamical untilted spatially non-flat XCDM model is compatible with the Dark Energy Survey (DES) limits on the current value of the rms mass fluctuation amplitude $\sigma_8$ as a function of the matter density parameter at the present epoch $\Omega_{m0}$ but it does not give such a good agreement with higher multipoles of CMB temperature anisotropy data as the standard spatially flat $\Lambda$CDM model.

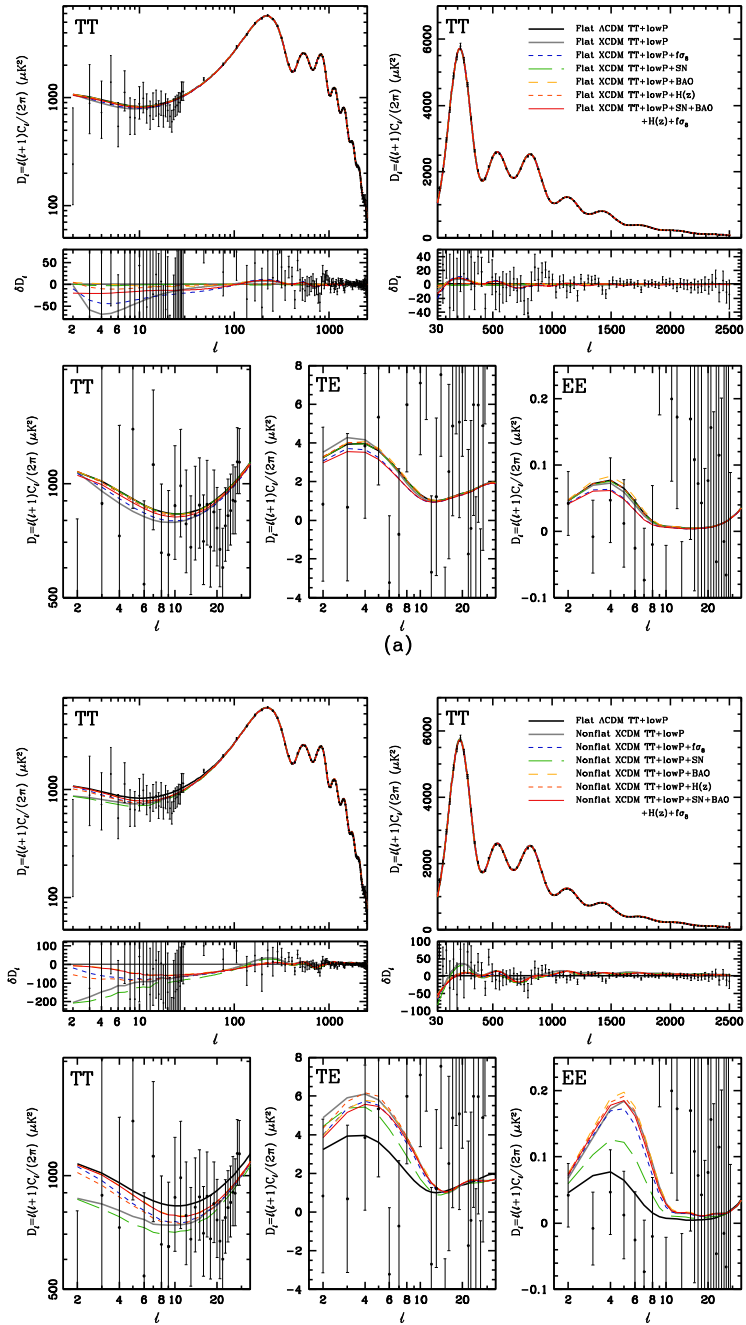

**Figure 34.** Best-fit CMB temperature anisotropy power spectra of (**a**) the tilted spatially flat XCDM model (top five panels) and (**b**) the untilted spatially non-flat XCDM model (bottom five panels) constrained by Plank CMB TT + lowP data (excluding lensing data) together with data on SNe Ia apparent magnitude, BAO peak length scale, $H(z)$, and LSS growth rate. The best-fit power spectra of the tilted spatially flat $\Lambda$CDM model are shown as black curves. The residual $\delta D_l$ of the TT power spectra are shown with respect to the spatially flat $\Lambda$CDM power spectrum that best fits the TT + lowP data. The high-$l$ region $C_l$ and residuals are shown on the bottom panels. The figure is adapted from [324].

*3.4. Baryon Acoustic Oscillations Data*

Samushia and Ratra [325] constrained the standard spatially flat ΛCDM, the XCDM, and the $\phi$CDM-RP models from BAO peak length scale measurements [17,20], in conjunction with WMAP measurements of the apparent acoustic horizon angle and galaxy cluster gas mass fraction measurements [283]. These constraints are presented in Figure 35. It is seen that the measurements of Percival et al. (2007) [17] constrain the $\phi$CDM model less effectively (left panel of Figure 35), while measurements of the joint BAO peak length scale and the galaxy cluster gas mass give consistent and more accurate constraints on the parameters of the $\phi$CDM model than those derived from other data, i.e., $\alpha < 3.5$ (right panel of Figure 35).

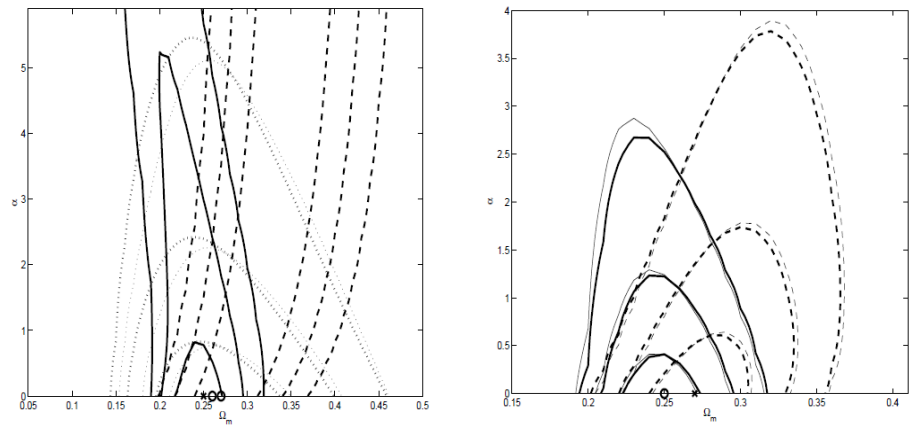

**Figure 35.** The $1\sigma$, $2\sigma$, and $3\sigma$ confidence level contour constraints on parameters of the scalar field $\phi$CDM model with the inverse power-law RP scalar field potential. The $\alpha = 0$ axis corresponds to the standard spatially flat ΛCDM model. (Left panel) Solid lines are constraints derived by Percival et al. (2007) [17] using BAO peak length scale data in conjunction with WMAP data on acoustic horizon angle. Dashed lines are constraints obtained by Eisenstein et al. (2005) [20] from BAO peak length scale data. The circle denotes the best-fit value. Two sets of dotted lines are constraints obtained from galaxy cluster gas mass fraction measurements of Samushia and Ratra (2008) [325]; thick dotted lines are derived using WMAP priors for $h$ and $\Omega_b h^2$ while thin dotted lines are obtained for alternate priors. The cross denotes the best-fit value. (Right panel) Solid lines are joint constraints obtained by Percival et al. (2007) from BAO peak length scale data in conjunction with WMAP data on acoustic horizon angle and galaxy cluster gas mass fraction measurements. The circle denotes the best-fit value with a suitable $\chi^2 \simeq 58$ for 42 degrees of freedom; dashed lines are joint constraints derived by Eisenstein et al. (2005) [20] using BAO peak length scale data. The cross denotes the best-fit value with a suitable $\chi^2 \simeq 52$ for 41 degrees of freedom. Thick lines are derived using the WMAP priors for $h$ and $\Omega_b h^2$, and thin lines are for alternate priors. Joint best-fit values for two prior sets overlap. Here, $\Omega_m$ and $\alpha$ ranges are smaller than those shown on the left panel. The figure is adapted from [325].

The above models were also constrained by Samushia et al. [326] using the lookback time versus redshift data [327], the passively evolving galaxies data [258], the current BAO peak length scale data, and the SNe Ia apparent magnitude measurements. Applying a Bayesian prior on the total age of the universe based on WMAP data, the authors obtained constraints on the $\phi$CDM model as shown in Figure 36. Constraints on the $\phi$CDM model by joint datasets consisting of measurements of the age of the universe, SNe Ia Union apparent magnitude, and BAO peak length scale are tighter than those obtained from datasets consisting of data on the lookback time and the age of the universe.

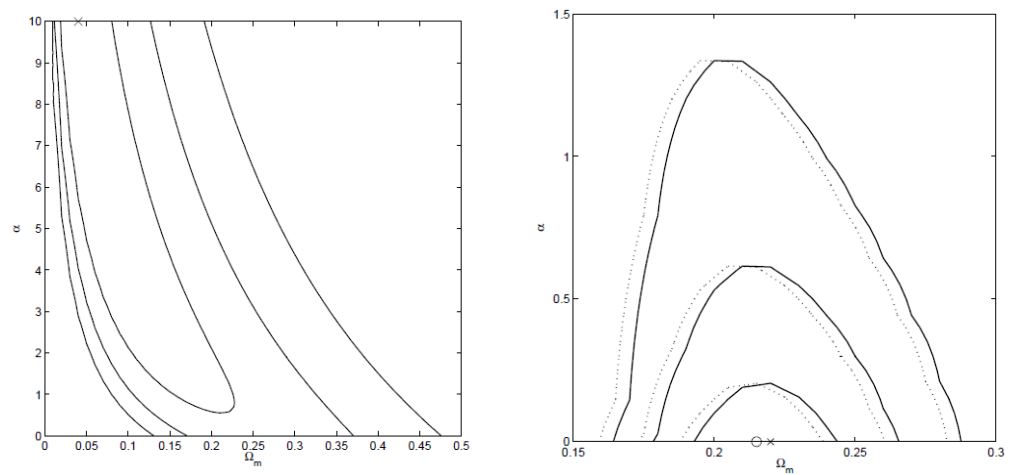

**Figure 36.** The 1 $\sigma$, 2$\sigma$, and 3$\sigma$ confidence level contour constraints on parameters of the scalar field $\phi$CDM model with the inverse power-law RP potential. The horizontal axis with $\alpha = 0$ corresponds to the standard spatially flat $\Lambda$CDM model. (Left panel) Dotted lines obtained from the lookback time data and measurements of the age of the universe. The cross denotes the best-fit parameters $\Omega_m = 0.04$ and $\alpha = 10$ with $\chi^2 = 22$, for $\alpha = 0$ with $\chi^2 = 359$ for 346 degrees of freedom derived using measurements of the lookback time, the age of the universe, SNe Ia apparent magnitude, and BAO peak length scale, while solid lines are derived using only SNe Ia apparent magnitude measurements and BAO peak length scale data. The cross denotes the best-fit point at $\Omega_m = 0.22$ and $\alpha = 0$ with $\chi^2 = 329$ for 307 degrees of freedom. The figure is adapted from [326].

The quintessential inflation model with the generalized exponential potential $V(\phi) \propto \exp(-\lambda \phi^n / M_{\rm pl}^n)$, $n > 1$ was studied by Geng et al. [178]. The authors extended this model including massive neutrinos that are non-minimally coupled to a scalar field, obtaining observational constraints on parameters from combinations of data: the CMB temperature anisotropy [288,288], the BAO peak length scale from BOSS [23,312], and the 11 SNe Ia apparent magnitudes from Supernova Legacy Survey (SNLS) [255]. It was found that the upper bound on possible values of the sum of neutrino masses $\sum m_\nu < 2.5$ eV is significantly larger than in the spatially flat $\Lambda$CDM model (Figure 37). The authors concluded that the model under consideration is in good agreement with observations and represents a successful scheme for the unification of the primordial inflaton field causing inflation in the very early universe and dark energy causing the accelerated expansion of the universe at the present epoch.

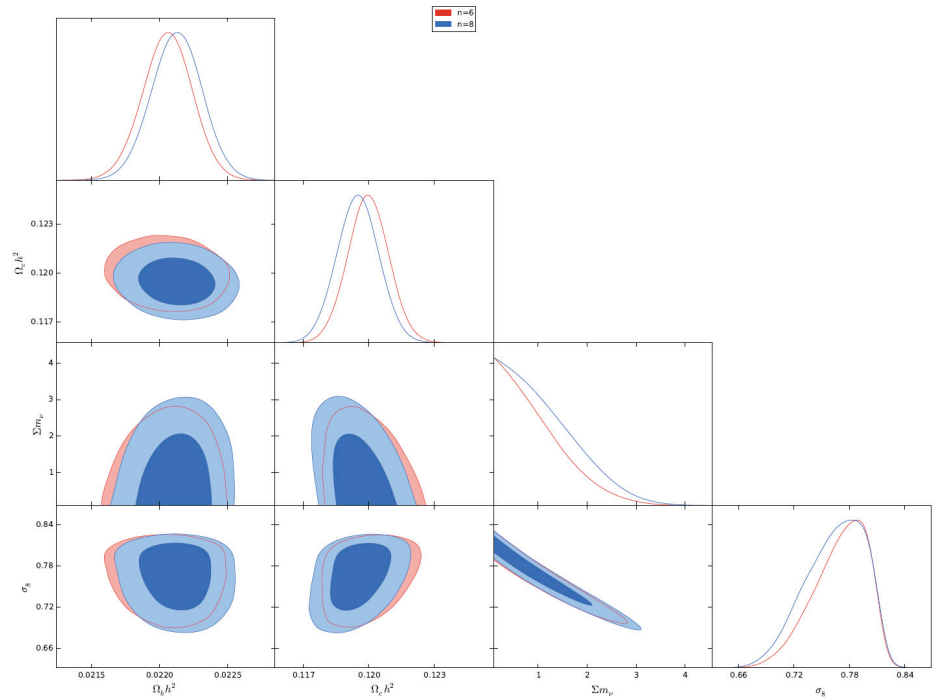

**Figure 37.** The 1 $\sigma$ and 2$\sigma$ confidence level contours of one- and two-dimensional distributions of $\Omega_b h^2$, $\Omega_m h^2$, $\sum m_\nu$, and $\sigma_8$ for the quintessential inflation model with the exponential potential $V(\phi) \propto \exp(-\lambda \phi^n / M_{\text{pl}}^n)$, $n = 6$ (orange line) and $n = 8$ (blue line). The figure is adapted from [178].

The compilation of CMB angular power spectrum data from the Planck 2015 mission [291] and BAO peak length scale measurements from the matter power spectra obtained by missions 6dFGS [22], BOSS, LOWZ and CMASS [23], and SDSS-MGS [24] were applied by Ooba et al. [328] to obtain constraints on the spatially non-flat quintessential inflation $\phi$CDM-RP model. The theoretical angular power spectra of the CMB temperature anisotropy were calculated using the Cosmic Linear Anisotropy Solving System (CLASS) code of Blas et al. [294] and the MCMC analysis was performed with Monte Python from Audren et al. [295]. The authors also used a physically consistent power spectrum for energy density inhomogeneities in the spatially non-flat (spatially closed) quintessential inflation $\phi$CDM model and found that the spatially closed $\phi$CDM model provides a better fit to the lower multipole region of CMB temperature anisotropy data compared to that provided by the tilted spatially flat $\Lambda$CDM model. The former reduces the tension between the Planck and the weak lensing $\sigma_8$ constraints, while the higher multipole region of the CMB temperature anisotropy data is in better agreement with the tilted spatially flat $\Lambda$CDM model than with the spatially closed $\phi$CDM model (Figure 38).

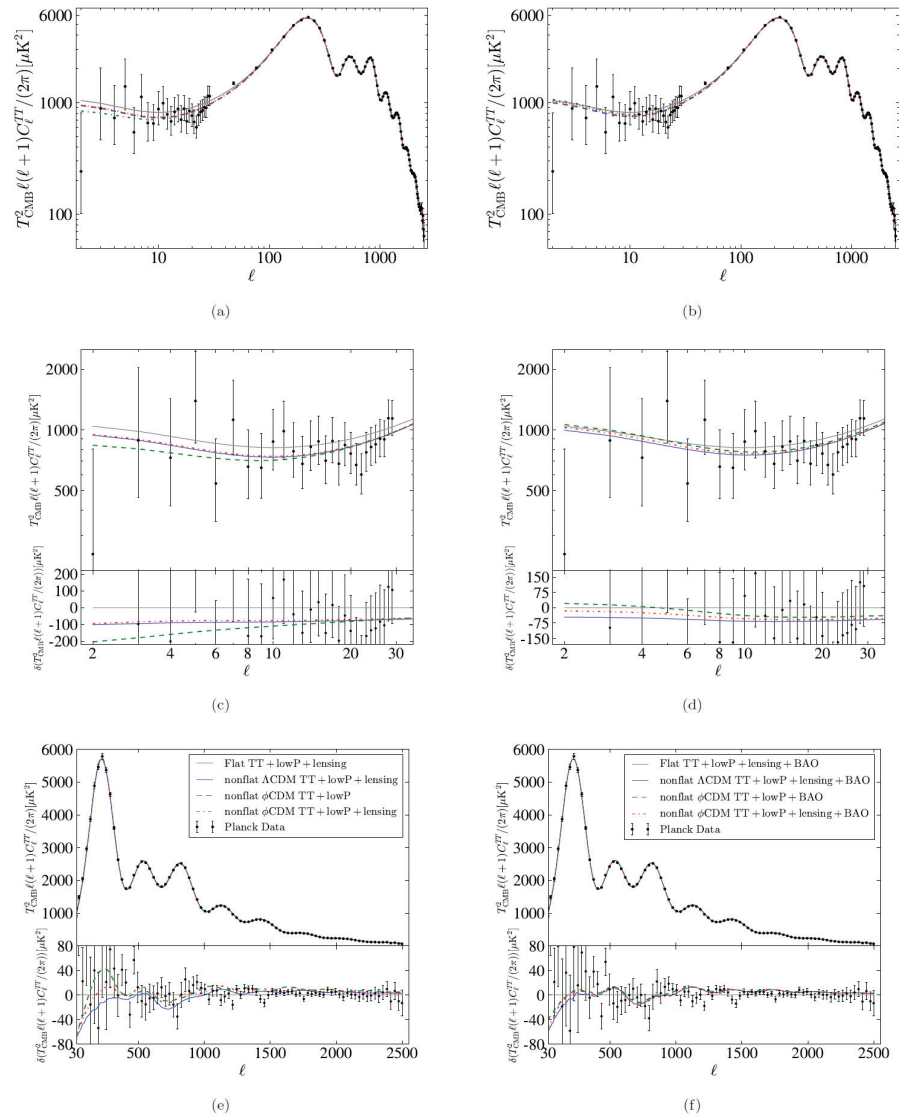

**Figure 38.** The $C_l$ for the best-fit spatially non-flat $\phi$CDM, spatially non−flat $\Lambda$CDM, and spatially flat tilted $\Lambda$CDM (gray solid line) models. (Left panels) (**a**,**c**,**e**) Results obtained from only CMB temperature anisotropy data. (Right panels) (**b**,**d**,**f**) results obtained only from CMB temperature anisotropy+BAO peak length scale data. All−$l$ regions are demonstrated in top panels. The low−$l$ region $C_l$ and residuals are shown in the middle panels. The high−$l$ region $C_l$ and residuals are presented in the bottom panels. The figure is adapted from [328].

Ryan et al. [329] constrained the parameters of the $\phi$CDM-RP, the XCDM, and the $\Lambda$CDM models from BAO peak length scale measurements [22,24–26,292] and the Hubble parameter $H(z)$ data [21,28,30–34,258]. The results obtained for the $\phi$CDM model are presented in Figure 39, which shows that this dataset is consistent with the standard spatially flat $\Lambda$CDM model. Depending on the value of the Hubble constant $H_0$ as a prior and the cosmological model under consideration, the data provides evidence in favor of the spatially non-flat scalar field $\phi$CDM model.

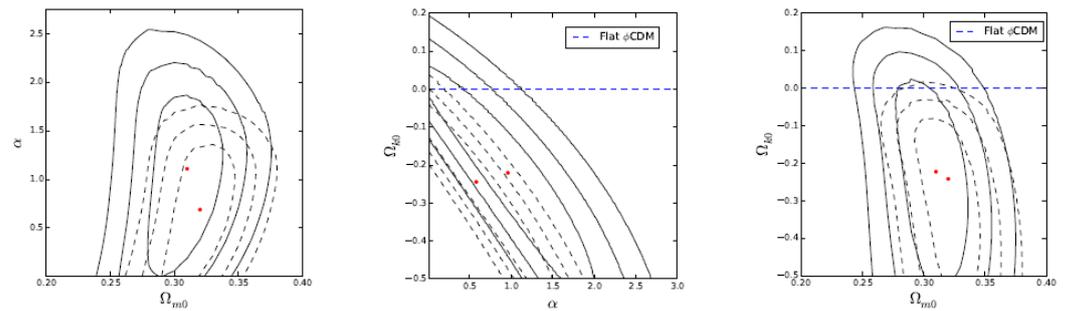

**Figure 39.** The 1 $\sigma$, 2$\sigma$, and 3$\sigma$ confidence level contour constraints on parameters of the spatially non-flat $\phi$CDM model with the RP potential. Solid (dashed) contours correspond to $H_0 = 68 \pm 2.8$ ($73.24 \pm 1.74$) km s$^{-1}$Mpc$^{-1}$ prior; the red dots indicate the location of the best-fit point in each prior case. The horizontal axis with $\alpha = 0$ denotes the spatially flat $\Lambda$CDM model. (Left panel) The results obtained for the $\Omega_{k0}$ marginalization. (Center panel) The results obtained for the $\Omega_{m0}$ marginalization. (Right panel) The results obtained for the parameter $\alpha$ marginalization. The figure is adapted from [329].

Chudaykin et al. [330] obtained constraints on the parameters of the $o$CDM, XCDM (here $w_0$CDM), and $w$CDM models by using the joint analysis from data on the BAO peak length scale, BBN, and SNe Ia apparent magnitude. The resulting constraints are completely independent of the CMB temperature anisotropy data but compete with the CMB temperature anisotropy constraints in terms of parameter error bars. The authors consequently obtained the value of the spatial curvature density parameter at the present epoch $\Omega_{k0} = -0.043^{+0.036}_{-0.036}$ at a 1$\sigma$ confidence level, which is consistent with the spatially flat universe; in the spatially flat XCDM model, the value of the dark energy EoS parameter at the present epoch $w_0 = -1.031^{+0.052}_{-0.048}$ at a 1$\sigma$ confidence level, which approximately equals the value of the EoS parameter for the $\Lambda$CDM model; values of the $w_0$ and $w_a$ in the CPL parameterization of the EoS parameter of the $w$CDM model $w_0 = -0.98^{+0.099}_{-0.11}$ and $w_a = -0.33^{+0.63}_{-0.48}$ at a 1$\sigma$ confidence level. The authors also found that the exclusion of the SNe Ia apparent magnitude data from the joint data analysis does not significantly weaken the resulting constraints. This means that, when using a single external BBN prior, full-shape and BAO peak length scale data can provide reliable constraints independent of CMB temperature anisotropy constraints. The authors also tightened the observational constraints on cosmological parameters with the inclusion of the hexadecapole ($l = 4$) moment of the redshift-space power spectrum.

Bernui et al. [67] investigated the effect of the BAO measurements on the IDE models that have significantly different dynamic behavior compared to the prediction of the standard $\Lambda$CDM model. The authors used the compilation of 15 transversal 2D BAO measurements [331,332] and CMB data [119] to constrain the IDE models. It was found that the transversal 2D BAO and traditional 3D BAO measurements can generate completely different observational constraints on the coupling parameter in the IDE models. Moreover, in contrast to the joint Planck + BAO analysis, where it is not possible to solve the Hubble constant $H_0$ tension, the joint Planck + BAO (transversal) analysis agrees well with the measurements made by the SH0ES team, and when applied to the IDE models, solves the Hubble constant $H_0$ tension. The 1$\sigma$ and 2$\sigma$ confidence level contour constraints on the coupling parameter $\xi$ in the IDE model using the 2D transversal 2D BAO are shown in Figure 40.

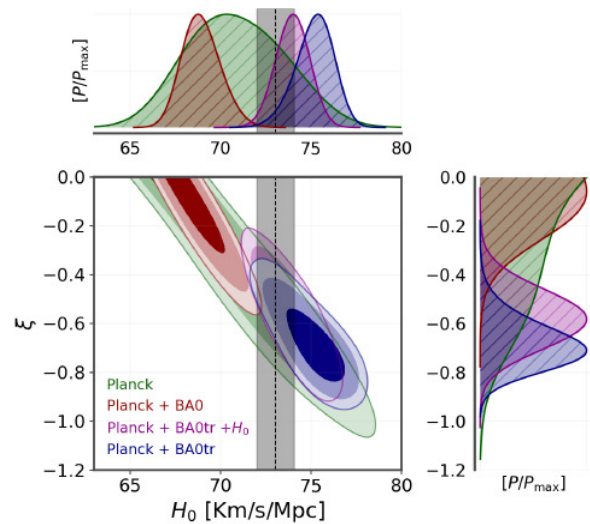

**Figure 40.** The 1σ and 2σ confidence level 2D contour constraints on the coupling parameter ξ in the IDE model and 1D posteriors for the cases without lensing. The grey vertical stripe refers to the value of $H_0$ measured by the SH0ES team ($H_0 = 73.04 \pm 1.04$ km s$^{-1}$ Mpc$^{-1}$ at 1σ confidence level). The figure is adapted from [67].

*3.5. Hubble Parameter Data*

Samushia and Ratra [333] used the Simon, Verde, and Jimenez (SVJ) [258] definition of the redshift dependence of the Hubble parameter $H(z)$ (so-called SVJ $H(z)$ data) to constrain cosmological parameters in the scalar field $\phi$CDM-RP model. According to the results obtained (Figure 41), using the $H(z)$ data, the constraints on the matter density parameter $\Omega_m$ are more stringent than those on the model parameter $\alpha$. Constraints on the matter density $\Omega_m$ are approximately as tight as the ones derived from the galaxy cluster gas mass fraction data [334] and from the SNe Ia apparent magnitude data [335].

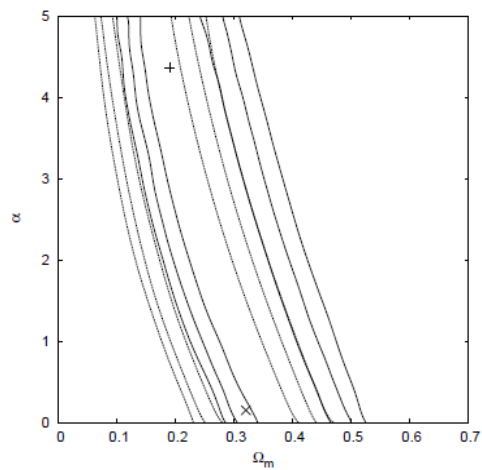

**Figure 41.** The 1 σ, 2σ, and 3σ confidence level contour constraints on parameters of the $\phi$CDM model with the RP potential. Solid lines correspond to $H_0 = 73 \pm 3$ km s$^{-1}$ Mpc$^{-1}$, while dashed lines correspond to $H_0 = 68 \pm 4$ km s$^{-1}$ Mpc$^{-1}$. The plus sign denotes the maximum likelihood at $\Omega_{m0} = 0.32$ and $\alpha = 0.15$ with reduced $\chi^2 = 1.8$. The cross denotes the maximum likelihood at $\Omega_{m0} = 0.19$ and $\alpha = 4.37$ with reduced $\chi^2 = 1.89$. The horizontal axis for which $\alpha = 0$ corresponds to the spatially flat $\Lambda$CDM model. The figure is adapted from [333].

Chen and Ratra [336] analyzed constraints on the model parameters of the $\phi$CDM-RP, the XCDM, and the $\Lambda$CDM models, using 13 Hubble parameter $H(z)$ data versus redshift [28,259]. The authors showed (see Figure 42) that the Hubble parameter $H(z)$ data yield quite strong constraints on the parameters of the $\phi$CDM model. The constraints derived from the $H(z)$ measurements are almost as restrictive as those implied by the currently available lookback time observations and the GRB luminosity data, but more stringent than those based on the currently available galaxy cluster angular size data. However, they are less restrictive than those following from the joint analysis of SNe Ia apparent magnitude and BAO peak length scale data. The joint analysis of the Hubble parameter $H(z)$ data with SNe Ia apparent magnitude and BAO peak length scale data favor the standard spatially flat $\Lambda$CDM model but do not exclude the dynamical scalar field $\phi$CDM model.

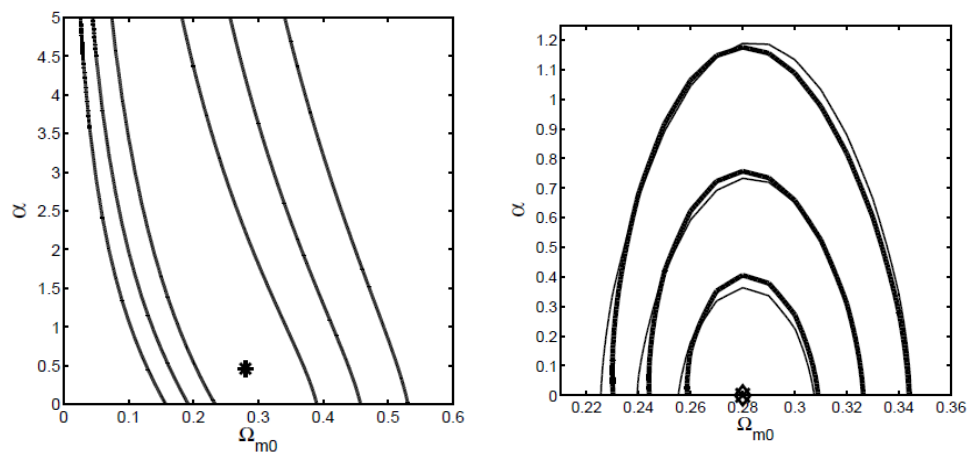

**Figure 42.** The 1 $\sigma$, 2$\sigma$, and 3$\sigma$ confidence level contour constraints on parameters of the $\phi$CDM model with the RP potential. The horizontal axis with $\alpha = 0$ corresponds to the standard spatially flat $\Lambda$CDM model. (Left panel) Contours obtained from $H(z)$ data. The star denotes the best-fit pair $(\Omega_{m0}, \alpha) = (0.28, 0.46)$, $\chi^2_{\min} = 10.1$. (Right panel) Contours were obtained from a joint analysis of the BAO peak length scale and SNe Ia apparent magnitude data (with systematic errors), with (and without) $H(z)$ data. The cross denotes the best-fit point determined from the joint sample with $H(z)$ data at $\Omega_{m0} = 0.28$ and $\alpha = 0$, with $\chi^2_{\min} = 531$. The diamond denotes the best-fit point obtained from the joint sample with $H(z)$ data at $\Omega_{m0} = 0.28$ and $\alpha = 0$, $\chi^2_{\min} = 541$. The figure is adapted from [336].

In [337], Farooq et al. obtained constraints on the parameters of the $\phi$CDM-RP, the XCDM, the $w$CDM, and the $\Lambda$CDM models from analysis of measurements of the BAO peak length scale, SNe Ia apparent magnitude [257], and 21 Hubble parameter $H(z)$ [28,30,258,259]. The results of this analysis are shown in Figure 43. Constraints are more restrictive with the inclusion of eight new $H(z)$ measurements [30] than those derived by Chen and Ratra [338]. This analysis favors the standard spatially flat $\Lambda$CDM model but does not exclude the scalar field $\phi$CDM model.

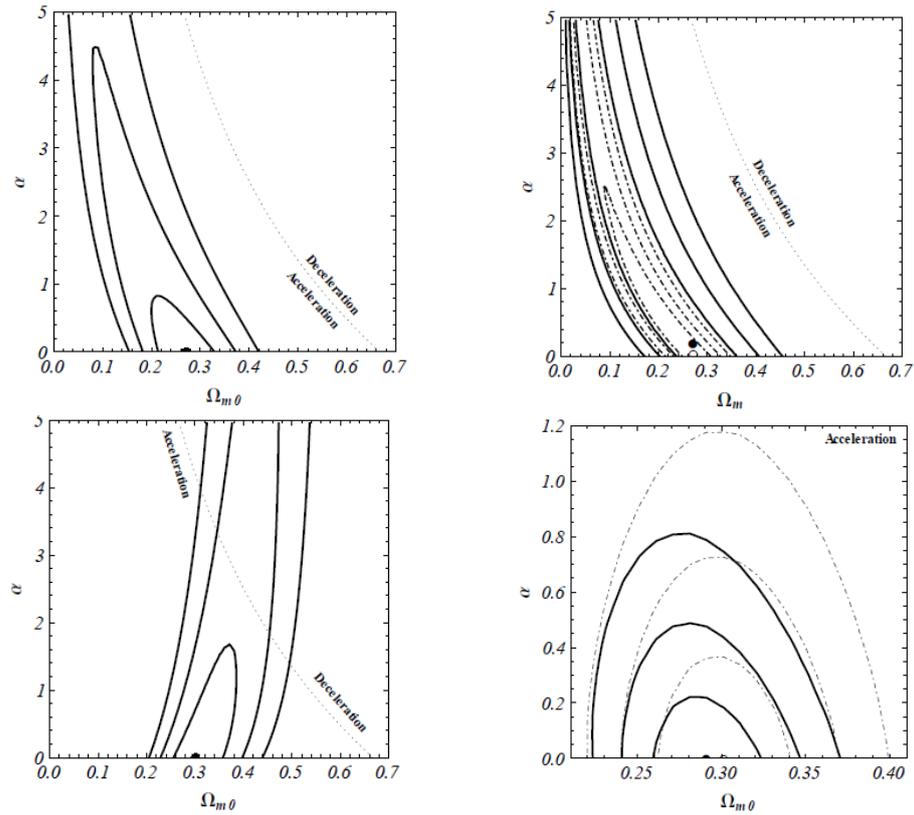

**Figure 43.** Thick solid lines are 1$\sigma$, 2$\sigma$, and 3$\sigma$ confidence level contour constraints on the parameters of the spatially flat $\phi$CDM model with the RP potential, for the prior $H_0 = 73.8 \pm 2.4$ km s$^{-1}$Mpc$^{-1}$. The horizontal axis with $\alpha = 0$ corresponds to the standard spatially flat $\Lambda$CDM model. (Left upper panel) Contours obtained from $H(z)$ data. Thin dot–dashed lines are 1$\sigma$, 2$\sigma$, and 3$\sigma$ confidence level contours reproduced from [338], where the prior is $H_0 = 68 \pm 3.5$ km s$^{-1}$Mpc$^{-1}$; the empty circle corresponds to the best-fit point. The curved dotted lines denote zero-acceleration models. The filled black circles correspond to best-fit points. (Right upper panel) Contours obtained from only SNe Ia apparent magnitude data with (without) systematic errors. Filled (open) circles denote likelihood maxima for the case of data with (without) systematic errors. (Left lower panel) Contours were obtained from only the BAO peak length scale data. Filled circles denote likelihood maxima. (Right lower panel) Contours obtained from data on the BAO peak length scale and SNe Ia apparent magnitude (with systematic errors), with (without) $H(z)$ data. The full (empty) circle denotes the best-fit point determined from a joint analysis with (without) $H(z)$ data. The figure is adapted from [337].

Farooq and Ratra [339] worked out constraints on the parameters of the $\phi$CDM-RP, the XCDM, and the $\Lambda$CDM models from measurements of the Hubble parameter $H(z)$ at redshift $z = 2.3$ [340] and 21 lower redshift measurements [28,30,258,259]. Constraints with the inclusion of the new $H(z)$ measurement of Busca et al. are more restrictive than those derived by Farooq et al. (Figure 44). As seen in this figure, the $H(z)$ constraints depend on the Hubble constant prior to $H_0$ used in the analysis. The resulting constraints are more stringent than those which follow from measurements of the SNe Ia apparent magnitude of Suzuki et al. (2012) [257]. This joint analysis consisting of measurements of $H(z)$, SNe Ia apparent magnitude, and BAO peak length scale favors the standard spatially flat $\Lambda$CDM model, but the dynamical scalar field $\phi$CDM model is not excluded either.

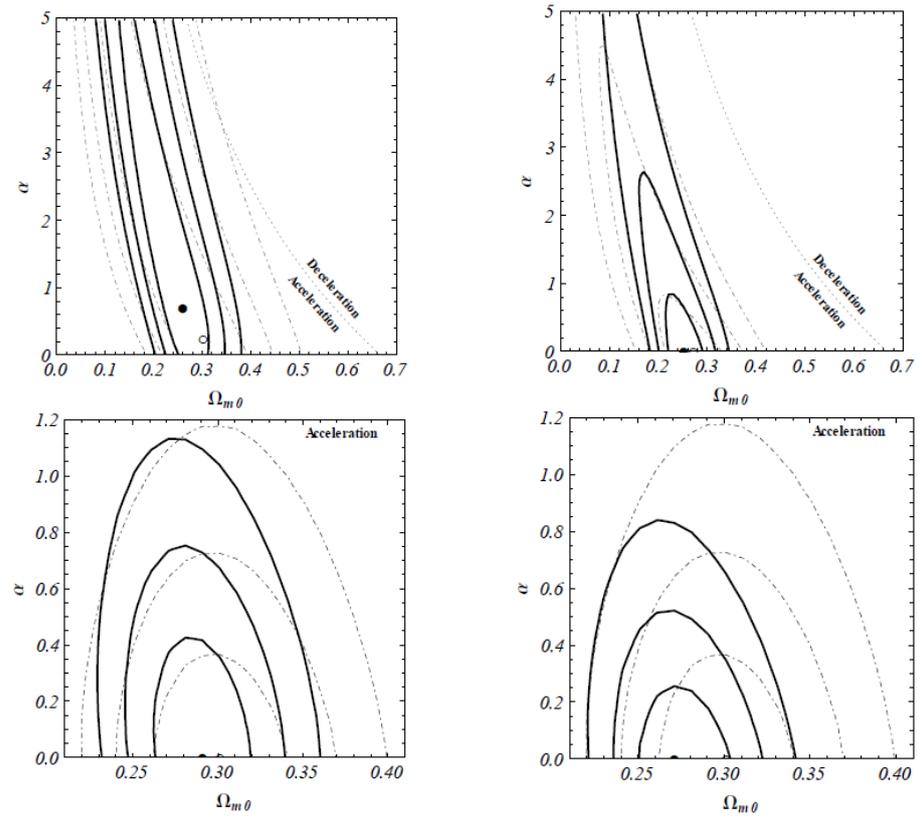

**Figure 44.** Thick solid (thin dot–dashed) lines are $1\sigma$, $2\sigma$, and $3\sigma$ confidence level contour constraints on the parameters of the spatially flat $\phi$CDM model with the RP potential from the new $H(z)$ data (old $H(z)$ data were used in [337]). The filled (empty) circle is the best-fit point from new (old) $H(z)$ measurements. The horizontal axis with $\alpha = 0$ corresponds to the standard spatially flat $\Lambda$CDM model. The curved dotted lines denote zero-acceleration models. (Left upper panel) Contours obtained for the $H_0 = 68 \pm 3.5$ kms$^{-1}$Mpc$^{-1}$ prior. The filled circles correspond to the best-fit pair $(\Omega_{m0}, \alpha) = (0.36, 0.70)$, $\chi^2_{\min} = 15.2$. The empty circles correspond to the best-fit pair $(\Omega_{m0}, \alpha) = (0.30, 0.25)$, $\chi^2_{\min} = 14.6$. (Right upper panel) Contours obtained for the $H_0 = 73.8 \pm 2.4$ kms$^{-1}$Mpc$^{-1}$ prior. The filled circles correspond to the best-fit pair $(\Omega_{m0}, \alpha) = (0.25, 0)$, $\chi^2_{\min} = 16.1$. Empty circles correspond to the best-fit pair $(\Omega_{m0}, \alpha) = (0.27, 0)$, $\chi^2_{\min} = 15.6$. (Left lower panel) Contours obtained from joint analysis with SNe Ia apparent magnitude data (with systematic errors) and BAO peak length scale data, with (without) $H(z)$ data. The full (empty) circle marks the best-fit point determined from a joint analysis with (without) $H(z)$ data. Contours obtained for $H_0 = 68 \pm 3.5$ kms$^{-1}$Mpc$^{-1}$ prior. The full circle indicates the best-fit pair $(\Omega_{m0}, \alpha) = (0.29, 0)$, $\chi^2_{\min} = 567$ while the empty circle corresponds to the best-fit pair $(\Omega_{m0}, \alpha) = (0.30, 0)$, $\chi^2_{\min} = 551$. (Right lower panel) Contours obtained for the $H_0 = 73.8 \pm 2.4$ kms$^{-1}$Mpc$^{-1}$ prior. The empty circle denotes the best-fit pair $(\Omega_{m0}, \alpha) = (0.30, 0)$, $\chi^2_{\min} = 551$ while the full circle denotes the best-fit pair $(\Omega_{m0}, \alpha) = (0.27, 0)$, $\chi^2_{\min} = 569$. The figure is adapted from [339].

Farooq and Ratra [341] found constraints on the parameters of the $\phi$CDM-RP model from the compilation of 28 independent measurements of the Hubble parameter $H(z)$ within the range of redshift $0.07 \leq z \leq 2.3$. Measurements of $H(z)$ require a currently accelerating cosmological expansion at a $3\sigma$ confidence level. The authors determined the deceleration–acceleration transition redshift $z_{da} = 0.74 \pm 0.05$. This result is in good agreement with the result obtained by Busca et al. [340], which is $z_{da} = 0.82 \pm 0.08$ based on 11 measurements of $H(z)$ from BAO peak length scale data within the range of redshift $0.2 \leq z \leq 2.3$. The resulting constraints with different priors of $H_0$ are demonstrated in Figure 45.

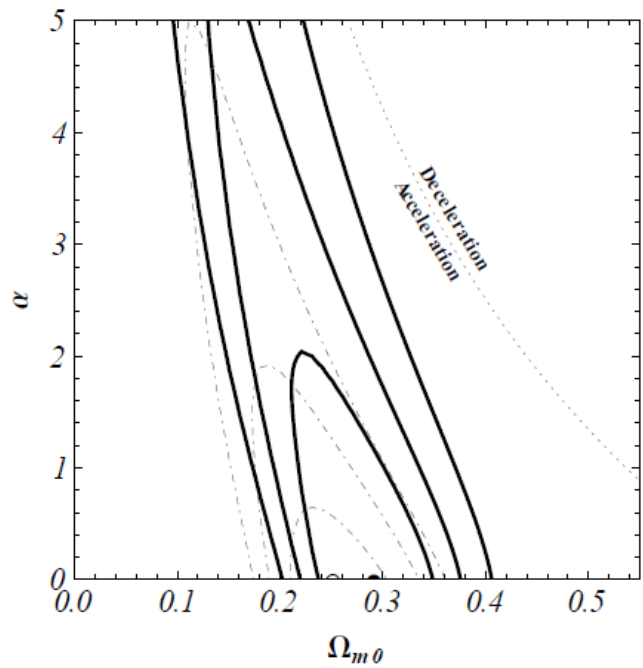

**Figure 45.** Thick solid and thin dot–dashed lines are $1\sigma$, $2\sigma$, and $3\sigma$ confidence level contour constraints on the parameters of the scalar field $\phi$CDM model with the RP potential from the compilation of $H(z)$ data for $H_0 = 68 \pm 3.5$ km s$^{-1}$Mpc$^{-1}$ and $H_0 = 73.8 \pm 2.4$ km s$^{-1}$Mpc$^{-1}$ priors, respectively. The horizontal axis with $\alpha = 0$ corresponds to the standard spatially flat $\Lambda$CDM model and the curved dotted line denotes zero-acceleration models. Filled and empty circles are best-fit points for which $(\Omega_{m0}, \alpha) = (0.29, 0)$, $\chi^2_{\min} = 18.24$ and $(\Omega_{m0}, \alpha) = (0.25, 0)$, $\chi^2_{\min} = 20.64$, respectively. The figure is adapted from [341].

Farooq et al. [261] analyzed constraints on the parameters of the spatially flat $\phi$CDM-RP, the XCDM, and the $\Lambda$CDM models from a compilation of measurements of the Hubble parameter $H(z)$. To obtain this compilation, the authors used weighted mean and median statistics techniques to combine 23 independent lower redshifts, $z < 1.04$, and Hubble parameter $H(z)$ measurements, and define binned forms of them. Then, this compilation was combined with 5 $H(z)$ measurements at the higher redshifts $1.3 \leq z \leq 2.3$. The resulting constraints are shown in Figure 46. As seen from the figure, the weighted mean binned data are almost identical to those derived from analysis using 28 independent measurements of $H(z)$. Binned weighted-mean values of $H(z)/(1+z)$ versus redshift data are presented in Figure 47. These results are consistent with a moment of the deceleration–acceleration transition at redshift $z_{da} = 0.74 \pm 0.05$ derived by Farooq and Ratra [341], which corresponds to the standard spatially flat $\Lambda$CDM model.

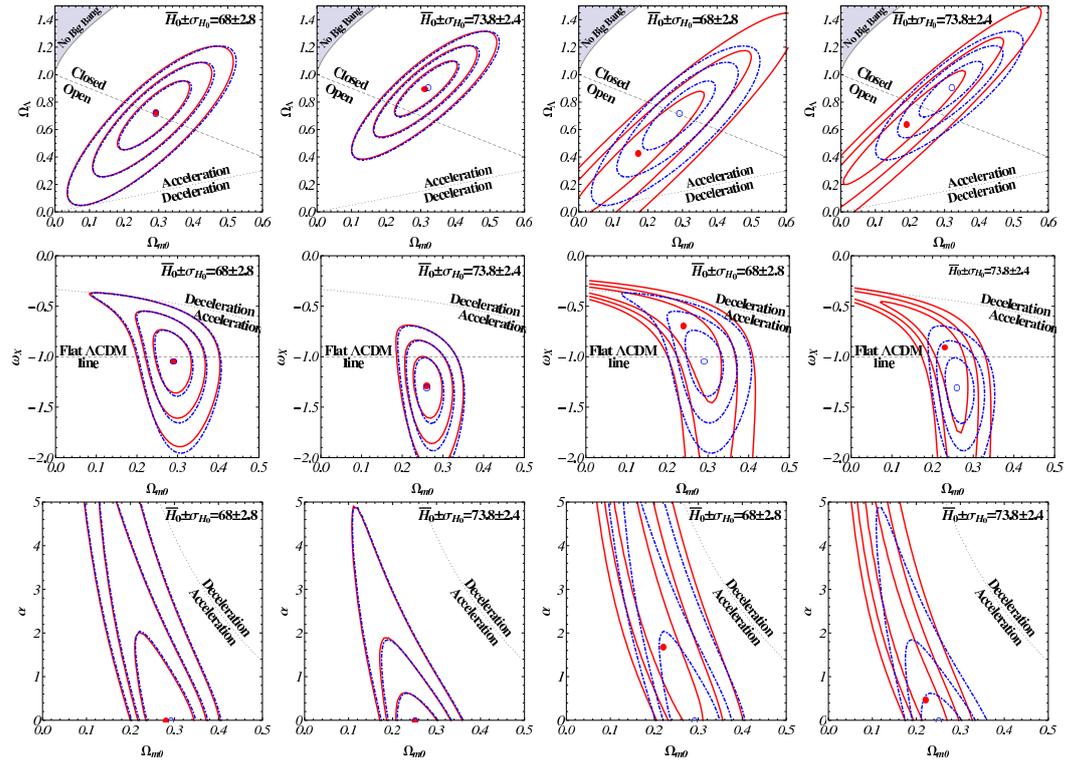

**Figure 46.** Thick solid and thin dot–dashed lines are 1$\sigma$, 2$\sigma$, and 3$\sigma$ confidence level contour constraints on the parameters of the $\phi$CDM model with the RP potential, the XCDM model, and the $\Lambda$CDM model from 7 or 9 measurements per bin data. In these three rows, the first two plots include red weighted-mean constraints while the second two include red median statistics. Filled red and empty blue circles correspond to the best-fit points. Dashed diagonal lines denote spatially flat models and dotted lines show zero-acceleration models. The figure is adapted from [261].

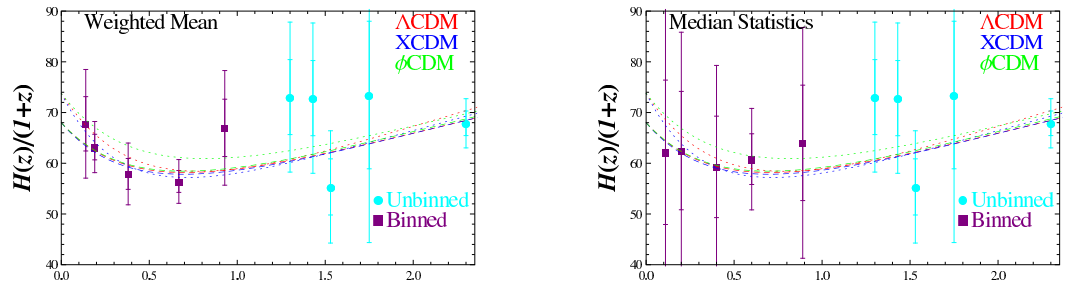

**Figure 47.** The $H(z)/(1+z)$ data binned with 7 or 9 measurements per bin, as well as 5 higher measurements of redshift, and Farooq and Ratra [261] best-fit model predictions. Dashed and dotted lines correspond to $H_0 = 68 \pm 3.5$ km s$^{-1}$Mpc$^{-1}$ and $H_0 = 73.8 \pm 2.4$ km s$^{-1}$Mpc$^{-1}$ priors, respectively. The figure is adapted from [341].

Chen et al. [342] used 28 measurements of the Hubble parameter $H(z)$ within the redshift range $0.07 \leq z \leq 2.3$ [21,28,30,31,258,340,343] to determine the value of the Hubble constant $H_0$ in the $\phi$CDM-RP, $w$CDM, and the spatially flat and spatially non-flat $\Lambda$CDM models. The result obtained for the $\phi$CDM-RP model is shown in Figure 48. The value of the Hubble constant $H_0$ is found as follows: for the spatially flat and spatially non-flat $\Lambda$CDM model, $H_0 = 68.3^{+2.7}_{-3.3}$ km s$^{-1}$Mpc$^{-1}$ and $H_0 = 68.4^{+2.9}_{-3.3}$ km s$^{-1}$Mpc$^{-1}$; for the $w$CDM model, $H_0 = 65.0^{+6.5}_{-6.6}$ km s$^{-1}$Mpc$^{-1}$; for the $\phi$CDM model, $H_0 = 67.9^{+2.4}_{-2.4}$ km s$^{-1}$Mpc$^{-1}$ (at a 1$\sigma$ confidence level). The obtained $H_0$ values are more consistent with the smaller values determined from the recent CMB temperature anisotropy and BAO peak length scale data, and with the values derived from the median statistics analysis of Huchra's compilation of $H_0$ data.

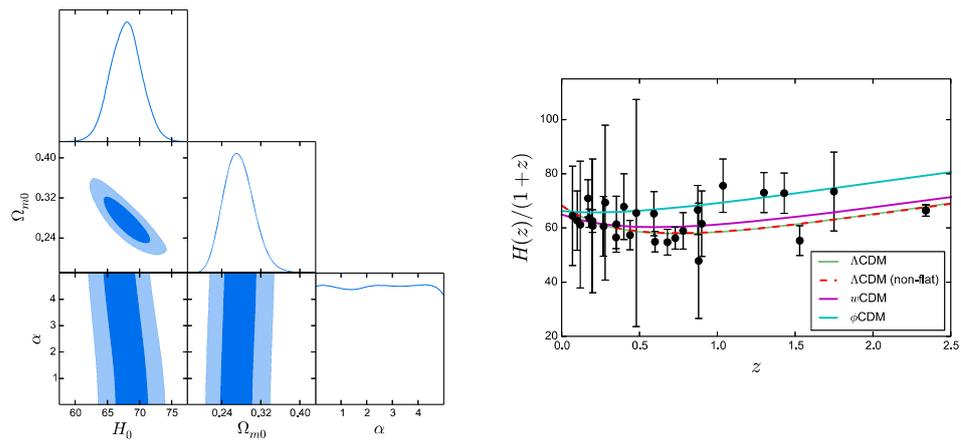

**Figure 48.** (Left panel) The 1$\sigma$ and 2$\sigma$ confidence level contour constraints on the parameters of the $\phi$CDM model with the RP potential. (Right panel) Best-fit model curves from the 28 $H(z)$ data points for the spatially flat $\phi$CDM model, $w$CDM model, and the spatially flat and spatially non-flat $\Lambda$CDM model. The figure is adapted from [342].

Farooq et al. [193] determined constraints on the parameters of the $\phi$CDM-RP, XCDM, $w$CDM, and the $\Lambda$CDM models in the spatially flat and spatially non-flat universe. The authors used the updated compilation of 38 measurements of the Hubble parameter $H(z)$ within the redshift range $0.07 \leq z \leq 2.36$ [21,25,28,30–34,258,292]. The result for these constraints is shown in Figure 49. The authors determined the redshift of the cosmological deceleration–acceleration transition, $z_{\rm da}$, and the value of the Hubble constant $H_0$ from the $H(z)$ measurements. The determined values of $z_{\rm da}$ are insensitive to the chosen model and depend only on the assumed value of the Hubble constant $H_0$. The weighted mean of these measurements is $z_{\rm da} = 0.72 \pm 0.05 \, (0.84 \pm 0.03)$ for $H_0 = 68 \pm 2.8 \, (73.24 \pm 1.74)$ km s$^{-1}$Mpc$^{-1}$. The authors proposed a model-independent method to determine the value of the Hubble constant $H_0$. The $H(z)$ data are consistent with the standard spatially flat $\Lambda$CDM model while they do not rule out the spatially non-flat XCDM and spatially non-flat $\phi$CDM models.

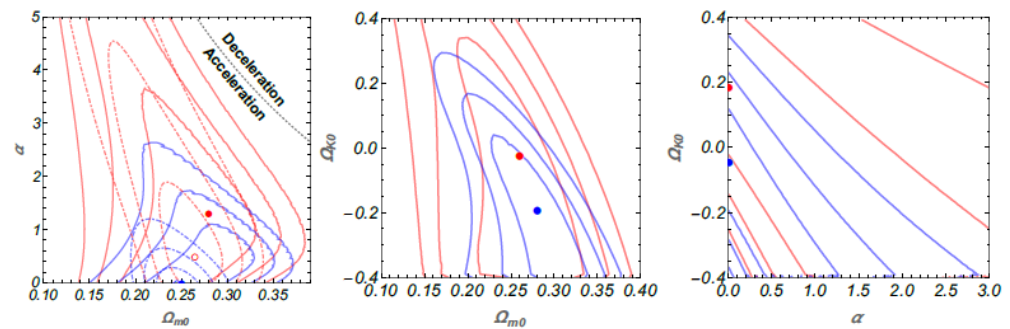

**Figure 49.** The 1$\sigma$, 2$\sigma$, and 3$\sigma$ confidence level contours constraints on the parameters of the spatially non-flat $\phi$CDM model with the RP potential. Red (blue) solid lines are for the lower (higher) $H_0$ prior. (Left, center, and right panels) The results obtained correspond to the marginalization over $\Omega_{\rm k0}$, $\alpha$, and $\Omega_{\rm m0}$, respectively. Red (blue) solid circles are the best-fit points for the lower (higher) $H_0$ prior. Red (blue) dot–dashed lines in the left panel are 1$\sigma$, 2$\sigma$, and 3$\sigma$ for the lower (higher) $H_0$ prior in the spatially flat $\phi$CDM model. The figure is adapted from [193].

*3.6. Quasar Angular Size Data*

Ryan et al. [194] determined constraints on the parameters of the spatially flat and spatially non-flat $\Lambda$CDM, XCDM, and $\phi$CDM-RP models using BAO peak length scale measurements [22,24–26,292], the Hubble parameter $H(z)$ data [21,30–34,258], and quasar (QSO) angular size data [344,345]. The 1$\sigma$, 2$\sigma$, and 3$\sigma$ confidence level contour constraints on the parameters of the spatially non-flat $\phi$CDM model with the RP potential from $H(z)$,

QSO, and BAO peak length scale datasets are presented in Figure 50. Depending on the chosen model and dataset, the observational data slightly favor both the spatially closed hypersurfaces with $\Omega_{k0} < 0$ at $1.7\sigma$ confidence level and the dynamical dark energy models over the standard spatially flat $\Lambda$CDM model at a slightly higher than $2\sigma$ confidence level. Furthermore, depending on the dataset and the model, the observational data favor a lower Hubble constant value over the one measured by the local data at a $1.8\sigma$ confidence level to $3.4\sigma$ confidence level.

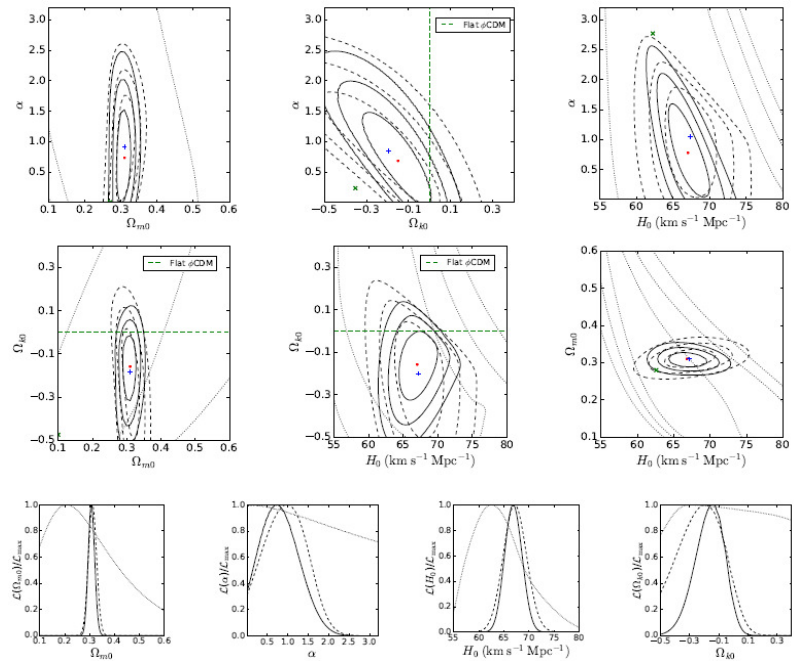

**Figure 50.** The 1 $\sigma$, 2$\sigma$, and 3$\sigma$ confidence level contour constraints on the parameters of the spatially non-flat $\phi$CDM model with the RP potential from data on $H(z)$, QSO, and BAO peak length scale. (Upper and middle panels) The vertical green dashed line in the upper center panel, and the horizontal green dashed lines in the middle left and middle center panels separate spatially closed models (with $\Omega_{k0} < 0$) from spatially open models (with $\Omega_{k0} > 0$). The horizontal line with $\alpha = 0$ in the upper panels corresponds to the spatially non-flat $\Lambda$CDM model. (Lower panel) One-dimensional likelihoods for $\Omega_{m0}, \alpha, H_0, \Omega_{k0}$. The figure is adapted from [194].

Cao et al. [346] found constraints on the parameters of the spatially flat and non-flat $\Lambda$CDM, XCDM, and $\phi$CDM-RP models using $H_{II}$ starburst galaxy apparent magnitude measurements [347,348], the compilation of 1598 X-ray and UV flux measurements of QSO 2015 data within the redshift range $0.036 \leq z \leq 5.1003$ and 2019 QSO data [349,350] alone and in conjunction with BAO peak length scale measurements [22,24–26,292], and Hubble parameter $H(z)$ data [21,28,30–34,258]. The constraints on the parameters of the spatially flat and spatially non-flat $\phi$CDM model with the RP potential obtained from the datasets mentioned above are shown in Figure 51. A combined analysis of all datasets leads to the relatively model-independent and restrictive estimates for the values of the matter density parameter at the present epoch $\Omega_{m0}$ and the Hubble constant $H_0$. Depending on the cosmological model, these estimates are consistent with a lower value of $H_0$ in the range of a $2.0\sigma$ to $3.4\sigma$ confidence level. Combined datasets favor the spatially flat $\Lambda$CDM, while at the same time they do not rule out dynamical dark energy models.

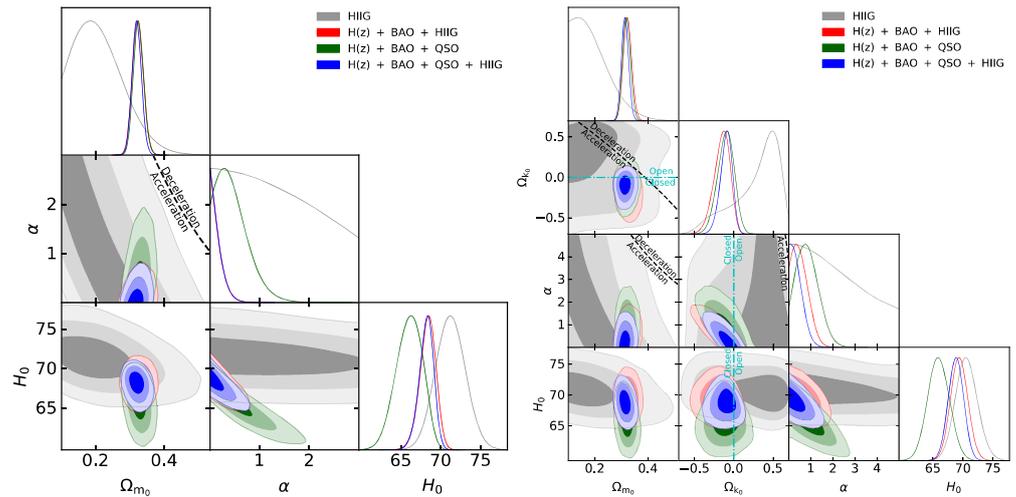

**Figure 51.** The 1 $\sigma$, 2$\sigma$, and 3$\sigma$ confidence level contour constraints on the parameters of the spatially flat $\phi$CDM model with the RP potential (left panel). The black dotted line splits the parameter space into accelerated and decelerated regions. The axis with $\alpha = 0$ denotes the spatially flat $\Lambda$CDM model. Constraints for the spatially non-flat $\phi$CDM model with the RP potential are depicted in the right panel. The figure is adapted from [346].

The compilation of 1598 X-ray and UV flux measurements of QSO 2015 data within the redshift range $0.036 \leq z \leq 5.1003$, and 2019 QSO data [349,350] alone and in conjunction with BAO peak length scale measurements [22,24–26,292], and Hubble parameter $H(z)$ data [21,28,30–34,258] was applied by Khadka and Ratra [196] to impose constraints on the parameters of the tilted spatially flat and untilted spatially non-flat $\Lambda$CDM, XCDM, and $\phi$CDM-RP quintessential inflation models. The obtained constraints for the untilted spatially non-flat $\phi$CDM-RP model from the combination of various datasets and extended QSO data only are presented in Figure 52. In most of the models, the QSO data favor the values of the matter density parameter at the present epoch $\Omega_{m0} \sim 0.5$–0.6, while, in a combined analysis of QSO data with $H(z)$ + BAO peak length scale dataset, the values of the matter density parameter at the present epoch $\Omega_{m0}$ are shifted slightly towards larger values. A combined set of data, QSO + BAO peak length scale + $H(z)$, is consistent with the standard spatially flat $\Lambda$CDM model, but favors slightly both the spatially closed hypersurfaces and the dynamical dark energy models.

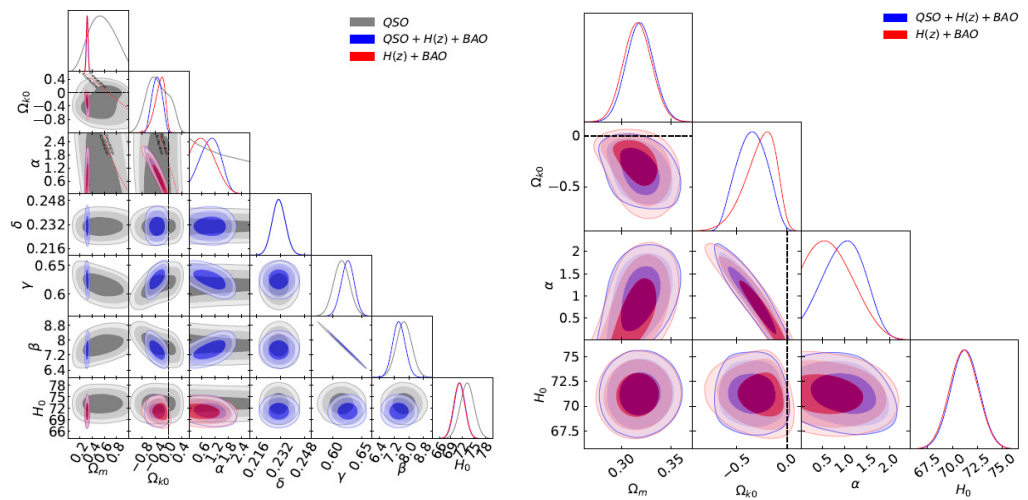

**Figure 52.** The 1$\sigma$, 2$\sigma$, and 3$\sigma$ confidence level contour constraints on the parameters of the untilted spatially non-flat $\phi$CDM model with RP potential using the combination of datasets: QSO (gray line), $H(z)$ + BAO peak length scale (red line), and QSO + $H(z)$ + BAO peak length scale (blue line). (Left panel) Contours and one-dimensional likelihoods for all free parameters. The red dotted curved lines denote zero-acceleration lines. (Right panel) Plots for $\Omega_{m0}$, $\Omega_{k0}$, $\alpha$, $H_0$ cosmological parameters, without constraints from QSO data. These plots are for $H_0 = 73.24 \pm 1.74$ km s$^{-1}$Mpc$^{-1}$ as a prior. The black dashed straight lines denote the flat hypersurface with $\Omega_{k0} = 0$. The figure is adapted from [196].

Khadka and Ratra [197] obtained constraints on the parameters of the tilted spatially flat and untilted spatially non-flat $\Lambda$CDM, XCDM, and $\phi$CDM-RP quintessential inflation models from a compilation of 808 X-ray and UV flux measurements of QSOs (quasi-stellar objects) within the redshift range $0.061 \leq z \leq 6.28$ alone [349] and in combination with the BAO peak length scale measurements [22,24–26,292], and the Hubble parameter $H(z)$ data [21,28,30–34,258]. The 1$\sigma$, 2$\sigma$, and 3$\sigma$ confidence level contours constraints on the parameters of the untilted spatially non-flat $\phi$CDM model with the RP potential from the combination of various datasets are presented in Figure 53. The constraints using only the QSO data are significantly weaker but consistent with those from the combination of the $H(z)$ + BAO peak length scale data. Combined analysis from QSO + $H(z)$ + BAO peak length scale data is consistent with the standard spatially flat $\Lambda$CDM model but slightly favors both closed spatial hypersurfaces and the untilted spatially non-flat $\phi$CDM model.

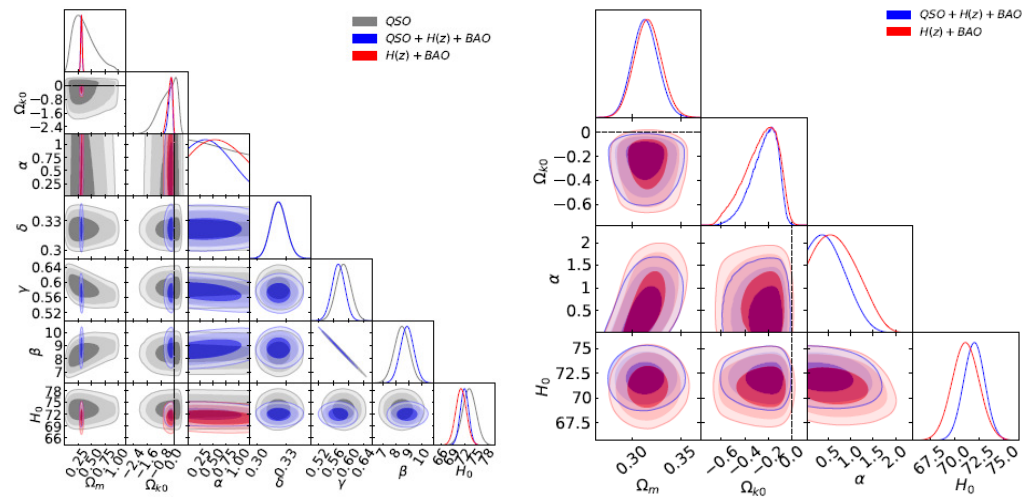

**Figure 53.** The 1$\sigma$, 2$\sigma$, and 3$\sigma$ confidence level contours constraints on the parameters of the untilted spatially non-flat $\phi$CDM model with the RP potential using the combination of datasets: QSO (gray line), $H(z)$ + BAO peak length scale (red line), and QSO + $H(z)$ + BAO peak length scale (blue line). (Left panel) Contours and one-dimensional likelihoods for all free parameters. (Right panel) Plots for only $\Omega_{m0}$, $\Omega_{k0}$, $\alpha$, $H_0$ cosmological parameters, without constraints only from QSO data. These plots are for $H_0 = 73.24 \pm 1.74$) km s$^{-1}$Mpc$^{-1}$ as a prior. Black dashed straight lines denote the spatially flat hypersurface with $\Omega_{k0} = 0$. The figure is adapted from [197].

Cao et al. [351] found constraints on the parameters of the spatially flat and non-flat $\Lambda$CDM, XCDM, and $\phi$CDM-RP models using the higher-redshift GRB data [352,353], the starburst galaxy ($H_{II}$G) measurements [347,348,354], and the QSO angular size (QSO-AS) data [344,345]. Constraints from the combined analysis of cosmological parameters of the spatially flat and non-flat $\phi$CDM-RP model are presented in Figure 54. The constraints from the combined analysis of these datasets are consistent with the currently accelerating cosmological expansion, as well as with the constraints obtained from analysis of the Hubble parameter $H(z)$ data and the measurements of the BAO peak length scale. From the analysis of the $H(z)$ + BAO peak length scale + QSO-AS + $H_{II}$G + GRB dataset, the model-independent values of the matter density parameter at the present epoch $\Omega_{m0} = 0.313 \pm 0.013$ and the Hubble constant $H_0 = 69.3 \pm 1.2$ km s$^{-1}$Mpc$^{-1}$ are obtained. This analysis favors the spatially flat $\Lambda$CDM model but also does not rule out dynamical dark energy models.

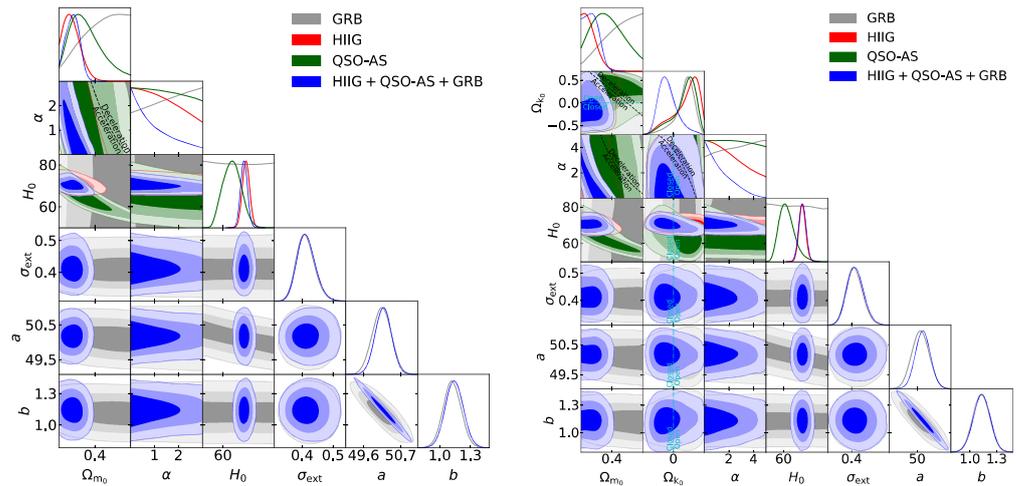

**Figure 54.** The 1 $\sigma$, 2$\sigma$, and 3$\sigma$ confidence level contours constraints on the parameters of the spatially flat (left panel) and spatially non-flat (right panel) $\phi$CDM models with the RP potential from various datasets. The black dotted line splits the parameter space into the regions of the currently decelerating and accelerating cosmological expansion. The axis with $\alpha = 0$ denotes the spatially flat $\Lambda$CDM model. The figure is adapted from [351].

Khadka and Ratra [198] determined constraints on the parameters of the spatially flat and non-flat $\Lambda$CDM, XCDM, and $\phi$CDM-RP models from compilation of the X-ray and UV flux measurements of 2038 QSOs which span the redshift range $0.009 \leq z \leq 7.5413$ [350,355]. The authors found that, for the full QSO dataset, parameters of the X-ray and UV luminosity $L_X - L_{UV}$ relation used to standardize these QSO data depends on the cosmological model, and therefore cannot be used to constrain the cosmological parameters in these models. Subsets of these QSOs, limited by redshift $z \leq 1.5 - 1.7$, obey the $L_X - L_{UV}$ relation in a way that is independent of the cosmological model and can therefore be used to constrain the cosmological parameters. Constraints from these smaller subsets of lower redshift QSO data are generally consistent, but much weaker than those inferred from the Hubble parameter $H(z)$ and the BAO peak length scale measurements (Figure 55).

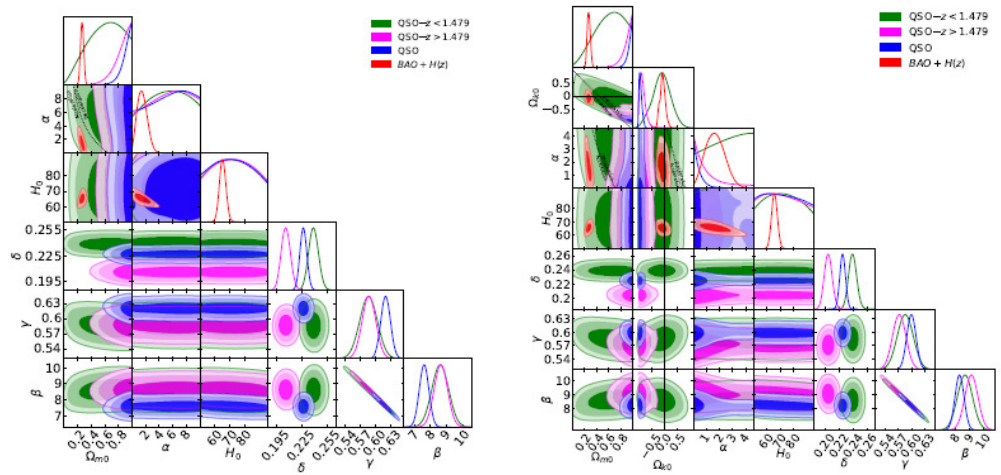

**Figure 55.** The 1 $\sigma$, 2$\sigma$, and 3$\sigma$ confidence level contour constraints on the parameters of the spatially flat (left panel) and non-flat (right panel) $\phi$CDM model with the RP potential, using QSO (blue) and the BAO peak length scale + $H(z)$ (red) datasets. The black dotted line in the $\alpha - \Omega_{m0}$ sub-panels is the line of zero acceleration, under which the accelerated cosmological expansion occurs. The axis with $\alpha = 0$ denotes the spatially flat $\Lambda$CDM model. The figure is adapted from [198].

Cao et al. [202] determined constraints on the parameters of the spatially flat and non-flat ΛCDM, XCDM, and $\phi$CDM-RP models by analyzing a total of 1383 measurements consisting of 1048 Pantheon SNe Ia apparent magnitude measurements from Scolnic et al. (2018) [213], and 20 binned SNe Ia apparent magnitude measurements of DES Collaboration [356,357], 120 QSO measurements [349,350,355], 153 $H_{II}$G data [347,348,354], 11 BAO peak length scale measurements [22,24–26,292], and 31 Hubble parameter $H(z)$ data [21,28,30–34,258]. Constraints on the parameters of the spatially non-flat $\phi$CDM model with the RP potential from that analysis of the data are shown in Figure 56. From the analysis of those datasets, the model-independent estimates of the Hubble constant, $H_0 = 68.8 \pm 1.8$ km s$^{-1}$Mpc$^{-1}$, as well as the matter density parameter at the present epoch, $\Omega_{m0} = 0.294 \pm 0.020$, are obtained. While the constraints favor dynamical dark energy and slightly spatially closed hypersurfaces, they do not preclude dark energy from being a cosmological constant and spatially flat hypersurfaces.

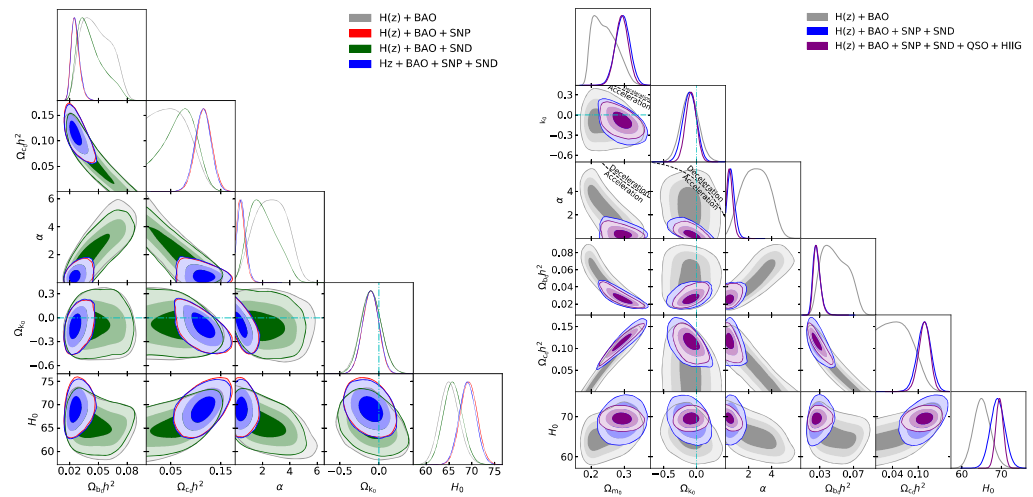

**Figure 56.** The 1 $\sigma$, 2$\sigma$, and 3$\sigma$ confidence level contour constraints on the parameters of the spatially non-flat $\phi$CDM model with the RP potential. The zero-acceleration line splits the parameter space into regions of the currently decelerating and accelerating cosmological expansion. The cyan dash–dot lines show the spatially flat $\phi$CDM model; regions with spatially closed geometry are located either below or to the left. The axis with $\alpha = 0$ denotes the spatially flat ΛCDM model. The figure is adapted from [202].

Khadka and Ratra [199] found constraints on the parameters of the spatially flat and non-flat ΛCDM, XCDM, and $\phi$CDM-RP models from 78 reverberation-measured $Mg_{II}$ time-lag QSOs within the redshift range $0.0033 \leq z \leq 1.89$ [358,359]. The authors applied the radius–luminosity or $R - L$ relation to standardized 78 $Mg_{II}$ QSOs data. In each model, the authors simultaneously determined the $R - L$ relation and parameters in these models, thus avoiding the problem of circularity. It was found that values of the $R - L$ relation parameter are independent of the model used in the analysis, which makes it possible to establish that current $Mg_{II}$ QSOs data are standardizable candles. Constraints derived from the QSO data only are significantly weaker than those derived from the combined set of the BAO peak length scale and the Hubble parameter $H(z)$ measurements, but are consistent with both of them. The constraints obtained from the $Mg_{II}$ QSOs data in conjunction with the BAO peak length scale + $H(z)$ measurements agree with the spatially flat ΛCDM model as well as with spatially non-flat dynamical dark energy models.

Khadka and Ratra [200] found that the recent compilation of the QSO X-ray and UV flux measurements [355] includes the QSO data that appear not to be standardized via the X-ray luminosity and the UV luminosity $L_X - L_{UV}$ relation parameters that are dependent on both the cosmological model and the redshift, so it should not be used to constrain the model parameters. These data include a compilation of seven different subsamples. The authors analyzed these subgroups and some combinations of subgroups to define which QSO subgroups are responsible for questions specified in the paper by Khadka and Ratra [198]. They considered that the largest of the seven subsamples in this compilation, SDSS-4XMM QSOs, which contributes about two-thirds of all QSOs, has the $L_X - L_{UV}$ ratios that depend on both the accepted cosmological model and the redshift, and thus is the source of a similar problem found earlier when collecting the full QSO data.

The second and third largest subsamples, SDSS-Chandra and XXL QSOs, which together account for about 30% of total QSO data, appear to be standardized. Constraints on the cosmological parameters from these subsamples are weak and consistent with the standard spatially flat $\Lambda$CDM model or with the constraints from the better-established cosmological probes. Constraints on the cosmological parameters of the spatially flat and spatially non-flat $\phi$CDM models with the RP potential, using SDSS-Chandra and XXL QSO data as well as $H(z)$ data and BAO peak length scale data, are presented in Figure 57.

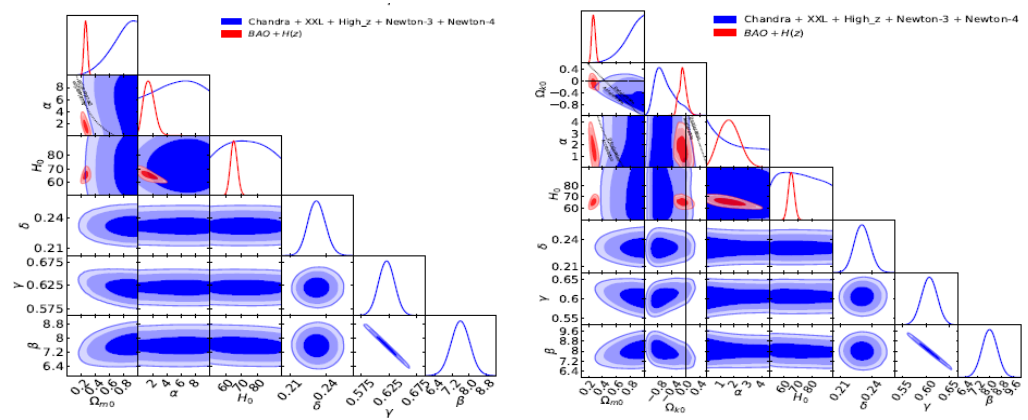

**Figure 57.** The $1\sigma$, $2\sigma$, and $3\sigma$ confidence level contour constraints on the parameters of the spatially flat (left panel) and spatially non-flat (right panel) $\phi$CDM models with the RP potential, using Chandra + XXL + High-$z$ + Newton-3 + Newton-4 (blue) and BAO peak length scale + $H(z)$ (red) datasets. In all plots, black dotted lines are zero-acceleration lines, which split the parameter space into the regions of current acceleration and deceleration. Black dashed line corresponds to $\Omega_{k0} = 0$. The axis with $\alpha = 0$ denotes the spatially flat $\Lambda$CDM model. The figure is adapted from [200].

Khadka et al. [201] used 118 $H\beta$ QSO measurements [360] within the redshift range $0.0023 \leq z \leq 0.89$ to simultaneously constrain cosmological model parameters and QSO two-parameter radius–luminosity $R - L$ relation parameters of the spatially flat and non-flat $\Lambda$CDM, XCDM, and $\phi$CDM-RP models. The authors found that the $R - L$ relation parameters for $H\beta$ QSO data are independent in models under investigation, therefore QSO data seem to be standardizable through $R - L$ relation parameters. The constraints derived using $H\beta$ QSO data are weak, slightly favoring the currently accelerating cosmological expansion, and are generally in the $2\sigma$ tension with the constraints derived from analysis of the measurements of the BAO peak length scale and the Hubble parameter $H(z)$. Constraints on the cosmological parameters of the spatially flat and non-flat $\phi$CDM-RP model, from the $H\beta$ QSO measurements, and the $H(z)$ and BAO peak length scale data are presented in Figure 58.

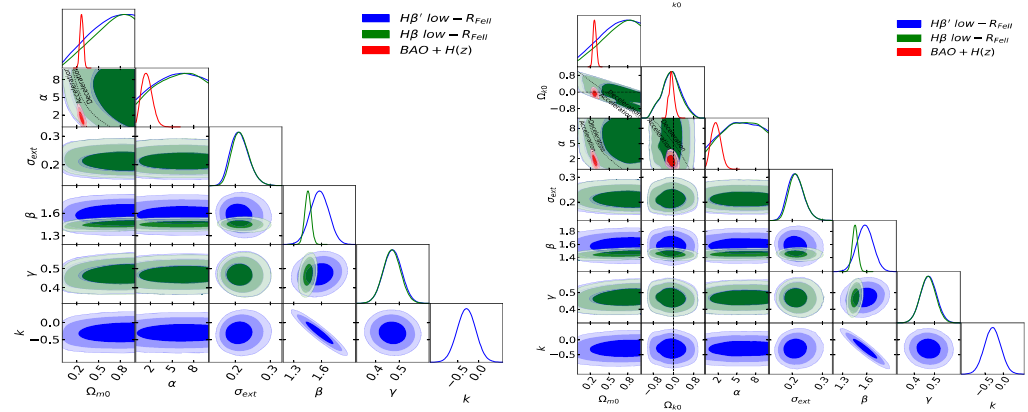

**Figure 58.** The $1\sigma$, $2\sigma$, and $3\sigma$ confidence level contour constraints on the parameters of the spatially flat (left panel) and spatially non-flat (right panel) $\phi$CDM model with the RP potential from 3-parameter $H\beta'$ high-$\Re_{FeII}$ (blue), 2-parameter $H\beta'$ high-$\Re_{FeII}$ (green), and $H(z)$ + BAO peak length scale (red) measurements. Black dotted lines correspond to zero-acceleration lines. Black dashed lines represent $\Omega_{k0} = 0$. The figure is adapted from [201].

Khadka et al. [361] determined constraints on the parameters of the spatially flat and spatially non-flat $\Lambda$CDM, XCDM, and $\phi$CDM-RP models using the observations of 66 reverberation-measured $Mg_{II}$ QSOs within the redshift range $0.36 \leq z \leq 1.686$. Constraints on the cosmological parameters of the spatially flat and spatially non-flat $\phi$CDM models with the RP potential from various QSO datasets are shown in Figure 59. The authors also studied the two- and three-parameter radius–luminosity $R - L$ relations [362,363] for $Mg_{II}$ QSO sources, and found that these relations do not depend on the assumed cosmological model; therefore, they can be used to standardize QSO data. The authors found for the two-parameter $R - L$ relation that the data subsets with low-$\Re_{FeII}$ and high-$\Re_{FeII}$ obey the same $R - L$ relation within the error bars. Extending the two-parameter $R - L$ relation to three parameters does not lead to the expected decrease in the intrinsic variance in the $R - L$ relation. None of the three-parameter $R - L$ relations provides a significantly better measurement fit than the two-parameter $R - L$ relation. The results obtained differ significantly from those found by Khadka et al. [201] from analysis of reverberation-measured $H\beta$ QSOs.

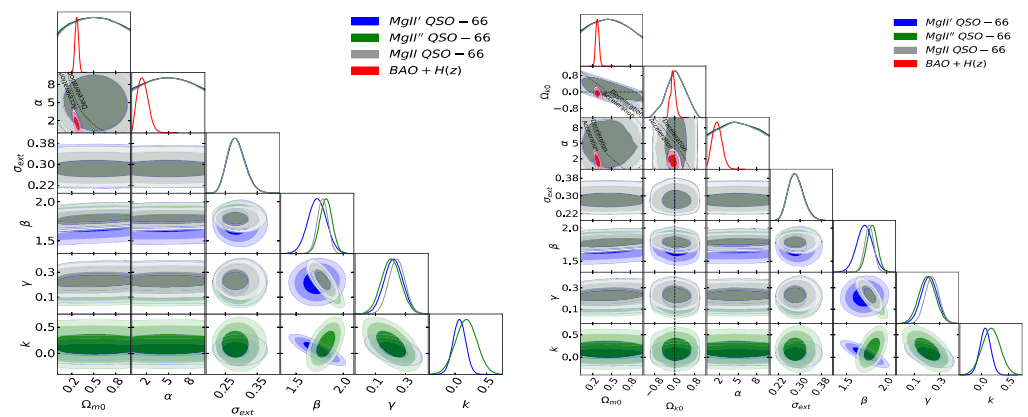

**Figure 59.** The $1\sigma$, $2\sigma$, and $3\sigma$ confidence level contour constraints on the parameters of the spatially flat (left panel) and spatially non-flat (right panel) $\phi$CDM models with the RP potential from the measurements of $Mg_{II'}$ high-$\Re_{FeII}$ (blue), MgII″ high-$\Re_{FeII}$ (green), $Mg_{II}$ high-$\Re_{FeII}$ (gray), and BAO peak length scale + $H(z)$. Black dotted lines correspond to zero-acceleration lines. Black dashed lines represent $\Omega_{k0} = 0$. The figure is adapted from [361].

Cao et al. [204] determined constraints on the parameters of the spatially flat and non-flat $\Lambda$CDM, XCDM, $\phi$CDM-RP models, as well as on the QSO radius–luminosity $R - L$ relation parameters from the 38 $C_{IV}$ QSO reverberation-measured data in the redshift range $0.001064 \leq z \leq 3.368$. An improved method is used that takes into account more accurately the asymmetric error bars for the time-delay measurements. The authors found that the parameters of the $R - L$ relation do not depend on the cosmological models considered and, therefore, the $R - L$ relation can be used to standardize the $C_{IV}$ QSO data. Mutually consistent constraints on the cosmological parameters from $C_{IV}$, $Mg_{II}$, and $H(z)$ + BAO peak length scale data allow conducting the analysis from $C_{IV} + Mg_{II}$ dataset as well as from the $H(z)$ + BAO peak length scale + $C_{IV} + Mg_{II}$ datasets. Although the $C_{IV} + Mg_{II}$ cosmological constraints are weak, they slightly (at a ∼0.1$\sigma$ confidence level) change the constraints from the $H(z)$ + BAO peak length scale + $C_{IV} + Mg_{II}$ datasets. The constraints on the cosmological parameters of the spatially flat and non-flat $\phi$CDM-RP from various QSO datasets are shown in Figure 60.

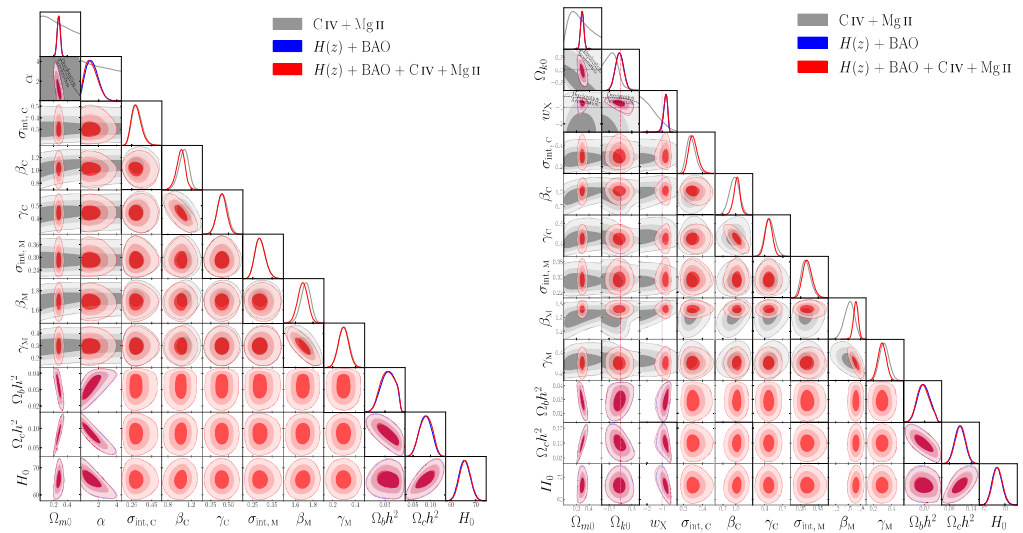

**Figure 60.** The 1$\sigma$, 2$\sigma$, and 3$\sigma$ confidence level contour constraints on the parameters of the spatially flat (left panel) and spatially non-flat (right panel) $\phi$CDM models with the RP potential from various combinations of datasets. The axis with $\alpha = 0$ denotes the spatially flat $\Lambda$CDM model. The black dash–dotted lines denote spatially flat hypersurfaces $\Omega_{k0} = 0$; closed spatial hypersurfaces are located either below or to the left. The black dotted lines correspond to the lines of zero acceleration and split the parameter space into currently accelerating (bottom left) and decelerating (top right) regions. The figure is adapted from [204].

*3.7. Gamma Ray Burst Distance Data*

Samushia and Ratra [364] derived constraints on the parameters of the spatially flat $\Lambda$CDM, XCDM, and $\phi$CDM-RP models using the observational datasets of SNe Ia Union apparent magnitude data [365] and BAO peak length scale data [17], and measurements of gamma-ray burst (GRB) distances [366,367]. The authors applied two methods for analyzing the GRB data-fitting luminosity relation of GRB, Wang's method [367] and Schaefer's method [366]. The constraints on the cosmological parameters of the $\phi$CDM model from analysis of the SNe Ia Union apparent magnitude data and the BAO peak length scale measurements, with and without the GRB measurements, are presented in Figure 61. The constraints from the GRB data obtained by two different methods disagree with each other at a more than 2$\sigma$ confidence level. The cosmological parameters of the $\phi$CDM model could not be tightly constrained only by the current GRB data.

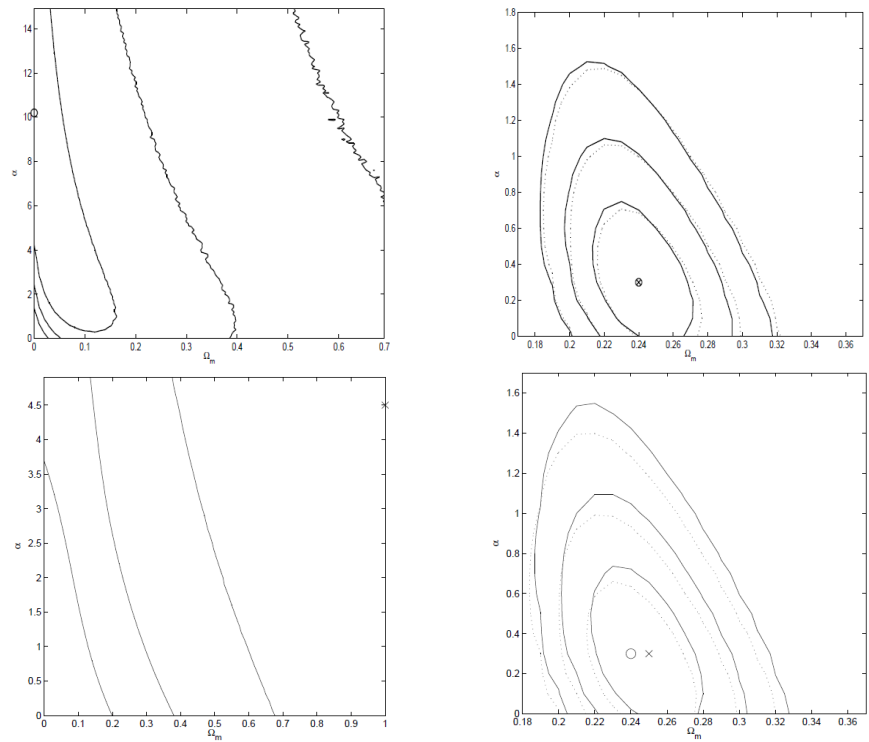

**Figure 61.** The 1 $\sigma$, 2$\sigma$, and 3$\sigma$ confidence level contours constraints on parameters of the $\phi$CDM model with the RP potential. The horizontal axis with $\alpha = 0$ corresponds to the standard spatially flat $\Lambda$CDM model. (Left upper panel) Contours are obtained by Wang's (2008) [367] method. The circle indicates the best-fit parameter values $\Omega_m = 0$, $\alpha = 10.2$ with $\chi^2 = 1.39$ for 4 degrees of freedom. (Right upper panel) Contours are derived using the GRB data by Wang's (2008) [367] method, the SNe Ia Union apparent magnitude data, and the BAO peak length scale measurements, while dotted lines (here the cross denotes the best-fit point) are derived using only the SNe Ia apparent magnitude data and the BAO peak length scale measurements. The best-fit parameters in both cases are $\Omega_m = 0.24$, $\alpha = 0.3$ with $\chi^2 = 326$ for 313 degrees of freedom (solid lines) and $\chi^2 = 321$ for 307 degrees of freedom. (Left lower panel) Contours are obtained using GRB data by Schaefer's (2007) method (here the cross indicates the best-fit parameter values): $\Omega_m = 1$, $\alpha = 4.5$ with $\chi^2 = 77.8$ for 67 degrees of freedom. (Right lower panel) Contours are obtained using Schaefer's (2007) [366] method, the SNe Ia Union apparent magnitude data, and the BAO peak length scale measurements, while dotted lines are obtained using the SNe Ia apparent magnitude data and the BAO peak length scale measurements only. Solid lines (circle denotes best-fit point) are derived using GRB data, here the cross denotes the best-fit point. The best-fit matter density parameters are $\Omega_m = 0.24$, $\alpha = 0.30$ with $\chi^2 = 401$ for 376 degrees of freedom (solid lines), and $\Omega_m = 0.25$, $\alpha = 0.3$ with $\chi^2 = 321$ for 307 degrees of freedom (dotted lines). The figure is adapted from [364].

Khadka and Ratra [368] performed an analysis of the constraints on the parameters of the spatially flat and non-flat $\Lambda$CDM, XCDM, and $\phi$CDM-RP models from measurements of the peak photon energy and bolometric fluence of 119 GRBs extending over the redshift range of $0.34 \leq z \leq 8.2$ [352,353], and Amati relation parameters [369], BAO peak length scale measurements [22,24–26,292], and Hubble parameter $H(z)$ data [21,28,30–34,258]. Resulting constraints on the parameters of the spatially flat and spatially non-flat $\phi$CDM models with the RP potential are presented in Figure 62.

The Amati relation between the peak photon energy of the GRB in the cosmological rest frame, $E_p$, and $E_{\rm iso}$ is given as

$$\log(E_{\rm iso}) = a + b \log(E_p), \tag{46}$$

where *a* and *b* are free parameters defined from data, representing points of intersection and slope in the Amati relation, respectively. $E_p$ and $E_{\rm iso}$ are specified as

$$E_{\rm iso} = \frac{4\pi D_{\rm L}^2(z,p) S_{\rm bolo}}{(1+z)}, \quad E_p = E_{p,{\rm obs}}(1+z), \tag{47}$$

where $D_{\rm L}(z,p)$ is the luminosity distance, $p$ is a cosmological parameter, $S_{\rm bolo}$ is the measured bolometric fluence, and $E_{p,{\rm obs}}$ is the measured peak energy of the GRB.

The resulting Amati relation parameters are almost identical in all considered cosmological models, which confirms the use of the Amati relation parameters to standardize these GRB data. The constraints on the cosmological parameters of the models under consideration from the GRB data are consistent with the constraints obtained from the analysis of the BAO peak length scale and the measurements of the Hubble parameter $H(z)$, but are less restrictive.

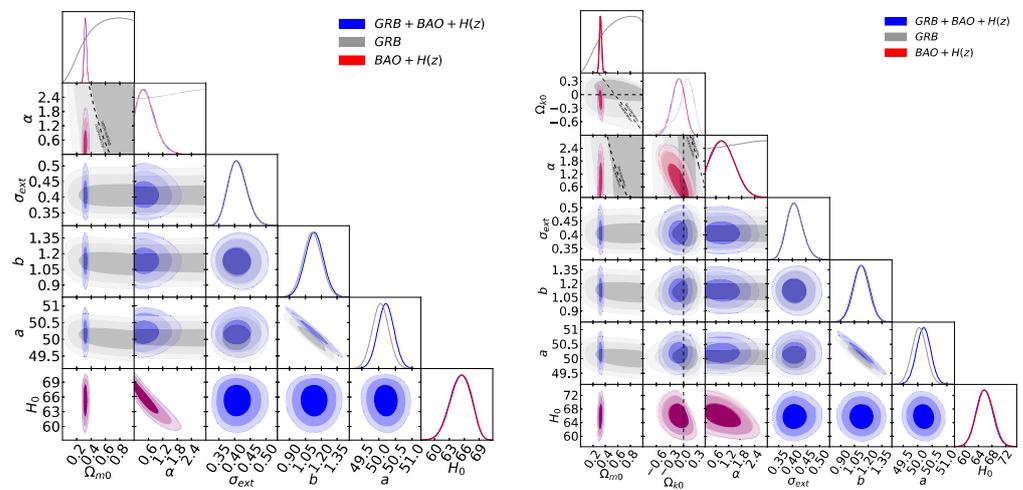

**Figure 62.** The 1 $\sigma$, 2$\sigma$, and 3$\sigma$ confidence level contour constraints on the parameters of the spatially flat (left panel) and spatially non-flat (right panel) $\phi$CDM models with the RP potential, using the combination of datasets: GRB (gray line), $H(z)$ + BAO peak length scale (red line), and GRB + $H(z)$ + BAO peak length scale (blue line). The black dotted line splits the parameter space into accelerating and decelerating regions. The axis with $\alpha = 0$ denotes the spatially flat $\Lambda$CDM model. The figure is adapted from [368].

Khadka et al. [370] analyzed constraints on the parameters of the spatially flat and non-flat $\Lambda$CDM, XCDM, and $\phi$CDM-RP models from the GRB data. The authors considered eight different GRB datasets to test whether the current GRB measurements, which probe a largely unexplored range of cosmological redshifts, can be used to reliably constrain the parameters of these models. The authors applied the MCMC analysis implemented in Monte Python to find the most appropriate correlations and cosmological parameters for the eight GRB samples, with and without the BAO peak length scale and the $H(z)$ data.

They applied three Amati correlation samples [369] and five Combo correlation samples [371] to obtain correlations and constraints on the model parameters. Constraints on the parameters of the spatially non-flat $\phi$CDM-RP model, using various datasets of GRB, as well as the BAO peak length scale + $H(z)$ data. The authors found that the cosmological constraints, determined from the A118 sample consisting of 118 bursts, agree but are much weaker than those following from the BAO peak length scale and the $H(z)$ data. These constraints are consistent with the spatially flat $\Lambda$CDM as well as with the spatially non-flat dynamical dark energy models.

Cao et al. [203] applied the spatially flat and non-flat $\Lambda$CDM, XCDM, and $\phi$CDM-RP models in the analysis of the three (ML, MS, and GL) ($L_0 - t_b$) Dainotti-correlated sets of GRB measurements collected by Wang et al. [372] and Hu et al. [373] that together explore the redshift range $0.35 \leq z \leq 5.91$. The authors found that each dataset, as well as the combinations of MS + GL, ML + GL, and ML + MS, obey the cosmological model-independent Dainotti correlations [374,375]) and therefore are standardized. The luminosity of the plateau phase for GRBs that obey the Dainotti correlation is defined as

$$L_0 = \frac{4\pi D_L^2 F_0}{(1+z)}^{1-\beta'}, \quad (48)$$

where $F_0$ is the GRB X-ray flux, $\beta'$ is the spectral index in the plateau phase, and $D_L$ is the luminosity distance.

The authors applied these GRB data in combination with the best currently available Amati-correlated GRB data of Amati [369] that explore the redshift range $0.3399 \leq z \leq 8.2$ to constrain the cosmological model parameters. As a result, constraints are weak, providing lower bounds on the matter density parameter at the present epoch $\Omega_{m0}$, moderately favoring the non-zero spatial curvature, and largely consistent with both the currently accelerated cosmological expansion and with constraints determined on the basis of more reliable data. Constraints of cosmological parameters of the spatially flat and non-flat $\phi$CDM-RP model, using the Dainotti-correlated sets of the GRB measurements as well as the $H(z)$ and BAO peak length scale data are presented in Figure 63.

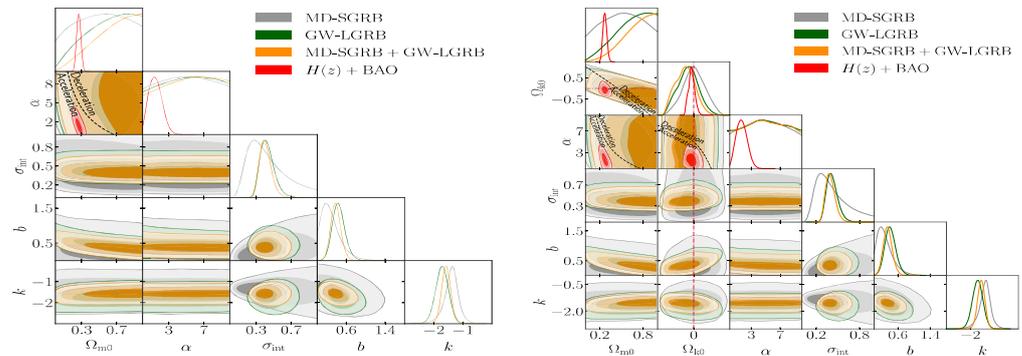

**Figure 63.** The 1$\sigma$, 2$\sigma$, and 3$\sigma$ confidence level contour constraints on the parameters of the spatially flat (left panel) and spatially non-flat (right panel) $\phi$CDM models with the RP potential from MD-SGRB (gray), GWLGRB (green), MD-SGRB + GW-LGRB (orange), and $H(z)$ + BAO peak length scale (red) data. Black dashed lines denote zero-acceleration lines, which split the parameter space into two regions of current acceleration and deceleration. Dash–dotted crimson lines correspond to spatially flat hypersurfaces with spatially closed hypersurfaces either below or to the left. The magenta lines correspond to the $\phi$CDM model; the closed spatial geometry are either below or to the left. The axis with $\alpha = 0$ denotes the spatially flat $\Lambda$CDM model. The figure is adapted from [203].

Cao et al. [205] used the spatially flat and non-flat $\Lambda$CDM, XCDM, and $\phi$CDM-RP models to analyze the compilation of data from 50 Platinum GRBs within the redshift range $0.553 \leq z \leq 5.0$. The authors found that these data obey the three-parameter fundamental plane or Dainotti correlation, independent of the cosmological model, and therefore they are amenable to standardization and can be used to constrain cosmological parameters. To improve the accuracy of the constraints for the GRB data only, the authors excluded ordinary GRB data from the larger Amati-correlated A118 dataset of 118 GRBs and analyzed the remaining 101 Amati-correlated GRBs with 50 Platinum GRB datasets. This joint dataset of 151 GRBs is being investigated within the little-studied redshift range $z \in (2.3, 8.2)$. Due to the consistency of cosmological constraints from the platinum GRB data with the $H(z)$ + BAO peak length scale dataset, the authors combined platinum GRB and the $H(z)$ + BAO peak length scale data to carry out the analysis and found small changes in the cosmological parameter constraints compared to the constraints from the $H(z)$ + BAO peak length scale data. The resulting constraints from the GRBs only are more stringent than those from the $H(z)$ + BAO peak length scale dataset but are less precise.

Cao et al. [206] proposed the constraints on the parameters of the spatially flat and non-flat ΛCDM, XCDM, and $\phi$CDM-RP models, using the extended set of the GRB data including the 50 platinum GRBs within the redshift range $0.553 \leq z \leq 5$ by Dainotti et al. [376], the LGRB95 data that contains 95 long GRBs measurements within the redshift range $0.297 \leq z \leq 9.4$ by Dainotti et al. [376]. The compilation of the 145 GRB datasets was also used. The constraints on the cosmological parameters of the spatially flat and spatially non-flat $\phi$CDM model with the RP potential, using various GRB datasets, are shown in Figure 64. The authors also examined which of two correlations, the two-dimensional Dainotti correlation [377] or the three-dimensional Dainotti correlation [378,379], fits better the GRB datasets. Based on the results of *AIC*, *BIC*, and Deviation Information Criterion (*DIC*) analysis, the authors found that the three-dimensional Dainotti correlation is much preferable to the two-dimensional one for the GRB datasets.

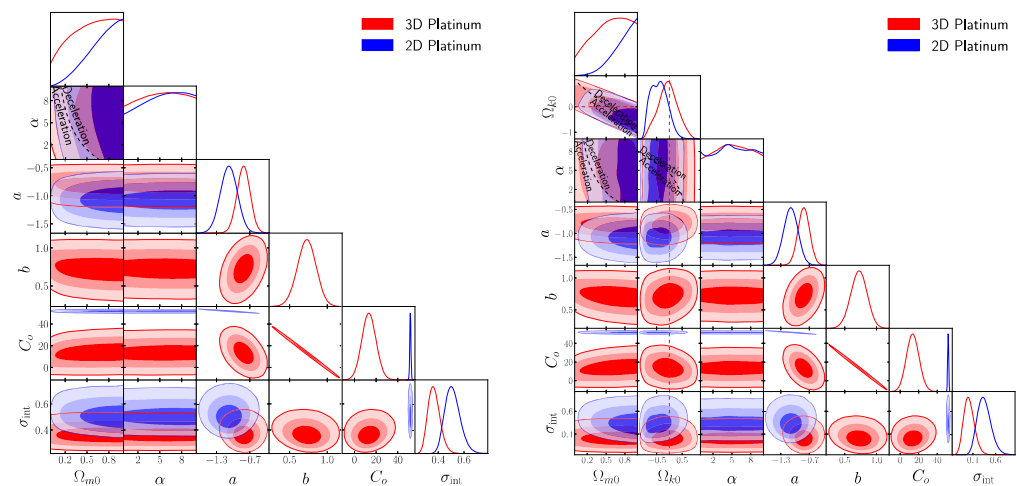

**Figure 64.** The 1 $\sigma$, 2$\sigma$, and 3$\sigma$ confidence level contours on parameters of the spatially flat (left panel) and spatially non-flat (right panel) $\phi$CDM models with the RP potential, using various combinations of GRB datasets. The axis with $\alpha = 0$ denotes the spatially flat ΛCDM model. The black dotted lines correspond to lines of zero acceleration and split the parameter space into currently accelerating (bottom left) and currently decelerating (top right) regions. The crimson dash–dot lines denote spatially flat hypersurfaces $\Omega_{k0} = 0$; closed spatial hypersurfaces are located either below or to the left. The figure is adapted from [206].

*3.8. Starburst Galaxy Data*

Mania and Ratra [380] analyzed constraints on the parameters of the $\phi$CDM-RP, the XCDM, and the ΛCDM models from the $H_{II}$ starburst galaxy apparent magnitude versus redshift data of Siegel et al. [381]. The authors followed the Percival et al. [260] procedure to obtain these constraints. The results are demonstrated in Figure 65. These constraints are largely consistent but not as restrictive as those derived from the measurements of the BAO peak length scale, the SNe Ia apparent magnitude, and the CMB temperature anisotropy.

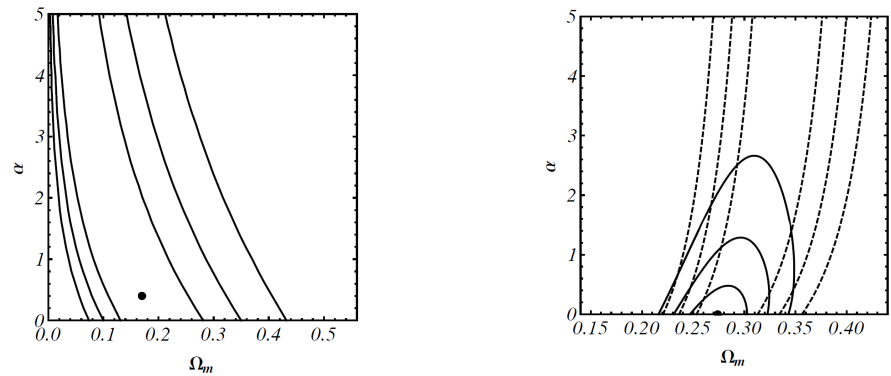

**Figure 65.** The 1 $\sigma$, 2$\sigma$, and 3$\sigma$ confidence level contour constraints on the parameters of the spatially flat $\phi$CDM model with the RP potential. The horizontal axis with $\alpha = 0$ corresponds to the standard spatially flat $\Lambda$CDM model. (Left panel) Contours obtained from $H_{II}$ galaxy data. The best-fit point with $\chi^2_{\min} = 53.3$ is indicated by the solid black circle at $\Omega_{m0} = 0.17$ and $\alpha = 0.39$. (Right panel) Contours obtained from joint $H_{II}$ galaxy and BAO peak length scale data (solid lines) and BAO peak length scale data only (dashed lines). The best-fit point with $-2\log(L_{\max}) = 55.6$ is indicated by the solid black circle at $\Omega_{m0} = 0.27$ and $\alpha = 0$. The figure is adapted from [380].

Cao et al. [382] derived constraints on the parameters of the spatially flat and non-flat $\Lambda$CDM, XCDM, and $\phi$CDM-RP models from the compilation of the $H_{II}$ starburst galaxy ($H_{II}$G) data of González-Morán et al. [354] and the $H_{II}$G data of González-Morán et al. [383]. The authors tested the model independence of the QSO angular size measurements. They found that the new compilation of 2019 $H_{II}$G data provides tighter constraints and favors lower values of the cosmological parameters than those from the 2019 $H_{II}$G data. The use of the QSO measurements gives model-independent constraints on the characteristic linear size $l_m$ of QSO within a sample. Analysis of the data on the $H(z)$, BAO peak length scale, the SNe Ia apparent magnitude-Pantheon, the SNe Ia apparent magnitude-DES, QSO, and the latest compilation of the $H_{II}$G data provides almost model-independent estimates of the Hubble constant, the matter density parameter at the present epoch, and the characteristic linear size, respectively, as $H_0 = 69.7 \pm 1.2$ km s$^{-1}$Mpc$^{-1}$, $\Omega_{m0} = 0.295 \pm 0.021$, and $l_m = 10.93 \pm 0.25$ pc.

Cao and Ratra [207] performed analysis of constraints on the parameters of the spatially flat and non-flat $\Lambda$CDM, XCDM, and $\phi$CDM-RP models from joint datasets consisting of data on the updated 32 $H(z)$ Hubble parameter, 12 BAO peak length scale, 1048 Pantheon SNe Ia apparent magnitudes, 20 binned DES-3yr SNe Ia apparent magnitudes, 120 QSO-AS and 78 $Mg_{II}$ reverberation-measured QSO, 181 $H_{II}$ starburst galaxy, and 50 Platinum Amati-correlated GRB. As a result, the authors found that constraints from each dataset are mutually consistent. There is a slight difference between constraints determined from the QSO-AS + $H_{II}$G + $Mg_{II}$ QSO + A118 dataset and those from QSO-AS + $H_{II}$G + $Mg_{II}$ QSO + Platinum + A101 dataset, so the authors considered only the cosmological constraints from the joint dataset $H(z)$ + BAO peak length scale + SNe Ia apparent magnitudes + QSO-AS + $H_{II}$G + $Mg_{II}$ QSO + A118 (HzBSNQHMA). The model-independent value of the Hubble constant, $H_0 = 69.7 \pm 1.2$ km s$^{-1}$Mpc$^{-1}$, and the matter density parameter at the present epoch, $\Omega_{m0} = 0.295 \pm 0.017$, were obtained by using the HzBSNQHMA dataset. The obtained value of the constraint for $H_0$ lies in the middle of the spatially flat $\Lambda$CDM model result of Planck Collaboration 2018 of Aghanim et al. [13] and the local expansion rate $H(z)$ result of Riess et al. [113], a bit closer to the former. Based on the $DIC$ analysis, the spatially flat $\Lambda$CDM model is the most preferable, but both dynamic dark energy models and space curvature are not ruled out.

*3.9. X-ray Gas Mass Fraction of Cluster Data*

Using Chandra measurements of X-ray gas mass fraction of 26 rich clusters obtained by Allen et al. [384], Chen and Ratra [334] constrained the parameters of the $\phi$CDM-RP,

ΛCDM, and the XCDM models. Resulting constraints are consistent with those derived from other cosmological tests but favor the spatially flat ΛCDM model more, Figure 14. Constraints on the parameters of the $\phi$CDM model are tighter than those derived from the SNe Ia apparent magnitude data of Podariu and Ratra [242], Waga and Frieman [385], redshift–angular size data of Chen and Ratra [386], Podariu et al. [387], and gravitational lensing statistics of Chae et al. [388], Figure 66 (left panel).

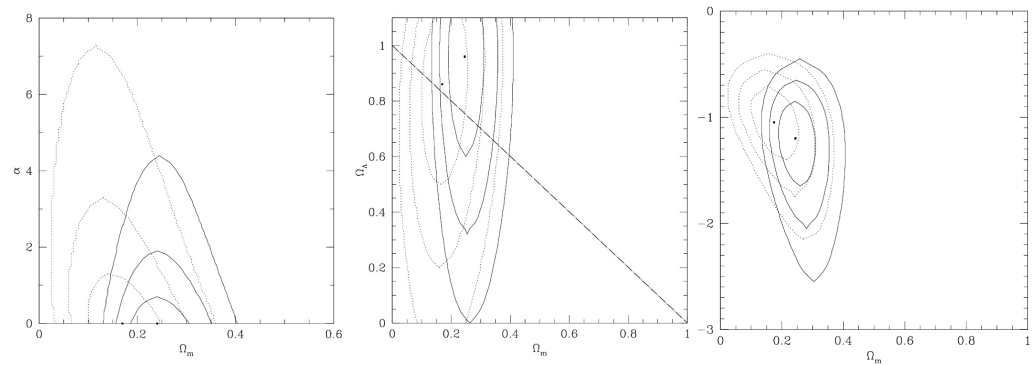

**Figure 66.** The 1$\sigma$, 2$\sigma$, and 3$\sigma$ confidence level contour constraints on parameters. (Left panel) For the spatially flat $\phi$CDM model with the RP potential and non-relativistic CDM. Continuous lines are obtained for $h = 0.72 \pm 0.08$ and $\Omega_b h^2 = 0.0214 \pm 0.002$ while dotted lines match $h = 0.68 \pm 0.04$ and $\Omega_b h^2 = 0.014 \pm 0.004$. (Middle panel) For the spatially flat ΛCDM model. Continuous lines are obtained for $h = 0.72 \pm 0.08$ and $\Omega_b h^2 = 0.0214 \pm 0.002$ while dotted lines obtained for $h = 0.68 \pm 0.04$ and $\Omega_b h^2 = 0.014 \pm 0.004$. The diagonal dash–dotted line delimits spatially flat models. (Right panel) For the XCDM model. Continuous lines are obtained for $h = 0.72 \pm 0.08$ and $\Omega_b h^2 = 0.0214 \pm 0.002$ while dotted lines are derived for $h = 0.68 \pm 0.04$ and $\Omega_b h^2 = 0.014 \pm 0.004$. In all pictures, two dots indicate the place of maximum probability. The figure is adapted from [334].

Wilson et al. [335] used the R04 gold SNe Ia apparent magnitude versus the redshift data of Riess et al. [212] and X-ray gas mass fraction of cluster data from Allen et al. [384] to constrain the $\phi$CDM-RP model; the results are given in Figure 67. According to these results, the standard spatially flat ΛCDM model is preferable, but the $\phi$CDM model is not ruled out either. The contours obtained from joint R04 gold SNe Ia apparent magnitude data and galaxy cluster gas mass fraction data are tighter constrained than those obtained by Podariu and Ratra [242] from earlier SNe Ia apparent magnitude versus redshift data.

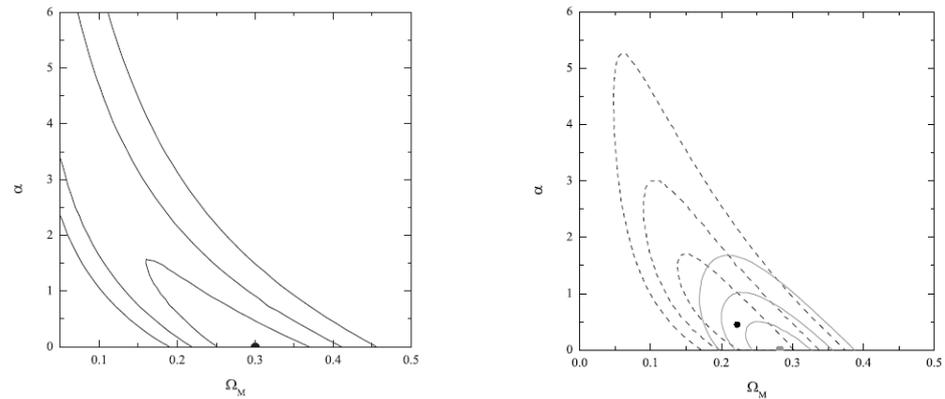

**Figure 67.** The $1\sigma$, $2\sigma$, and $3\sigma$ confidence level contour constraints on the parameters of the $\phi$CDM model with the RP potential. (Left panel) For the R04 gold SNe Ia apparent magnitude sample. The dot indicates the maximum likelihood for which $\Omega_{m0} = 0.30$ and $\alpha = 0$. (Right panel) For the joint R04 gold SNe Ia apparent magnitude sample and galaxy cluster gas mass fraction data. The solid gray lines are computed for $h = 0.72 \pm 0.08$ and $\Omega_b h^2 = 0.0214 \pm 0.002$ with maximum likelihood at $\Omega_{m0} = 0.28$ and $\alpha = 0$. The black dotted lines are computed for $h = 0.68 \pm 0.04$ and $\Omega_b h^2 = 0.014 \pm 0.004$, with maximum likelihood at $\Omega_{m0} = 0.22$ and $\alpha = 0.45$. The figure is derived from constraints on the parameters of the spatially flat $\Lambda$CDM, the XCDM, and the $\phi$CDM models with the RP potential adapted from [335].

Constraints on the model parameters $w_0$ and $w_a$ of the $w$CDM model using the X-ray temperature data on massive galaxy clusters within the redshift range $0.05 \leq z \leq 0.83$ with massive galaxy clusters ($M_{\text{cluster}} > 8 \times 10^{14} \, h^{-1} M_\odot$ within a comoving radius of $R_{\text{cluster}} = 1.5 \, h^{-1}$Mpc), were determined by Campanelli et al. [389]. The results are presented in Figure 68. Current data on massive clusters weakly constrain $w_0$ and $w_a$ parameters around the $(w_0, w_a) = (-1, 0)$ values corresponding to the $\Lambda$CDM model. In the analysis including data from the galaxy cluster number count, Hubble parameter $H(z)$, CMB temperature anisotropy, BAO peak length scale, and the SNe Ia apparent magnitude, the values of $w_0 = -1.14^{+0.14}_{-0.16}$ and $w_a = 0.85^{+0.42}_{-0.60}$ were obtained at a $1\sigma$ confidence level.

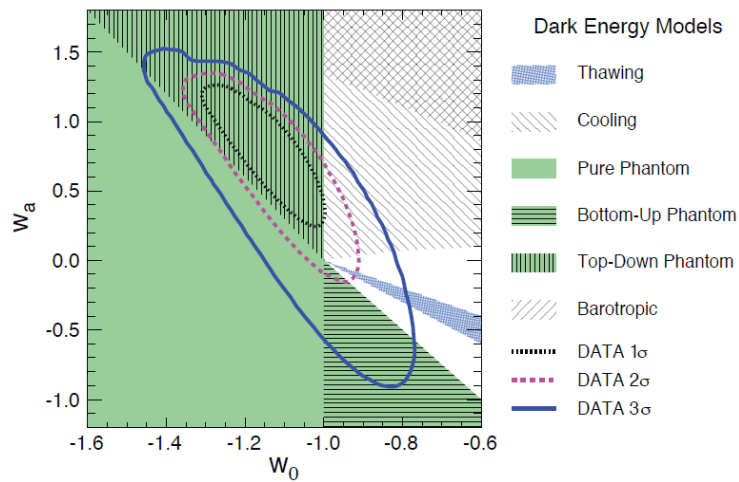

**Figure 68.** The $1\sigma$, $2\sigma$, and $3\sigma$ confidence level contours in the $w_0$-$w_a$ plane from the data analysis of the galaxy cluster number count, Hubble parameter $H(z)$, CMB temperature anisotropy, BAO peak length scale, and the SNe Ia apparent magnitude. The shaded areas represent various types of dynamical dark energy models. The figure is adapted from [389].

Chen and Ratra [390] applied angular size versus redshift measurements for galaxy clusters from Bonamente et al. [391] to constraint parameters of the $\phi$CDM-RP, the XCDM, and the $\Lambda$CDM models. X-ray observations of the intracluster medium in combination with radio observations of the Sunyaev–Zel'dovich effect of galaxy clusters make it possible to estimate the distance from the angular diameter $d_A$ of galaxy clusters. The authors applied the 38 angular diameter distance $d_A$ from Bonamente et al. (2006) [391] to constrain the cosmological parameters of the models presented above. The results are demonstrated in Figure 69. The analysis of the angular size measurements along with the more restrictive BAO peak length scale data and the SNe Ia apparent magnitude measurements favors the spatially flat $\Lambda$CDM model but does not exclude the $\phi$CDM model.

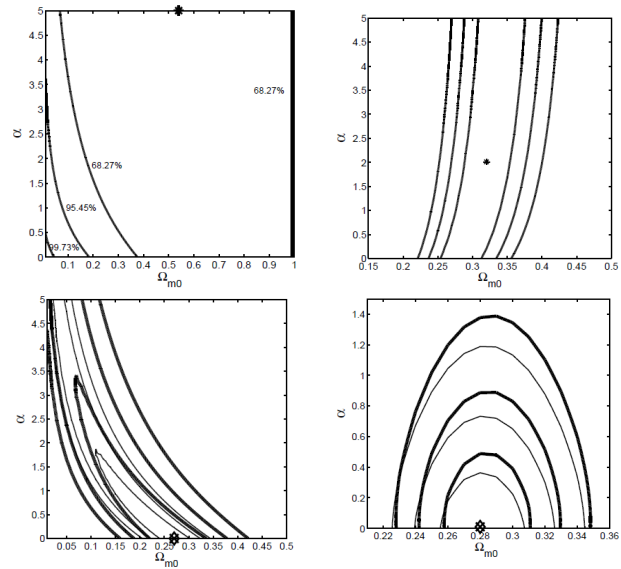

**Figure 69.** The 1 $\sigma$, 2$\sigma$, and 3$\sigma$ confidence level contour constraints on the parameters of the $\phi$CDM model with the RP potential. The horizontal axis with $\alpha = 0$ corresponds to the standard spatially flat $\Lambda$CDM model. (Left upper panel) Contours obtained from angular diameter distance $d_A$ data. The star denotes the best-fit point $(\Omega_{m0}, \alpha) = (0.54, 5)$, $\chi^2_{\min} = 37.3$. (Right upper panel) Contours obtained using BAO peak length scale data. The star marks the best-fit point $(\Omega_{m0}, \alpha) = (0.32, 2.01)$, $\chi^2_{\min} = 0.169$. (Left lower panel) Contours obtained from SNe Ia apparent magnitude data. Thin solid lines (best-fit pair at $(\Omega_{m0}, \alpha) = (0.27, 0.00)$, $\chi^2_{\min} = 543$, marked by cross "x" ) exclude systematic errors, while thick solid lines (best-fit pair at $(\Omega_{m0}, \alpha) = (0.27, 0.00)$, $\chi^2_{\min} = 531$, marked by diamond "$\diamond$") calculated for systematics. (Right lower panel) Thick (thin) solid lines are contours obtained from a joint analysis of BAO peak length scale and SNe Ia apparent magnitude (with systematic errors) data, with (and without) angular diameter distance $d_A$. The cross "x" means the best-fit point defined from the joint sample without the $d_A$ data at $(\Omega_{m0}, \alpha) = (0.28, 0.00)$ with $\chi^2_{\min} = 531$. The diamond "$\diamond$ denotes the best-fit point determined from the joint sample with the $d_A$ data at $(\Omega_{m0}, \alpha) = (0.28, 0.01)$ with $\chi^2_{\min} = 572$. The figure is adapted from [390].

*3.10. Reionization Data*

Mitra et al. [195] studied the influence of dynamical dark energy and spatial curvature on cosmic reionization. With this aim, the authors examined reionization in the tilted spatially flat and untilted spatially non-flat XCDM, and $\phi$CDM-RP quintessential inflation models. Statistical analysis was performed based on a principal component analysis and the MCMC analysis using a compilation of the lower-redshift reionization data by Wyithe and Bolton [296] and Becker and Bolton [297] to estimate uncertainties in the model reionization histories. The authors found that, regardless of the nature of dark energy, there are significant differences between the reionization histories of the spatially flat and spatially non-flat cosmological models. Although both the flat and non-flat models fit well the low-redshift $z \leq 6$ reionization observations, there is a discrepancy between high-redshift $z \geq 7$ Lyman-$\alpha$ emitter (LAE) data from Songaila and Cowie [298] and Prochaska et al. [299], and the predictions from spatially non-flat models. Unlike spatially flat models, the spatially non-flat ones have a much earlier and more extended reionization scenario that is completed around $z \approx 7$. The authors found that the higher the electron scattering optical depths $\tau_{el}$ the more extended is the reionization process. Moreover, such models predict a much lower fraction of neutral hydrogen at higher redshifts, $7 \lesssim z \lesssim 13$ (from $2\sigma$ limits for $Q_{HII} \sim 1$), namely, $\tau_{el} < 1$ at $z \sim 8$, is clearly contradictory to most current observation limits from distant QSO, GRB, and LAE data. Also, a serious disadvantage of spatially non-flat models can be seen from the results obtained on the evolution of the photon escape fraction $f_{esc}(z)$: in spatially non-flat models $f_{esc}(z) \gtrsim 1$ even at $z \gtrsim 8$, given its $2\sigma$ limits. Such non-physical $f_{esc}(z)$ values indicate the possibility of excluding these models. However, the Planck 2018 [13] reduction in the value of the $\tau_{el}$ in the six-parameter tilted flat $\Lambda$CDM inflation model by about $0.9\sigma$ reconciles the predictions of the non-flat model with observations.

*3.11. Gravitational Lensing Data*

Constraints on the parameters of two $\phi$CDM models, with the RP and the pNGb potentials, were analyzed by Waga and Frieman [385]. These models predict radically different futures for our universe. In the model with the RP potential, the expansion of the universe will continue to accelerate. In the model with the pNGb potential, the present epoch of the expansion of the universe with acceleration will be followed by a return to the matter-dominated epoch. For these observational tests, the authors used the compilation of measurements: gravitational lensing statistics [392–396] and the high-$z$ SNe Ia apparent magnitudes [212,397,398]. The results of these studies are presented in Figure 70, where it is shown that a large region of parameter space for the considered models is consistent with the SNe Ia apparent magnitude data if $\Omega_{m0} > 0.15$. The authors obtained the constraint on the model parameter $\alpha$ of the RP potential, $\alpha < 5$. The $\phi$CDM model with the pNGb potential is constrained by the SNe Ia apparent magnitude and lensing measurements at a $2\sigma$ confidence level.

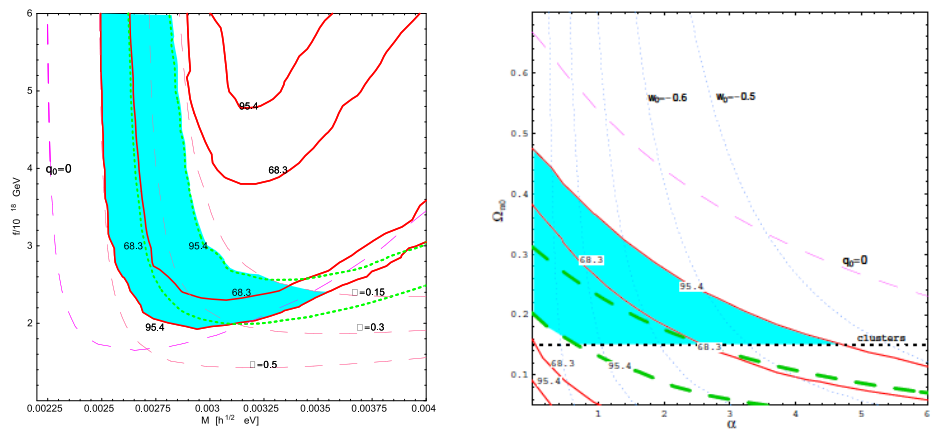

**Figure 70.** The $1\sigma$ and $2\sigma$ confidence level contours arising from lensing statistics and SNe Ia apparent magnitude versus redshift data (solid curves). (Left panel) The $\phi$CDM model with the pNGb potential. Solid curves correspond to constraints from SNe Ia apparent magnitude data. Contours of the constant matter energy parameter at the present epoch $\Omega_{m0}$ and the limit for the acceleration parameter at the present epoch $q_0 = 0$ are depicted. (Right panel) The $\phi$CDM model with the RP potential. The lower bound of $\Omega_{m0} = 0.15$ from clusters and curves for the constant value of the EoS parameter at the present epoch $w_0$ are shown. The figure is adapted from [385].

Statistics on strong gravitational lensing based on Cosmic Lens All-Sky Survey data from Myers et al. [399] and Browne et al. [400] was applied by Chae et al. [388] to constrain parameters of the $\phi$CDM-RP model. The results are presented in Figure 71. The maximum of the likelihood accords to the values of the matter density parameter at the present epoch $\Omega_{m0} = 0.34$ and the model parameter $\alpha = 0$, i.e., to the standard spatially flat $\Lambda$CDM model. For the 68% confidence level, $0.18 < \Omega_{m0} < 0.62$ and $\alpha < 2.7$, while, for the 95% confidence level, $\Omega_{m0} = 1$ and $\alpha = 8$. Strong gravitational lensing constraints are favorable for the standard spatially flat $\Lambda$CDM model, and consistent with Chen and Ratra [386] constraints from the SNe Ia apparent magnitude data, but are weaker.

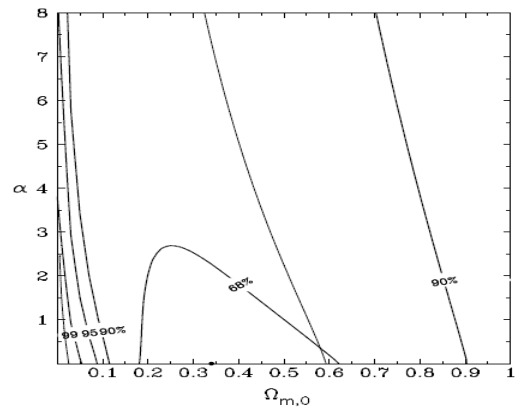

**Figure 71.** The 68%, 90%, 95%, and 99% confidence level contour constraints on the parameters of the $\phi$CDM model with the RP potential from the strong gravitational lensing data. The thin line represents the 68% confidence level derived from the SNe Ia apparent magnitude versus redshift data by Chen and Ratra [386]. The horizontal axis for which $\alpha = 0$ corresponds to the spatially flat $\Lambda$CDM model. The figure is adapted from [388].

*3.12. Compact Radio Source Data*

The compact radio source angular size versus redshift data of Gurvits et al. [401] were used by Chen and Ratra [386] to derive constraints on the parameters of the $\phi$CDM-RP model. These constraints are consistent with the results obtained from the SNe Ia apparent magnitude data of [2,212] but they are less restrictive, Figure 72.

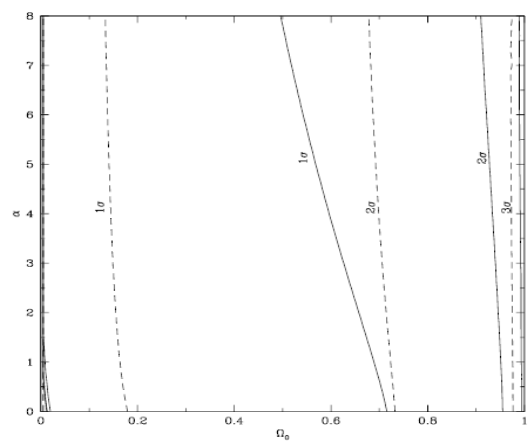

**Figure 72.** The $1\sigma$, $2\sigma$, and $3\sigma$ confidence level contours constraints on the parameters of the spatially flat $\phi$CDM model with the RP potential. Solid lines are contours computed for the uniform prior $p(\Omega_0) = 1$. Short dashed lines are obtained for the logarithmic prior $p(\Omega_0) = 1/\Omega_0$. The figure is adapted from [386].

Podariu et al. [387] used the redshift–angular size data from double radio galaxies called FRIIb sources to constrain the parameters of the $\phi$CDM-RP model in the spatially flat universe. These constraints are consistent both with the results obtained from the SNe Ia apparent magnitude data of [2,212] and with the results obtained from the compact radio source redshift–angular size data of Chen and Ratra in [386], but they are less restrictive, Figure 73.

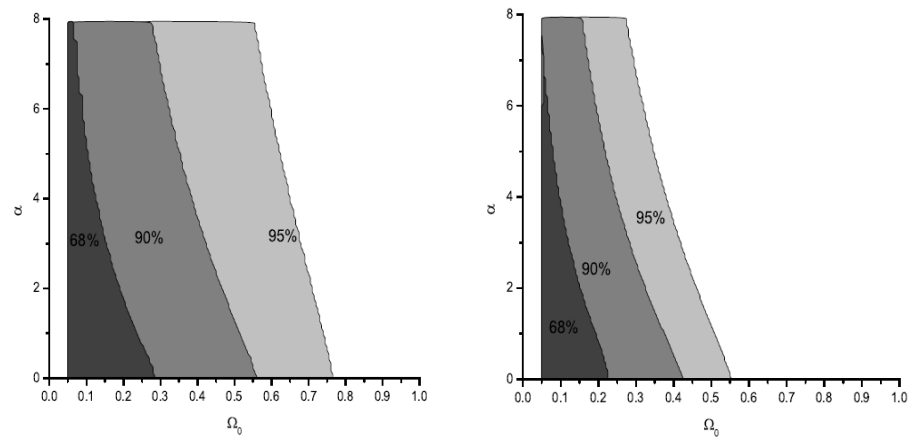

**Figure 73.** The $1\sigma$, $2\sigma$, and $3\sigma$ confidence level contour constraints on the parameters of the spatially flat $\phi$CDM model with the RP potential. (Left panel) Constraints were obtained using all twenty radio galaxies (including 3C 427.1). (Right panel) Constraints were obtained using only nineteen radio galaxies (excluding 3C 427.1). The figure is adapted from [387].

## 4. Summary and Results

The results of research based on the papers considered in the review are summarized and grouped into the following topics: results of constraints on the parameters of dynamical dark energy models, alleviation and resolving of the $\Lambda$CDM model tensions, data preferences, disadvantages of models to data, failure and incompatibility of data, sensibility of various data, consistency of constraints results by various data, comparing constraints with various data, model-independent estimate of the Hubble constant $H_0$ and matter density parameter at the present epoch $\Omega_{m0}$, and problems with QSO available data. The main research results are presented in more detail below.

- Results of constraints on the parameters of dynamical dark energy models
  1. For both the extended and ordinary quintessence $\phi$CDM-RP models, $1\sigma$ constraints were obtained of $\alpha < 0.8$ and $\alpha < 0.6$, using the SNe Ia+SNAP data, Caresia et al. [215].
  2. Constraints on the spatial curvature density parameter today to be $|\Omega_{k0}| \leq 0.15$ at a $1\sigma$ confidence level in the spatially non-flat $\phi$CDM-RP model as well as the XCDM model, from SNe Ia +$H(z)$+ BAO data. More precise data are required to tighten the bounds on the parameters, Farooq et al. [256].
  3. In constraints on the model parameters of the $\Lambda$CDM model, the XCDM model, and the $\phi$CDM-RP model using galaxy cluster gas mass fraction data, $\Omega_m$ is better constrained than $\alpha$, whose best-fit value is $\alpha = 0$, corresponding to the standard spatially flat $\Lambda$CDM model; however, the scalar field $\phi$CDM model is not excluded [282].
  4. The deceleration–acceleration transition redshift $z_{da} = 0.74 \pm 0.05$ was obtained as a result of the constraints on the parameters of the $\phi$CDM-RP model from $H(z)$ data [341].
  5. A likelihood analysis of the COBE-DMR sky maps to normalize the spatially flat $\phi$CDM-RP model shows that this model remains an observationally viable alternative to the standard spatially flat $\Lambda$CDM model [281].
  6. The $2\sigma$ upper bounds of $\sum m_\nu < 0.165\ (0.299)$ eV and $\sum m_\nu < 0.164\ (0.301)$ eV, respectively, for the spatially flat (spatially non-flat) $\Lambda$CDM model and the spatially flat (spatially non-flat) $\phi$CDM model were defined using CMB + BAO +SNe Ia and the Hubble Space Telescope $H_0$ prior observations. The inclusion of spatial curvature as a free parameter leads to a significant expansion of the confidence regions for $\sum m_\nu$ and other parameters in spatially flat $\phi$CDM models, but the corresponding differences are larger for both the spatially non-flat $\Lambda$CDM and spatially non-flat $\phi$CDM models [287].

7. When the bispectrum component is included in the BAO + LSS data for the $\phi$CDM model, a significant dynamical dark energy signal was achieved at a $2.5 - 3\sigma$ confidence level. Thus, the bispectrum can be a very useful tool for tracking and examining the possible dynamical features of dark energy and their influence on the LSS formation in the linear regime [319]. (The bispectrum component has been used by Solà et al. [188] before to study the running cosmic vacuum in the RVMs!)

8. As a result of constraints on the parameters of the $o$CDM, XCDM (here $w_0$CDM), and $w$CDM models by using the BAO+BBN +SNe Ia data the value of epoch $\Omega_{k0} = -0.043^{+0.036}_{-0.036}$ at a $1\sigma$ confidence level, which is consistent with the spatially flat universe; in the spatially flat XCDM model, the value of the dark energy EoS parameter at the present epoch $w_0 = -1.031^{+0.052}_{-0.048}$ at a $1\sigma$ confidence level, which approximately equals the value of the EoS parameter for the $\Lambda$CDM model; and values of the $w_0$ and $w_a$ in the CPL parameterization of the EoS parameter of the $w$CDM model $w_0 = -0.98^{+0.099}_{-0.11}$ and $w_a = -0.33^{+0.63}_{-0.48}$ at $1\sigma$ confidence level were obtained. The exclusion of the SNe Ia data from the joint data analysis does not significantly weaken the resulting constraints. This means that, when using a single external BBN prior, full-shape and BAO peak length scale data can provide reliable constraints independent of CMB temperature anisotropy constraints [330].

9. Current X-ray temperature data on massive galaxies weakly constrain the $w_0$ and $w_a$ parameters of the $w$CDM model around the $(w_0, w_a) = (-1, 0)$ values of the $w$CDM model corresponding to the $\Lambda$CDM model. In the analysis including data from the galaxy cluster number count+ $H(z)$ + CMB temperature anisotropy + BAO + SNe Ia, the values of $w_0 = -1.14^{+0.14}_{-0.16}$ and $w_a = 0.85^{+0.42}_{0.60}$ were obtained at a $1\sigma$ confidence level [389].

10. In constraints on parameters in the spatially flat $\Lambda$CDM model, the spatially closed $\phi$CDM models with the RP and Sugra potentials using SNe Ia data, values of $\Omega_{m0}$ and $\Omega_{\phi 0}$, are quite different from those for the $\Lambda$CDM. The quintessence scalar field creates more structures outside the filaments, lighter halos with higher internal velocity dispersion, as seen from N-body simulations performed by the authors to study the influence of quintessence on the distribution of matter on large scales [254].

11. In the $\phi$CDM-RP model in a spacetime with non-zero spatial curvature, the dynamical scalar field has an attractor solution in the curvature dominated epoch, while the energy density of the scalar field increases relative to that of the spatial curvature [253].

12. In constraints on $H_0$ in the $\phi$CDM-RP, the $w$CDM, and the spatially flat and spatially non-flat $\Lambda$CDM models from measurements of $H(z)$, the value of the $H_0$ is found as follows: for the spatially flat and spatially non-flat $\Lambda$CDM model, $H_0 = 68.3^{+2.7}_{-3.3}$ km s$^{-1}$Mpc$^{-1}$ and $H_0 = 68.4^{+2.9}_{-3.3}$ km s$^{-1}$Mpc$^{-1}$; for the $w$CDM model, $H_0 = 65.0^{+6.5}_{-6.6}$ km s$^{-1}$Mpc$^{-1}$; for the $\phi$CDM model, $H_0 = 67.9^{+2.4}_{-2.4}$ km s$^{-1}$Mpc$^{-1}$ (at a $1\sigma$ confidence level) [342].

13. In constraints on the parameters of the spatially flat and non-flat $\Lambda$CDM, XCDM, and $\phi$CDM-RP models, as well as on the QSO radius–luminosity $R - L$ relation parameters from QSO reverberation measured, the parameters of the $R - L$ relation do not depend on the cosmological models considered and, therefore, the $R - L$ relation can be used to standardize the $C_{IV}$ QSO data. Mutually consistent constraints on the cosmological parameters from $C_{IV}$, $Mg_{II}$, and $H(z)$ + BAO peak length scale data allow conducting the analysis from the $C_{IV} + Mg_{II}$ dataset as well as from the $H(z)$ + BAO peak length scale + $C_{IV} + Mg_{II}$ datasets.

           Although the $C_{IV}$ + $Mg_{II}$ cosmological constraints are weak, they slightly (at a $\sim 0.1\sigma$ confidence level) change the constraints from the $H(z)$ + BAO peak length scale + $C_{IV}$ + $Mg_{II}$ datasets [204].

14. The quintessential inflation model with the generalized exponential potential including massive neutrinos that are non-minimally coupled with a scalar field obtains observational constraints on parameters using combinations of data: CMB + BAO (BOSS) + SNe Ia (SNLS). The upper bound on possible values of the sum of neutrino masses $\sum m_\nu < 2.5$ eV is significantly larger than in the spatially flat ΛCDM model [178].

- Alleviation and resolving of the ΛCDM model tensions
    1. The joint Planck + BAO (transversal) analysis agrees well with the measurements made by the SH0ES team and, applied to the IDE models, solves the Hubble constant $H_0$ tension [67].
    2. A larger value of the Hubble constant, i.e., alleviation of the Hubble constant tension (with a significance of $3.6\sigma$), has been obtained for the spatially non-flat IDE models. Searches for other forms of the interaction function and the EoS for the dark energy component in IDE models are needed, which may further ease the tension of the Hubble constant [121].
    3. The lower multipole region of CMB + BAO (6dFGS, SDSS-MGS) in the spatially closed quintessential inflation $\phi$CDM model reduces the tension between the Planck and the weak lensing $\sigma_8$ constraints [328].
    4. The maximum of the likelihood in the constraint parameters in the $\phi$CDM-RP model from the strong gravitational lensing data accords to the values of the matter density parameter at the present epoch $\Omega_{m0} = 0.34$ and the model parameter $\alpha = 0$, i.e., to the standard spatially flat ΛCDM model. For the 68% confidence level, $0.18 < \Omega_{m0} < 0.62$ and $\alpha < 2.7$, while, for 95% confidence level, $\Omega_{m0} = 1$ and $\alpha = 8$ [388].
    5. In extended $\phi$CDM- RP models with exponential coupling to the Ricci scalar, the projection of the ISW effect on the CMB temperature anisotropy is found to be considerably larger in the exponential case with respect to a quadratic non-minimal coupling. This reflects the fact that the effective gravitational constant depends exponentially on the dynamics of the scalar field [216].
    6. The value of the cosmological deceleration–acceleration transition $z_{da}$ is insensitive to the chosen model from the spatially flat and spatially non-flat $\phi$CDM-RP, the XCDM, and the $w$CDM using $H(z)$ data, and depend only on the assumed value of the Hubble constant $H_0$. The weighted mean of these measurements is $z_{da} = 0.72 \pm 0.05$ ($0.84 \pm 0.03$) for $H_0 = 68 \pm 2.8$ ($73.24 \pm 1.74$) km s$^{-1}$Mpc$^{-1}$ [193].
    7. In contrast to the joint Planck + BAO analysis, where it is not possible to solve the Hubble constant $H_0$ tension, the joint Planck + BAO (transversal) analysis agrees well with the measurements made by the SH0ES team and, applied to the IDE models, solves the Hubble constant $H_0$ tension [67].

- Data preferences
    1. Planck 2018 CMB data favor spatially closed hypersurfaces in spatially non-flat IDE models at more than 99% CL (with a significance of $5\sigma$) [121].
    2. The higher multipole region of the CMB temperature anisotropy data is in better agreement with the tilted spatially flat ΛCDM model than with the spatially closed $\phi$CDM model [328].
    3. Depending on the value of the Hubble constant $H_0$ as a prior and the cosmological model under consideration, the data provides evidence in favor of the spatially non-flat scalar field $\phi$CDM model [329].
    4. The spatially closed quintessential inflation $\phi$CDM model provides a better fit to the lower multipole region of CMB temperature anisotropy data +BAO (6dFGS,

5.  SDSS-MGS) data compared to that provided by the tilted spatially flat ΛCDM model [328].
6.  In most of the tilted spatially flat and untilted spatially non-flat ΛCDM, XCDM, and $\phi$CDM-RP quintessential inflation models, the QSO data favor $\Omega_{m0} \sim$ 0.5–0.6, while, in a combined analysis of QSO + $H(z)$ + BAO, the values of the $\Omega_{m0}$ are shifted slightly towards larger values. A combined QSO + BAO peak length scale + $H(z)$ dataset is consistent with the standard spatially flat ΛCDM model, but favors slightly both the spatially closed hypersurfaces and the dynamical dark energy models [196].
7.  Depending on the chosen model (from spatially flat and spatially non-flat ΛCDM, XCDM, and $\phi$CDM-RP models) and dataset (from BAO + $H(z)$ +QSO), the data slightly favor both the spatially closed hypersurfaces with $\Omega_{k0} < 0$ at a $1.7\sigma$ confidence level and the dynamical dark energy models over the standard spatially flat ΛCDM model at a slightly higher than $2\sigma$ confidence level. Furthermore, depending on the dataset and the model, the observational data favor a lower Hubble constant value than the one measured by the local data at a $1.8\sigma$ confidence level to $3.4\sigma$ confidence level [194].
8.  The analysis with the $H(z)$ + BAO + QSO-AS + $H_{II}$G + GRB dataset favors the spatially flat ΛCDM model but also does not rule out dynamical dark energy models [351].
9.  The Hubble constant $H_0$ value is constrained in the spatially flat and spatially non-flat ΛCDM, XCDM, and $\phi$CDM-RP models using various combinations of datasets: BAO +SNe Ia $H(z)$. The BAO +SNe Ia $H(z)$ dataset slightly favors the untilted spatially non-flat dynamical XCDM and $\phi$CDM quintessential inflation models, as well as smaller Hubble constant $H_0$ values [402].
10. Smaller angular scale SPTpol measurements (used jointly with only Planck CMB temperature anisotropy data or with the combination of Planck CMB temperature anisotropy data and non-CMB temperature anisotropy data) favor the untilted spatially closed models [301].
11. The spatially flat $\phi$CDM scalar field models could not be unambiguously preferred, from the DESI predictive data ($H(z)$ + $H(z)$ + angular diameter distance $d_A$), over the standard ΛCDM spatially flat model, the latter still being the most preferred dark energy model [318].
12. CMB (Planck 2015) + BAO + SNe Ia +$H(z)$ + LSS growth data slightly favor the spatially closed XCDM model over the spatially flat ΛCDM model at a $1.2\sigma$ confidence level, while also being in better agreement with the untilted spatially flat XCDM model than with the spatially flat ΛCDM model at the $0.3\sigma$ confidence level [324].
13. The analysis of the BAO + SNe Ia+ angular diameter distance $d_A$ (using X-ray observations of the intracluster medium + radio observations of the Sunyaev–Zel'dovich effect of galaxy clusters) data favors the spatially flat ΛCDM model but does not exclude the spatially flat $\phi$CDM-RP model [390].
14. SNe Ia + X-ray gas mass fraction of cluster data is preferable to the standard spatially flat ΛCDM model, but the $\phi$CDM model is not ruled out either [335].
15. The spatially flat ΛCDM model is the most preferable, but both dynamic dark energy models and space curvature are not ruled out [207].
16. Combined analysis from QSO + $H(z)$ + BAO data is consistent with the standard spatially flat ΛCDM model, but slightly favors both closed spatial hypersurfaces and the untilted spatially non-flat $\phi$CDM model [197].
17. Constraints on the parameters in the spatially flat and non-flat ΛCDM, XCDM, and $\phi$CDM-RP models from three (ML, MS, and GL) ($L_0 - t_b$) Dainotti-correlated sets of GRB measurements are weak, providing lower bounds on parameter $\Omega_{m0}$, moderately favoring the non-zero spatial curvature, and largely consistent with both the currently accelerated cosmological expansion and with constraints determined on the basis of more reliable data [203].

17. In constraints on the parameters in the spatially flat and non-flat ΛCDM, XCDM, and $\phi$CDM-RP models by GRB data, the three-dimensional Dainotti correlation [378,379] is much preferable to the two-dimensional [377] one for the GRB datasets [206].

18. The $R - L$ relation parameters for $H\beta$ QSO data are independent in models under investigation, from the spatially flat and non-flat ΛCDM, XCDM, and $\phi$CDM-RP models; therefore, QSO data seem to be standardizable through $R - L$ relation parameters. The constraints derived using $H\beta$ QSO data are weak, slightly favoring the currently accelerating cosmological expansion, and are generally in the $2\sigma$ tension with the constraints derived from analysis of the measurements of the BAO peak length scale and the Hubble parameter $H(z)$ [201].

19. Constraints on the parameters of the spatially non-flat untilted $\phi$CDM-RP inflation model were improved from a $1.8\sigma$ to a more than $3.1\sigma$ confidence level by combining by CMB (Planck 2015) + BAO + SNe Ia +$H(z)$ + LSS data. CMB (Planck 2015) + BAO + SNe Ia + $H(z)$ + LSS data favor a spatially closed universe with the spatial curvature contributing about two-thirds of a percent of the current total cosmological energy budget. The spatially flat tilted $\phi$CDM inflation model is a $0.4\sigma$ better fit to the observational data than is the standard spatially flat tilted ΛCDM model, i.e., current observational data allow for the possibility of dynamical dark energy in the universe. The spatially non-flat tilted $\phi$CDM model better fits the DES bounds on the rms amplitude of mass fluctuations $\sigma_8$ as a function of the parameter $\Omega_{m0}$ [290].

20. The ΛCDM model has a strong advantage, investigating both the minimally coupled with gravity scalar field spatially flat $\phi$CDM-RP model and non-minimally coupled scalar field extended quintessence model with gravity (with the Ricci scalar), applying the dataset: the Pantheon SNe +BAO (6dFGS, SDSSLRG, BOSS-MGS, BOSS-LOWZ, WiggleZ, BOSS-CMASS, BOSS-DR12)+CMB+$H(z)$+RSD, when local measurements of the Hubble constant $H_0$ [13] are not taken into account and, conversely, this statement is weakened when local measurements of $H_0$ are included in the data analysis [222].

- Disadvantages of models to data

    1. Spatially non-flat $\phi$CDM-RP quintessential inflation models predict a much lower fraction of neutral hydrogen at higher redshifts $7 \lesssim z \lesssim 13$ (from $2\sigma$ limits for $Q_{HII}{\sim}1$), namely, $\tau_{el} < 1$ at $z{\sim}8$, are clearly contradictory to most current observation limits from distant QSO + GRB + LAE data [195].

    2. A serious disadvantage of spatially non-flat $\phi$CDM-RP quintessential inflation models can be seen from the results obtained from the evolution of the photon escape fraction $f_{esc}(z)$: in spatially non-flat models $f_{esc}(z) \gtrsim 1$ even at $z \gtrsim 8$, given its $2\sigma$ limits. Such non-physical $f_{esc}(z)$ values indicate the possibility of excluding these models. (However, the Planck 2018 [13] reduction in the value of the $\tau_{el}$ in the six-parameter tilted flat ΛCDM inflation model by about $0.9\sigma$ reconciles the predictions of the non-flat model with observations) [195].

    3. $H_{II}$ starburst galaxy apparent magnitude + QSO only(or) + BAO datasets favor the spatially flat ΛCDM model, while at the same time do not rule out dynamical spatially flat and non-flat $\phi$CDM-RP models [346].

    4. Constraints on the parameters of the spatially flat and non-flat ΛCDM, XCDM, and $\phi$CDM-RP models using SNe Ia (Pantheon + DES) + QSO + $H_{II}$G data + BAO + $H(z)$ favor dynamical dark energy and slightly spatially closed hypersurfaces; they do not preclude dark energy from being a cosmological constant and spatially flat hypersurfaces [202].

    5. Constraints on the parameters of the spatially non-flat untilted $\phi$CDM-RP inflation model by CMB (Planck 2015) + BAO + SNe Ia + $H(z)$ + LSS data do not

provide such good agreement with the larger multipoles of CMB (Planck 2015) data as the spatially flat tilted ΛCDM model [290].

- Failure and incompatibility of data
    1. CMB (Planck 2015) + BAO + SNe Ia + $H(z)$ + LSS growth data are unable to rule out dynamical scalar field spatially flat $\phi$CDM models [324].
    2. The dynamical untilted spatially non-flat XCDM model is not compatible with with higher multipoles of CMB temperature anisotropy data, as is the standard spatially flat ΛCDM model [324].
    3. The parameters of the spatially flat $\phi$CDM model could not be tightly constrained only by the current GRB data [364].
    4. There is a strong degeneracy between the model parameters $\Omega_m$ and $\alpha$ in the spatially flat $\phi$CDM-RP model applying only LSS data. According to constraints from LSS growth rate + BAO + CMB data, $\Omega_m = 0.30 \pm 0.04$ and $0 \leq \alpha \leq 1.30$ at a $1\sigma$ confidence level (the best-fit value for the model parameter $\alpha$ is $\alpha = 0$) [315].

- Sensibility of various data
    1. Studying dark energy in the early universe using SNe Ia + WMAP + CBI + VSA + SDSS + HST data, the values $w_0 < -0.8$ and density parameter in the early universe $\Omega_e < 0.03$ at the $1\sigma$ confidence level are found. SNe Ia data are most sensitive to $w_0$, while CMB temperature anisotropies and LSS growth rate are the best constraints of $\Omega_e$, Doran et al. [248].
    2. Expansion history data are not particularly sensitive to the dynamic effects of dark energy, while the data compilation BAO + LSS + CMB anisotropy is more sensitive [319].

- Consistency of constraint results with various data
    1. Constraints on the parameters of the $\phi$CDM-RP model from compact radio source angular size versus redshift data are consistent with the results obtained from the SNe Ia apparent magnitude data of [2,212], but they are less restrictive [386].
    2. Constraints of the spatially flat $\phi$CDM-RP model from radio galaxies FRIIb sources+redshift–angular size data are consistent both with the results obtained from the SNe Ia apparent magnitude data of [2,212] and with the results obtained from the compact radio source redshift–angular size [386], but they are less restrictive [387].
    3. Constraints on the parameters of the $\phi$CDM-RP, the XCDM, and the ΛCDM models from BAO + $H(z)$ data. The BAO + $H(z)$ data dataset is consistent with the standard spatially flat ΛCDM model [329].
    4. Constraints on the parameters of the spatially flat $\phi$CDM-RP, the XCDM, and the ΛCDM models from measurements of $H(z)$ are consistent with a moment of the deceleration–acceleration transition at redshift $z_{da} = 0.74 \pm 0.05$ derived by Farooq and Ratra [341], which corresponds to the standard spatially flat ΛCDM model [261].
    5. Constraints on the parameters of the spatially flat and non-flat ΛCDM, XCDM, and $\phi$CDM-RP models using the higher-redshift GRB $H_{II}$G + QSO are consistent with the currently accelerating cosmological expansion, as well as with the constraints obtained from the analysis of the $H(z)$ + BAO peak length scale. From the analysis of the $H(z)$ + BAO + QSO-AS + $H_{II}$G + GRB dataset, the model-independent values of epoch $\Omega_{m0} = 0.313 \pm 0.013$ and $H_0 = 69.3 \pm 1.2$ km s$^{-1}$Mpc$^{-1}$ are obtained [351].
    6. In each dark energy model (from the spatially flat and untilted spatially non-flat ΛCDM, XCDM, and scalar field $\phi$CDM-RP quintessential inflation models), constraints on cosmological parameters from SPTpol measurements+ CMB temperature anisotropy and non-CMB temperature anisotropy measurements are largely consistent with one another [301].

7. The dynamical untilted spatially non-flat XCDM model is compatible with the DES limits on the current value of the rms mass fluctuation amplitude $\sigma_8$ as a $\Omega_{m0}$ [324].
8. A large region of parameter space for the $\phi$CDM models, with the RP and the pNGb potential models, is consistent with the SNe Ia data if $\Omega_{m0} > 0.15$, wherein the constraints on the model parameter $\alpha$ of the RP potential is $\alpha < 5$. The $\phi$CDM model with the pNGb potential is constrained by the SNe Ia apparent magnitude + lensing measurements at a $2\sigma$ confidence level [385].
9. The constraints obtained from the $Mg_{II}$ QSOs + BAO + $H(z)$ agree with the spatially flat $\Lambda$CDM model as well as with spatially non-flat dynamical dark energy models [199].
10. The $H(z)$ data are consistent with the standard spatially flat $\Lambda$CDM model while they do not rule out the spatially non-flat XCDM and spatially non-flat $\phi$CDM models [193].
11. The obtained $H_0$ values, as a result of constraints in $\phi$CDM-RP, the $w$CDM, and the spatially flat and spatially non-flat $\Lambda$CDM models by $H(z)$ measurements, are more consistent with the smaller values determined from the recent CMB temperature anisotropy and BAO peak length scale data and with the values derived from the median statistics analysis of Huchra's compilation of $H_0$ data [342].
12. Constraints on spatially flat and non-flat $\Lambda$CDM, XCDM, and $\phi$CDM-RP models from GRB are consistent with the spatially flat $\Lambda$CDM as well as with the spatially non-flat dynamical dark energy models [370].
13. The second and third largest subsamples, SDSS-Chandra and XXL QSOs, which together account for about 30% of total QSO data, appear to be standardized. Constraints on the cosmological parameters from these subsamples are weak and consistent with the standard spatially flat $\Lambda$CDM model or with the constraints from the better-established cosmological probes [200].
14. The quintessential inflation model with the generalized exponential potential is in good agreement with observations and represents a successful scheme for the unification of the primordial inflaton field causing inflation in the very early universe and dark energy causing the accelerated expansion of the universe at the present epoch [178].
15. In quintessential inflation models, the early quintessence is characterized by a suppressed ability to cluster at small scales, as suggested by the compilation of data from WMAP + CBI + ACBAR + 2dFGRS + $L_{y-\alpha}$. Quintessential inflation models are compatible with these data for a constant spectral index of primordial density perturbations [270].

- Comparing constraints with various data
  1. Constraints on cosmological parameters in the spatially flat $\phi$CDM model by joint datasets consisting of measurements of the age of the universe+SNe Ia + BAO are tighter than those obtained from datasets consisting of data on the lookback time + age of the universe [326].
  2. Constraints on cosmological parameters in the scalar field $\phi$CDM-RP model from SVJ $H(z)$ data [333]. Using the $H(z)$ data, the constraints on the $\Omega_m$ are more stringent than those on the model parameter $\alpha$. Constraints on the matter density $\Omega_m$ are approximately as tight as the ones derived from the galaxy cluster gas mass fraction data [334] and from the SNe Ia apparent magnitude data [335].
  3. Constraints on the parameters of the $\phi$CDM-RP, the XCDM, the $w$CDM, and the $\Lambda$CDM models using BAO + SNe Ia data are more restrictive with the inclusion of eight new $H(z)$ measurements than those derived by Chen and Ratra [338]. This analysis favors the standard spatially flat $\Lambda$CDM model but does not exclude the scalar field $\phi$CDM model [337].
  4. Constraints on the parameters of the $\phi$CDM-RP, the XCDM, and the $\Lambda$CDM models using $H(z)$ data. $H(z)$ data yield quite strong constraints on the parameters of the $\phi$CDM model. The constraints derived from the $H(z)$ measurements are almost as restrictive as those implied by the currently available lookback time observations and the GRB luminosity data, but more stringent than those based on the currently available galaxy cluster angular size data. However, they are less restrictive than those following from the joint analysis of SNe Ia + BAO. The joint analysis of the $H(z)$ + SNe Ia + BAO favors the standard spatially flat $\Lambda$CDM model but does not exclude the dynamical scalar field $\phi$CDM model [338].

5. Constraints on the parameters of the $\phi$CDM-RP, the XCDM, and the $\Lambda$CDM models with the inclusion of new $H(z)$ measurement of Busca et al. are more restrictive than those derived by Farooq et al. The $H(z)$ constraints depend on the Hubble constant prior to $H_0$ used in the analysis. The resulting constraints are more stringent than those which follow from measurements of the SNe Ia apparent magnitude of Suzuki et al. (2012). This joint analysis consisting of measurements of $H(z)$+SNe Ia + BAO favors the standard spatially flat $\Lambda$CDM model but the dynamical scalar field $\phi$CDM model is not excluded either [339].

6. SNe Ia + $H(z)$ + LSS growth rate data are consistent with the standard spatially flat $\Lambda$CDM model, as well as with the spatially flat $\phi$CDM-RP model [313].

7. Strong gravitational lensing constraints are favorable for the standard spatially flat $\Lambda$CDM model and consistent with [386] constraints from the SNe Ia apparent magnitude data, but are weaker [388].

8. Constraints on the parameters of the spatially flat $\phi$CDM model obtained from joint R04 gold SNe Ia apparent magnitude data and galaxy cluster gas mass fraction data are tighter than those obtained by Podariu and Ratra [242] from earlier SNe Ia apparent magnitude versus redshift data [335].

9. Constraints obtained in [334] on the parameters of the $\phi$CDM model using Chandra measurements of X-ray gas mass fraction of the clusters are tighter than those derived from the SNe Ia apparent magnitude data [242,385], redshift–angular size data of [386,387], and gravitational lensing statistics [388]

10. Constraints on the parameters of the spatially flat and non-flat $\Lambda$CDM, XCDM, and $\phi$CDM-RP models derived from the QSO data only are significantly weaker than those derived from the combined set of the BAO + $H(z)$, but are consistent with both of them [199].

11. QSO data are significantly weaker but consistent with those from the combination of the $H(z)$ + BAO data in tilted spatially flat and untilted spatially non-flat $\Lambda$CDM, XCDM, and $\phi$CDM-RP quintessential inflation models [197].

12. The constraints in spatially flat $\Lambda$CDM, XCDM, and $\phi$CDM-RP models from the GRB data obtained by Wang's method [367] and Schaefer's method [366] disagree with each other at a more than $2\sigma$ confidence level [364].

13. Constraints on spatially flat and non-flat $\Lambda$CDM, XCDM, and $\phi$CDM-RP models from GRB data agree but are much weaker than those following from the BA + $H(z)$ data [370].

14. Constraints on the parameters of the spatially flat and non-flat $\Lambda$CDM, XCDM, and $\phi$CDM-RP models from the GRB data are consistent with the constraints obtained from the analysis of the BAO + $H(z)$ but are less restrictive [368].

15. Constraints on parameters in spatially flat and non-flat $\Lambda$CDM, XCDM, and $\phi$CDM-RP models from GRB + $H(z)$ + BAO data take small changes in parameter constraints compared to the constraints from the $H(z)$ + BAO data. The constraints from the GRBs only are more stringent than those from the $H(z)$ + BAO dataset but are less precise [205].

16. Constraints on the parameters of the $\phi$CDM-RP, the XCDM, and the $\Lambda$CDM models from the $H_{II}$G are largely consistent but not as restrictive as those derived from the measurements of the BAO + SNe Ia + CMB temperature anisotropy [380].

17. Subsets of full QSO data, limited by redshift $z \leq 1.5$–1.7, obey the $L_X - L_{UV}$ relation in a way that is independent of the cosmological model (from the spatially flat and non-flat ΛCDM, XCDM, and $\phi$CDM-RP models) and can therefore be used to constrain the cosmological parameters. Constraints from these smaller subsets of lower redshift QSO data are generally consistent but much weaker than those inferred from the Hubble parameter $H(z)$+ BAO measurements [198].
18. WMAP + BAO + galaxy cluster gas mass fraction measurements give consistent and more accurate constraints on the parameters of the spatially flat $\phi$CDM model than those derived from other data, wherein, constraints on the parameter $\alpha$, $\alpha < 3.5$ [325].
19. Future measurements of the LSS growth rate in the near future will be able to constrain the spatially flat $\phi$CDM-RP models with an accuracy of about 10%, considering the fiducial spatially flat ΛCDM model, an improvement of almost an order of magnitude compared to those from currently available datasets. Constraints on the growth index parameter $\gamma$ are more restrictive in the ΛCDM model than in other models. In the $\phi$CDM model, constraints on the growth index parameter $\gamma$ are about a third tighter than in the $w$CDM and XCDM models [304].

- Model-independent estimate of the Hubble constant $H_0$ and matter density parameter at the present epoch $\Omega_{m0}$
    1. Constraints on the parameters of the spatially flat and non-flat ΛCDM, XCDM, and $\phi$CDM-RP models from the $H(z)$+BAO + SNe Ia (Pantheon, DES, QSO) + $H_{II}$G data provides almost model-independent estimates of the Hubble constant, the matter density parameter at the present epoch, and the characteristic linear size, respectively, as $H_0 = 69.7 \pm 1.2$ km s$^{-1}$Mpc$^{-1}$, $\Omega_{m0} = 0.295 \pm 0.021$, and $l_m = 10.93 \pm 0.25$ pc. [382].
    2. The model-independent value of the Hubble constant $H_0 = 69.7 \pm 1.2$ km s$^{-1}$Mpc$^{-1}$ and the parameter $\Omega_{m0} = 0.295 \pm 0.017$ were obtained by using the $H(z)$ + BAO + SNe Ia + QSO-AS + $H_{II}$G + $Mg_{II}$ QSO + A118 (HzBSNQHMA) data in the spatially flat and non-flat ΛCDM, XCDM, and $\phi$CDM-RP models [207].
    3. An analysis of all $H_{II}$ starburst galaxy apparent magnitude + QSO only(or) + BAO datasets in the spatially flat and non-flat ΛCDM, XCDM, and $\phi$CDM-RP models leads to the relatively model-independent and restrictive estimates for the values of the parameter $\Omega_{m0}$ and the Hubble constant $H_0$. Depending on the cosmological model, these estimates are consistent with a lower value of $H_0$ in the range of a $2.0\sigma$ to $3.4\sigma$ confidence level [346].

- Problems with QSO available data
    1. In constraints on the parameters of the spatially flat and spatially non-flat ΛCDM, XCDM, and $\phi$CDM-RP models using $Mg_{II}$ QSOs data, two- and three-parameter radius–luminosity $R - L$ relations do not depend on the assumed cosmological model; therefore, they can be used to standardize QSO data. The authors found for the two-parameter $R - L$ relation that the data subsets with low-$\Re_{FeII}$ and high-$\Re_{FeII}$ obey the same $R - L$ relation within the error bars. Extending the two-parameter $R - L$ relation to three parameters does not lead to the expected decrease in the intrinsic variance of the $R - L$ relation. None of the three-parameter $R - L$ relations provides a significantly better measurement fit than the two-parameter $R - L$ relation. The results obtained differ significantly from those found by Khadka et al. [201] from analysis of reverberation-measured $H\beta$ QSOs [361].
    2. Constraints on the parameters of the spatially flat and non-flat ΛCDM, XCDM, and $\phi$CDM-RP models for the full QSO dataset; parameters of the X-ray and UV luminosity $L_X - L_{UV}$ relation used to standardize these QSO data depends on the

cosmological model and therefore cannot be used to constrain the cosmological parameters in these models [198].

3. A compilation of the QSO X-ray and UV flux measurements [355] includes the QSO data that appear not to be standardized via the X-ray luminosity and the UV luminosity $L_X - L_{UV}$ relation parameters that are dependent on both the cosmological model and the redshift, so it should not be used to constrain the model parameters [200].

## 5. Ongoing and Upcoming Cosmological Missions

We look forward to more precise various cosmological data which will be obtained by the ongoing and upcoming cosmological missions (a list of which are presented below): GRB, X-ray temperature of massive galaxy clusters, the expansion rate of the universe $H(z)$, the BAO peak length scale, the CMB radiation, the angular diameter distance, and the LSS growth rate of matter density fluctuations. We hope that these observational data will allow cosmologists to obtain tighter constraints on the bounds of cosmological parameters in the dynamical dark energy models than have been obtained to date.

To study GRB and X-ray for investigating the early universe, the following are scheduled to launch: the Space-based multiband astronomical Variable Objects Monitor (SVOM) mission [403] (will be access on 14 January 2024), the Transient High-Energy Sky and Early Universe Surveyor (THESEUS) mission [404] (will be access on 2037), the Hydrogen Intensity and Real-time Analysis Experiment (HIRAX) (https://hirax.ukzn.ac.za) (will be accessed on 2024). To investigate the weak lensing survey BAO peak length scale and RSD, it is planned to launch the following: the 4-metre Multi-Object Spectroscopic Telescope (4MOST) (https://www.eso.org/sci/facilities/develop/instruments/4MOST.html) (accessed on 2023) and the BAO from the Integrated Neutral Gas Observations (BINGO) radio telescope (https://bingotelescope.org) (currently under construction). To investigate the acceleration expansion of the universe, the nature of the dark universe, the dynamics and evolution of the universe, and the growth of LSS in the universe, the Euclidean Space Telescope (Euclid) (https://www.euclid-ec.org/) (will be accessed on 1 July 2023; the following are scheduled to launch: the Spectro-Photometer for the History of the universe (SPHEREx) (https://spherex) (will be accessed on April 2025 ), the Nancy Grace Roman Space Telescope (https://www.jpl.nasa.gov/missions/the-nancy-grace-roman-space-telescope) (will be accessed on May 2027), the ArmazoNes high Dispersion Echelle Spectrograph (ANDES) (https://elt.eso.org/instrument/ANDES/) (will be accessed on 2030), the Extremely Large Telescope (ELT) (https://elt.eso.org) (will be accessed on 2028), the Rubin Observatory Legacy Survey of Space and Time (Rubin/LSST) (https://lsst.org) (will be access on 2025), and the Simons Observatory (SO) (https://simonsobservatory.org) (will be accessed on 2024). To explore CMB radiation, it is planned to launch the following: the Cosmic Microwave Background—Stage IV (CMB-S4) experiment (https://cmb-s4.org (will be accessed on 2029)) and the Lite Satellite for the studies of B-mode polarization and Inflation from Cosmic Background Radiation Detection (LiteBIRD) (https://litebird.html (will be accessed on 2032); to study the fingerprint of primordial gravitational waves in CMB, the SPIDER experiment (https://spider.princeton.edu (will be accessed on December 2024)).

## 6. Conclusions

In this review we analyzed and summarized the current research effort to constrain the parameters of the dynamical dark energy models through cosmological observations. Our review does not claim to be a complete presentation of all research conducted by scientists in this field. We did our best to account for the results of the papers that seem to us the most relevant, among numerous ones, where the authors applied different types of methods and observational constraints with various observational datasets to investigate the dynamical dark energy models. The main results of the papers considered in this review are summarized and grouped in Section 4 into the following topics: results of constraints on the parameters of the dynamical dark energy models, alleviation and resolving of the ΛCDM model tensions, data preferences, disadvantages of models to data, failure and incompatibility of data, sensibility of various data, consistency and comparisons of the constraint results with various data, model-independent estimate of the Hubble constant $H_0$ and the matter density parameter at the present epoch $\Omega_{m0}$, and problems with the QSO available data.

In most of papers presented in this review the quintessence scalar field $\phi$CDM model with the inverse power-law RP potential has been studied. This model is the simplest one, and it is a typical representative of the large family of the tracker quintessence scalar field models. It grasps the general properties of the tracker models. Due to its simplicity, this model has attracted and will continue to attract the community in its effort to study it.

This review is a kind of historical cross-section of the study of dynamical dark energy models, in which it is clearly seen that the complication, refinement, and increase in the diversity of cosmological data and methods for studying dynamical dark energy models lead to more precise constraints on the values of cosmological parameters. According to the results of the papers considered in this review, the accuracy of the available cosmological data is insufficient to tighten the bounds on the parameters of the dynamical dark energy models and more precise data are required (the main ongoing and upcoming cosmological experiments are presented in Section 5). At the same time, despite the refinement of observational data and complications of methods for studying dark energy in the universe, current observational data favor the standard spatially flat ΛCDM model, while not excluding dynamical dark energy models and spatially closed hyperspaces.

Since the current observational constraints seem to leave room for $\phi$CDM as a viable alternative to the true minimalist ΛCDM model, it is appropriate to point out some salient features which can naturally complement the big picture, if the idea that a scalar field is responsible for the observed dark energy is taken seriously.

One of the important problems to be addressed is the *coincidence problem*—at the present epoch, the energy scales of dark matter and dark energy are of comparable magnitudes, and these are again comparable to that of the cosmological neutrinos.

A proposal for the explanation of the neutrino mass due to coupling through the dark energy field (i.e., beyond the Higgs mechanism of the Standard Model) was put forward by Fardon, Nelson, and Weiner [405], followed by Peccei [406]. These works predict a time-dependent mass for dark matter particles and neutrinos. It has been shown in the following work [240,241] that it is possible to simultaneously obtain the values of the late-time evolution of the universe ($w \to -1$, $V(\phi) \sim M^4 \sim \Lambda$), asymptotically close to that of the ΛCDM model, as well as the dark-energy-generated neutrino masses consistent with the observations for several dark energy potentials coupled to the fermionic field.

The models analyzed in [240,241] are toy models since only a single fermionic species coupled to the scalar field is considered. An important and promising direction of the future work, aiming to reveal the origins of the dark sector of the universe and the neutrino mass generation, appears to be an approach unifying the scalar field(s) coupled to the matter fields of the Standard Model. Various proposals to incorporate the dark matter component of the universe along with dark energy and neutrinos within a common framework were put forward [407–409]. More detailed quantitative analysis needs to be done to fully explore the observable effects predicted by these models with multiple fermionic flavors and different DE potentials on the expansion history of the universe.

It has been pointed out many times in the literature that too many scalar fields are not very natural. There have been efforts to relate, e.g., the Higgs field to inflation [410,411]. A quite plausible conjecture is that the inflation and quintessence represent the same physical field analyzed at different regimes of the universe evolution [412]. Thus, the objective is to advance such unifying theories back in time to incorporate the inflation.

**Funding:** O.A. acknowledges partial support from the Shota Rustaveli Georgian NSF grant YS-22-998. T.K. acknowledges partial support from the NASA ATP award 80NSSC22K0825.

**Data Availability Statement:** No new data are used.

**Acknowledgments:** We greatly appreciate helpful comments and discussions of some of the issues and research findings with Bharat Ratra and Chan-Gyung Park.

**Conflicts of Interest:** The authors declare no conflict of interest.

## Abbreviations

| Abbreviation | Full Form |
| --- | --- |
| ACBAR | Arcminute Cosmology Bolometer Array Receiver |
| AIC | Akaike Information Criterion |
| BIC | Bayesian Information Criterion |
| BOSS | Baryon Oscillation Spectroscopic Survey |
| BAO | Baryon Acoustic Oscillations |
| BBN | Big Bang Nucleosynthesis |
| GRB | Gamma-ray Bursts |
| CBI | Cosmic Background Imager |
| CDM | Cold Dark Matter |
| CLASS | Cosmic Linear Anisotropy Solving System |
| COBE-DMR | Cosmic Background Explorer—Differential Microwave Radiometers |
| CPL | Chevallier–Polarsky–Linder |
| CMB | Cosmic Microwave Background Radiation |
| DES | Dark Energy Survey |
| DESI | Dark Energy Spectroscopic Instrument |
| DIC | Deviation Information Criterion |
| EoS | Equation of State |
| Euclid | Euclidean Space Telescope |
| FRII | Fanaroff–Riley Type II |
| FLRW | Friedmann–Lemaître–Robertson–Walker |
| HDM | Hot Dark Matter |
| HST | Hubble Space Telescope |
| HzBSNQHMA | $H(z)$ + BAO + SNe Ia + QSO-AS + $H_{II}$G + $Mg_{II}$ QSO + A118 |
| $H_{II}$S | $H_{II}$ Starburst Galaxy |
| IDE | Interacting Dark Energy |
| ISW | Integrated Sachs–Wolfe |
| JBD | Jordan–Brans–Dicke |
| JLA | Joint Light-Curve Analysis |
| LAE | Lyman-$\alpha$ Emitter |
| LSS | Large-Scale Structure |
| MCMC | Markov Chain Monte Carlo |
| MGS | Main Galaxy Sample |
| pNGb | Pseudo-Nambu–Goldstone Boson |
| QFT | Quantum Field Theory |
| QSO | Quasar |
| PR4 | Last Planck Data Release |
| RP | Ratra–Peebles |
| rms | root mean square |
| RSD | Redshift Space Distortion |
| RVM | Running Vacuum Model |
| SDSS | Sloan Digital Sky Survey |
| SNe Ia | Supernovae Ia |
| SNLS | Supernova Legacy Survey |
| SPTpol | South Pole Telescope Polarization |
| SVJ$H(z)$ data | Simon, Verde, and Jimenez $H(z)$ Data |
| UV | Ultraviolet |
| VSA | Very Small Array |
| $w$CDM | $w$ Cold Dark Matter |
| WMAP | Wilkinson Microwave Anisotropy Probe |
| WDM | Warm Dark Matter |
| WFIRST | Wide-Field Infrared Survey Telescope |
| XCDM | X Cold Dark Matter |
| $\Lambda$CDM | Lambda Cold Dark Matter |
| $\phi$CDM | Phi Cold Dark Matter |
| $o$CDM | $\Lambda$CDM Extension to Non-Flat Hypersurfaces |
| 2dFGRS | Two-Degree Field Galaxy Redshift Survey |
| 6dFGS | Six-Degree Field Galaxy Survey |

## Notes

1. For the latter model, the first Friedmann's equation and the Klein–Gordon scalar field equation for these models are defined, respectively, as

$$H^2(a) = \frac{1}{3F}\left(a^2\rho_{\text{fluid}} + \frac{1}{2}\dot{\phi}^2 + a^2 V(\phi) - 3H(a)\dot{F}(\phi)\right), \tag{49}$$

$$\ddot{\phi} + 2H\dot{\phi} = \frac{a^2}{2}F'(\phi)R - a^2 V'(a), \tag{50}$$

 where the dot now denotes the derivative with respect to the conformal time and the prime denotes a derivative with respect to the scalar field $\phi$. The quantity $\rho_{\text{fluid}}$ is the energy density associated with all components of the universe except for the quintessential scalar field, and the function $F(\phi)$ defines the non-minimal coupling between gravity and the scalar field $\phi$, with the form [244]

$$F(\phi) = 1/8\pi G + \tilde{F}(\phi) - \tilde{F}(\phi_0), \tag{51}$$

 where $\tilde{F}(\phi) = \vartheta \phi^2$, $\vartheta$ is a dimensionless constant, and $\phi_0$ is a value of the scalar field at the present epoch.

2. The observational constraints on a projection of the Integrated Sachs–Wolfe (ISW) effect on the CMB temperature anisotropy was obtained for a fixed value of the Jordan–Brans–Dicke (JBD) parameter at the present epoch $\omega_{JBD0}$, the latter being defined as

$$\omega_{\text{JBD}} = F\left(\frac{dF}{d\phi}\right)^{-2} = \frac{8\pi}{\vartheta^2}\exp\left[-\frac{\vartheta(\phi-\phi_0)}{M_{\text{pl}}}\right], \quad \omega_{\text{JBD0}} = 8\pi/\vartheta^2, \tag{52}$$

 where $\xi$ is a dimensionless coupling, $\phi_0$ is the present value for the scalar field, and $F = \frac{1}{16\pi G}\exp\left(\frac{\vartheta}{M_{\text{pl}}}(\phi-\phi_0)\right)$ is a generalized function of the gravity term $R/16\pi G$.

3. This is performed by fixing at the present epoch the amplitude of the initial energy density fluctuations generated in the early inflation epoch for this model and comparing the model predictions of the large angular scale spatial anisotropy in the CMB radiation with observational data. The authors computed model predictions as a function of the model parameter $\alpha$, as well as other cosmological parameters, following Brax et al. [413], and then determined the normalization amplitude by comparing these predictions with the COBE-DMR CMB temperature anisotropy measurements of Bennett [5] and Gorski et al. [414].